%% file: merge.tex
\documentclass[12pt]{article}
\usepackage{epsf}
\usepackage{graphicx}
\usepackage{dcolumn}
\usepackage{bm}
\usepackage{amsmath}
\usepackage{amssymb}
\usepackage{subfigure}

\newlength{\dinwidth}
\newlength{\dinmargin}
\newcommand{\beq}{\begin{equation}}
\newcommand{\eeq}{\end{equation}}
\newcommand{\bea}{\begin{eqnarray}}
\newcommand{\eea}{\end{eqnarray}}
\newcommand{\bdm}{\begin{displaymath}}
\newcommand{\edm}{\end{displaymath}}

\newcommand{\be}{\begin{equation}}
\newcommand{\ee}{\end{equation}}
\newcommand{\bit}{\begin{itemize}}
\newcommand{\eit}{\end{itemize}}
\newcommand{\ben}{\begin{enumerate}}
\newcommand{\een}{\end{enumerate}}

\def\NIMA{{\em Nucl. Instrum. Methods} A}

\def\PRL{\em Phys. Rev. Lett.}
\def\PRD{{\em Phys. Rev.} D}
\def\ZPC{{\em Z. Phys.} C}

\def\as{\alpha_s}

\def\m{{\cal M}}

\def\ord{{\cal O}}

\def\dy{\Delta y}

\def\lapproxeq{\lower .7ex\hbox{$\;\stackrel{\textstyle                                                    
<}{\sim}\;$}}                                                    
\def\gapproxeq{\lower .7ex\hbox{$\;\stackrel{\textstyle                                                    
>}{\sim}\;$}}                                                    
\def\be{\begin{equation}}                                                    
\def\ee{\end{equation}}                                                    
\def\bea{\begin{eqnarray}}                                                    
\def\eea{\end{eqnarray}}

\begin{document}                                                    
\titlepage                                                    
\begin{flushright}                                                    
\today \\                                                    
\end{flushright}       

\input{frontpage.tex}\newpage
\input{u.tex}\newpage
\input{marzaniUCLproc.tex}\newpage
\input{london09_proc.tex}\newpage
\input{watt_smlhc.tex}\newpage
\input{LamiLondon09.tex}\newpage
\input{Pilkington-Andrew-LondonSM2009.tex}\newpage
\input{white_chris.tex}\newpage
\input{sm_ucl.tex}\newpage
\input{dasgupta.tex}

\end{document}

%% file: frontpage.tex
\begin{center}
{\Large \bf Proceedings of the workshop}\\
{\Large \bf ``Standard Model at the LHC''} \\
{\Large \bf University College London}\\
{\Large \bf 30 March - 1 April 2009}
\vspace*{1cm}

M.Campanelli$^{a,*}$,
M.Dasgupta$^b$,
Y.Delenda$^c$,
J.Forshaw$^b$,
D.Kar$^d$ ,
J.Keats$^b$,
V.A. Khoze$^{e,f}$,
S.Lami$^g$,
A.D.Martin$^e$,
S.Marzani$^b$,
A.Pilkington$^b$,
M.G. Ryskin$^{e,f}$,
S.Sapeta$^h$,
G.Watt$^e$,
C.White$^i$ \\

\vspace*{0.5cm}
$^a$ University College London Gower Street  London UK, WC1E 6BT.\\
$^b$ School of Physics and Astronomy, University of Manchester, Oxford Road, Manchester, UK, M13 9PL.\\
$^c$ D\'epartement de Physique, Facult\'e des Sciences Universit\'e de Batna, Algeria.\\
$^d$ Inst. fuer Kern- und Teilchenphysik (IKTP) Technische Universitaet Dresden Germany\\
$^e$ Institute for Particle Physics Phenomenology, University of Durham, Durham, DH1 3LE\\
$^f$ Petersburg Nuclear Physics Institute, Gatchina, St.~Petersburg, 188300, Russia\\     
$^g$ INFN Pisa, Largo Pontecorvo, 3 - 56127 Pisa, Italy\\
$^h$ LPTHE, UPMC -- Paris 6, CNRS UMR 7589, Paris, France\\
$^i$ Nikhef, Science Park 105, 1098 XG Amsterdam, The Netherlands\\
$^*$ editor
\vspace*{1cm}
\end{center}

%% file: u.tex
\begin{center}                                                    
{\Large \bf Soft physics and exclusive Higgs production at the LHC\footnote{Based on a talk by Alan Martin at the UCL Workshop on ``Standard Model discoveries with early LHC data", 30 March - 1 April, 2009}}                                                                                                        
                                                    
\vspace*{1cm}                                                    
A.D. Martin$^a$, M.G. Ryskin$^{a,b}$ and V.A. Khoze$^{a,b}$ \\                                                    
                                                   
\vspace*{0.5cm}                                                    
$^a$ Institute for Particle Physics Phenomenology, University of Durham, Durham, DH1 3LE \\                                                   
$^b$ Petersburg Nuclear Physics Institute, Gatchina, St.~Petersburg, 188300, Russia            
\end{center}                                                    
                                                    
\vspace*{1cm}                                                    
                                                    
\begin{abstract}
We discuss two inter-related topics: a multi-component $s$- and $t$-channel model of `soft' high-energy $pp$ interactions and the properties of the exclusive Higgs signal at the LHC.
\end{abstract}    

\section{Introduction}

We begin by drawing attention to the exciting possibility of studying the Higgs sector via the exclusive process $pp \to p+H+p$ at the LHC, where the $+$ signs denote the presence of large rapidity gaps. The prediction of the event rate of such a process depends on an interesting mixture of `soft' and `hard' physics. We explain why the former requires the development of a multi-component $s$- and $t$-channel model of high-energy `soft' processes, in which absorptive effects play a key role. We describe how the model may be used to estimate the survival probability of the large rapidity gaps to eikonal and enhanced soft rescattering. We comment on other models used to calculate the survival factors.

We note that CDF experiments at the Tevatron have already measured the rate of similar exclusive processes, namely $pp \to p+A+p$ where $A=\gamma\gamma$ or dijet or $\chi_c$ \cite{albrow}. These processes are driven by the same theoretical mechanism used to estimate the exclusive Higgs signal. The agreement of the CDF experimental rates with the model predictions leads to optimism of the use of very forward proton taggers to explore the Higgs sector at the LHC \cite{FP420}.

The discussion here is brief, with a minimum of references. More details, and references, can be found in two recent reviews covering the same material \cite{KMR1,KMR2}.

\section{Advantages of the exclusive Higgs signal with $H\to b\bar{b}$}

The exclusive process $pp\to p+H+p$ for the production of a Higgs at the LHC with mass $M_H \lapproxeq 140$ GeV, where the dominant decay mode is $H \to b\bar{b}$, has the following advantages:
\begin{itemize}
\item
The mass of the Higgs boson (and in some cases the width) can be measured
with high accuracy (with mass resolution $\sigma(M)\sim 1$ GeV) by measuring the
missing mass to the forward outgoing protons, {\it provided} that they can be accurately tagged some 400 m from the interaction point.
\item
It offers a unique chance to study $H \to b\bar{b}$, since the leading order $b\bar b$
  QCD background is suppressed  by the $P$-even $J_z=0$ selection
rule, where the $z$ axis is along the direction of the proton beam.
Indeed, at LO, this background vanishes in the limit of massless $b$ quarks and forward outgoing protons.
 Moreover, a measurement of the mass of the decay products must match the `missing mass' measurement.
For a SM Higgs the signal-to-background ratio $S/B \sim O(1)$ 
\item
The quantum numbers of the central object (in particular, the
$C$- and $P$-parities) can be analysed by studying the azimuthal angle
distribution of the tagged protons. Due to the selection
rules, the production of $0^{++}$ states is strongly favoured.
\item
There is a clean environment for the
exclusive process --- this is even possible with overlapping interactions (pile-up) using fast timing detectors with very good resolution: 10 ps or better.  
\item
For SUSY Higgs there are regions of SUSY parameter space were the
signal is enhanced by a factor of 10 or more, while the background remains unaltered. Moreover,
there are domains of parameter space where Higgs boson production via the conventional inclusive processes
 is suppressed whereas the exclusive signal is
enhanced, and even such, that both the $h$ and $H$ $0^{++}$ bosons may be detected.   
\end{itemize}

\section{Is the exclusive Higgs cross section large enough?}

\begin{figure} [t]
\begin{center}
\includegraphics[height=4cm]{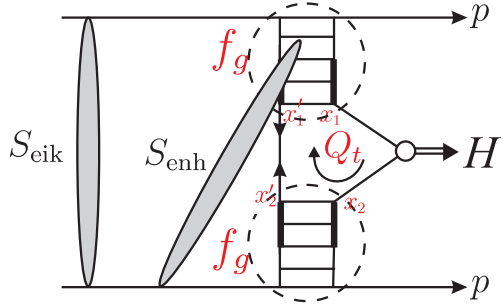}
\caption{The mechanism for the exclusive process $pp \to p+H+p$, with the eikonal and enhanced survival factors shown symbolically. The thick lines on the Pomeron ladders, either side of the subprocess ($gg \to H$), indicate the rapidity interval $\Delta y$ where enhanced absorption is not permitted, see Section \ref{sec:sigma}. }
\label{fig:pAp}
\end{center}
\end{figure}

What is the price that we pay for the large rapidity gaps? How do we calculate the cross section for the exclusive process $pp\to p+H+p$?
The calculation of the exclusive production of a heavy system is an interesting mixture of {\it soft} and {\it hard} QCD effects.  
The basic mechanism is
shown in Fig.~\ref{fig:pAp}. The $t$-integrated cross section is of the form
\begin{equation}
\sigma ~\simeq ~\frac{S^2}{B^2} ~\left|~N\int\frac{dQ^2_t}{Q^4_t}\: f_g(x_1, x_1', Q_t^2, \mu^2)f_g(x_2,x_2',Q_t^2,\mu^2)~\right|^2, 
\label{eq:M}
\end{equation}
where $B/2$ is the $t$-slope of the proton-Pomeron vertex, and $N$ is given in terms of the $H\to gg$ decay width.
The probability amplitudes, $f_g$, to find the appropriate pairs of $t$-channel gluons $(x_1,x'_1)$ and $(x_2,x'_2)$, are given by the skewed unintegrated gluon densities at a hard scale $\mu \sim M_H/2$. Since $(x'\sim Q_t/\sqrt s)\ll (x\sim M_H/\sqrt s)\ll 1$, it is possible
to express $f_g(x,x',Q_t^2,\mu^2)$, to single log accuracy, in
terms of the conventional integrated density $g(x)$, together with a known Sudakov suppression factor $T$, which ensures that the active gluons do not radiate in the
evolution from $Q_t$ up to the hard scale $\mu \sim M_H/2$, and
so preserves the rapidity gaps. The factor $T$ ensures that the integral is infrared stable, and may be calculated by perturbative QCD.

If we were to neglect the rapidity gaps survival factor, $S^2$ in (\ref{eq:M}), then QCD predicts that the exclusive cross for producing a SM Higgs of mass 120 GeV would be more than 100 fb at an LHC energy of $\sqrt{s}=$14 TeV.

The factor $S^2$ in (\ref{eq:M}) is the probability that the secondaries, which are produced by soft rescattering, do not populate the rapidity gaps.  As written, the cross section assumes soft-hard factorization. In other words, the survival factor, denoted by $S_{\rm eik}$ in Fig.~\ref{fig:pAp}, and calculated from an eikonal model of soft interactions, does not depend on the structure of the perturbative QCD amplitude embraced by the modulus signs in (\ref{eq:M}). Actually the situation is more complicated. There is the possibility of enhanced rescattering which involves intermediate partons, and which breaks soft-hard factorization. To calculate the corresponding survival factors,  $S_{\rm eik}$ and $S_{\rm enh}$, we need, first, a model for soft high-energy $pp$ interactions. We come back to the estimation of $\sigma(pp\to p+H+p)$ in Section \ref{sec:sigma}.

\section{Requirements of a model of soft interactions} 

Besides the need for calculating the rapidity gap survival factors, it is valuable to revisit the `soft' domain at this time because of the intrinsic interest in obtaining a reliable self-consistent model of high-energy soft interactions which may soon be illuminated by data from the LHC. Moreover, we need a reliable model so as to be able to predict the gross features of soft interactions; in particular to understand the structure of the underlying events at the LHC.

What are the requirements of such a high-energy model? It should be self-consistent theoretically -- it should satisfy unitarity; absorptive corrections are large and imply the importance of multi-Pomeron contributions. The model should describe all the available soft data in the CERN-ISR to Tevatron energy range. Finally, the model should include Pomeron components of different size so that we can allow for the effects of soft-hard factorization breaking.

\begin{figure} [t]
\begin{center}
\includegraphics[height=8cm]{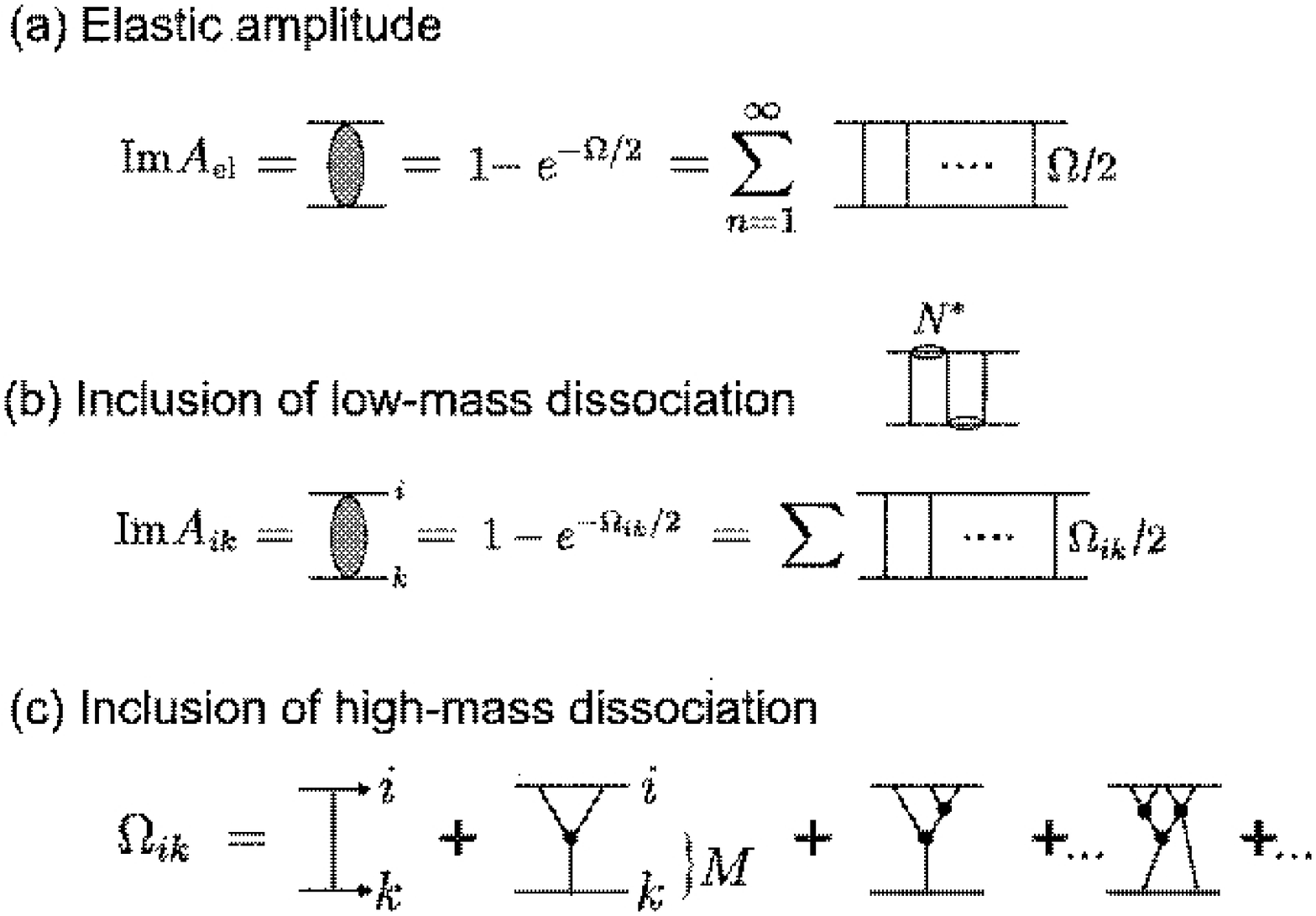}
\caption{(a) The single-channel eikonal description of elastic scattering; (b) the multichannel eikonal formula which allows for low-mass proton dissociations in terms of diffractive eigenstates $|\phi_i\rangle,~|\phi_k\rangle$; and (c) the inclusion of the multi-Pomeron-Pomeron diagrams which allow for high-mass dissociation.  }
\label{fig:epip}
\end{center}
\end{figure}
The total and elastic proton-proton cross sections
  are usually described in terms of an eikonal model, which automatically
satisfies  $s$-channel
elastic unitarity. The unitarity relation is diagonal in impact parameter $b$, and so these reactions can be described in terms of the opacity $\Omega(s,b) \ge 0$
\be
d\sigma_{\rm tot}/d^2b=2(1-e^{-\Omega /2}),~~~~~~~~d\sigma_{\rm el}/d^2b=(1-e^{-\Omega /2})^2,
\ee
see Fig. \ref{fig:epip}(a). The Good-Walker
formalism \cite{GW} is used to account for the possibility of excitation of the initial proton, 
that is for two-particle intermediate states with the
proton replaced by $N^*$ resonances, Fig. \ref{fig:epip}(b). Diffractive eigenstates $|\phi_i\rangle$ are introduced which only undergo `elastic' scattering. That is, we go from a single elastic channel to a multi-channel eikonal, $\Omega_{ik}$.  Already at Tevatron energies the absorptive
correction to the elastic amplitude, due to elastic eikonal
rescattering, gives about a 20\% reduction of
simple one-Pomeron exchange. After accounting for low-mass proton
excitations, the correction becomes twice
larger (that is, up to a 40\% reduction). 

At first sight, by enlarging the number of eigenstates $|\phi_i\rangle$ it seems we may even allow for high-mass proton dissociation. However, here, we face the problem of double counting when the partons originating from dissociation of the beam and `target' initial protons overlap in rapidities. For this reason, high-mass ($M$) dissociation is usually described by ``enhanced'' multi-Pomeron diagrams. The first, and simplest, such contribution to single proton dissociation $d\sigma_{\rm SD}/dM^2$, is the triple-Pomeron graph, see Fig. \ref{fig:epip}(c). The absorptive effects
in the triple-Regge domain are expected to be quite large ($\lapproxeq$80$\%$), since there is an extra factor of 2  from the AGK cutting rules \cite{AGK}. Recent triple-Regge analyses \cite{LKMR}, which include screening effects, of the available data find that the {\it bare} triple-Pomeron coupling is indeed much larger than the ({\rm effective}) value found in the original (unscreened) analyses. This can be anticipated by simply noting that since the original
triple-Regge analyses did not include
absorptive corrections, the resulting triple-Regge couplings must be
regarded, not as bare vertices, but as effective couplings
embodying the absorptive effects. That is,
\be
g_{3P}^{\rm effective}~\simeq~S^2~g_{3P}^{\rm bare},
\ee
where $S^2$ is the survival probability of the rapidity gap. Due to the large bare triple-Pomeron coupling ($g_{3P}=\lambda g_N$ with $\lambda \simeq 0.25$, where $g_N$ is the Pomeron-proton coupling), we need a model of soft high-energy processes which includes multi-Pomeron interactions, see, for example, the final diagrams in Fig. \ref{fig:epip}(c).

\begin{figure}
\begin{center}
\includegraphics[height=3cm]{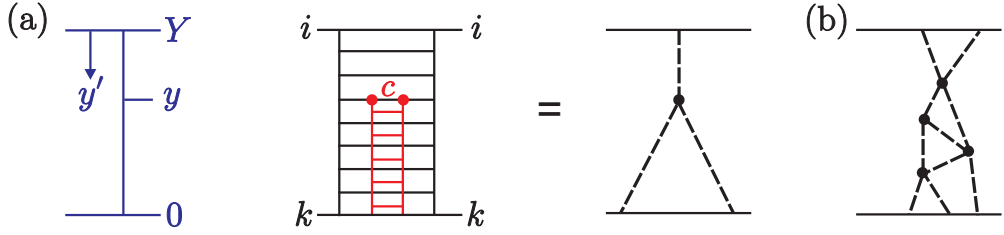}
\caption{(a) The ladder structure of the triple-Pomeron amplitude between diffractive eigenstates $|\phi_i\rangle,|\phi_k\rangle$ of the proton; the rapidity $y$ spans an interval 0 to $Y={\rm ln}s$. (b) A multi-Pomeron diagram.} 
\label{fig:2lad}
\end{center}
\end{figure}

\section{Multi-component $s$- and $t$-channel model}

Here we follow a {\it partonic} approach to obtain a model high-energy soft interactions \cite{KMRnns}.
While the eikonal formalism describes the rescattering of the incoming
fast particles,  the enhanced multi-Pomeron diagrams
represent the rescattering of the intermediate partons in the ladder
(Feynman diagram) which describes the Pomeron-exchange amplitude.
We refer to Fig.~\ref{fig:2lad}. The multi-Pomeron effects are included by the following equation describing the evolution in rapidity $y$ of the opacity $\Omega_k$ starting from the `target' diffractive eigenstate $|\phi_k\rangle$:
\begin{equation}
\frac{d\Omega_k(y,b)}{dy}\,=\,e^{-\lambda\Omega_i(y',b)/2}~~~e^{-\lambda
\Omega_k(y,b)/2}~~
\left(\Delta+\alpha'\frac{d^2}{d^2b}\right)\Omega_k(y,b)\; ,
\label{eq:evol1}
\end{equation}
where $y'=\ln s -y$.   Let us explain the meanings of the three factors on the right-hand-side of (\ref{eq:evol1}). If only the last factor, (...)$\Omega_k$, is present then the evolution generates the ladder-type structure of the bare Pomeron exchange amplitude, where the Pomeron trajectory $\alpha_P=1+\Delta+\alpha't$. The inclusion of the preceding factor allows for rescatterings of an intermediate parton $c$ with the ``target'' proton $k$; Fig.~\ref{fig:2lad}(a) shows the simplest (single) rescattering which generates the triple-Pomeron diagram. Finally, the first factor allows for rescatterings with the beam $i$. In this way the absorptive effects generated by all multi-Pomeron diagrams are included, like the one shown in Fig.~\ref{fig:2lad}(b). 
There is an analogous equation for the evolution in rapidity $y'$ of $\Omega_i(y',b)$ starting from the `beam' diffractive eigenstate $|\phi_i\rangle$. The two equations may be solved iteratively.

As we are dealing with elastic {\it amplitudes} we use $e^{-\lambda\Omega/2}$ and not $e^{-\lambda\Omega}$. The coefficient $\lambda$ in the exponents arises since  parton $c$ will have a different absorption cross section from that of the diffractive eigenstates. Naively, we may assume that the states $i,k$ contains a number $1/\lambda$ of partons.  The factors $e^{-\lambda\Omega/2}$  generate multi-Pomeron vertices of the form 
\begin{equation}
g^n_m\,=\, n~m~\lambda^{n+m-2}g_N/2\;\;\;\;\;\;\;\;\;
\mbox{for~~ $n+m\geq 3$}\, ,
\label{eq:g}
\end{equation}
where a factor $1/n!$, which comes from the expansion of the exponent, accounts for the
identity of the Pomerons. The factors $n(m)$ allow for the $n(m)$ possibilities to select the Pomeron $\Omega_i(\Omega_k)$ which enters the evolution (\ref{eq:evol1}) from the $n(m)$ identical Pomerons. In principle, the vertices $g^n_m$ are unknown. However, the above ansatz is physically motivated and is certainly better than to assume only a triple-Pomeron coupling, that is, to assume that $g^n_m=0$ for $n+m>3$. 

Even though $\lambda \simeq 0.2-0.25$, the role of factors $e^{-\lambda\Omega/2}$ is not negligible, since the suppression effect is accumulated throughout the evolution. For instance, if $\lambda \ll 1$ the full absorptive correction is given by the product $\lambda \Omega Y/2$, where the small value of $\lambda$ is compensated by the large rapidity interval $Y$.

So far, we have allowed multi-components in the $s$-channel via a multichannel eikonal. However,
a novel feature of the model of Ref.~\cite{KMRnns} is that four different $t$-channel states are included. One for the
secondary Reggeon ($R$) trajectory, and three Pomeron states ($P_1, P_2, P_3$) to mimic the BFKL
diffusion in the logarithm of parton transverse momentum,
$\ln(k_t)$. Recall that the
BFKL Pomeron is not a pole in the complex $j$-plane, but a
branch cut. Here the cut is approximated by three $t$-channel states of a different size.
The typical values of $k_t$ are $k_{t1}\sim 0.5$ GeV, $k_{t2}\sim 1.5$ GeV and $k_{t3}\sim 5$ GeV for the large-, intermediate- and small-size components of the Pomeron, respectively.
Thus (\ref{eq:evol1}) is rewritten as a four-dimensional matrix equation for $\Omega^a_k$ in $t$-channel space ($a=P_1,P_2,P_3,R$), as well as being a three-channel eikonal in diffractive eigenstate $|\phi_k\rangle$ space. The transition terms, added to the equations, which couple the different $t$-channel components, are fixed by the properties of the BFKL equation. So, in principle, we have the possibility to explore the matching of the soft Pomeron (approximated by the large-size component $P_1$) to the QCD Pomeron (approximated by the small-size component $P_3$). The key parameters which drive the evolution in rapidity are the intercepts $1+\Delta^a$ and the slopes $\alpha'_a$ of the $t$-channel exchanges.

The model is tuned to describe all the available soft data in the CERN-ISR to Tevatron energy range. In principle, it may be used to predict all features of soft interactions at the LHC.  All components of the Pomeron are taken to have a {\it bare} intercept $\Delta\equiv\alpha_P(0)-1=0.3$, consistent with resummed NLL BFKL. However, the large-size Pomeron component is heavily screened by the effect of `enhanced' multi-Pomeron diagrams, that is, by high-mass dissociation, which results in $\Delta_{\rm eff} \sim 0.08$ and $\alpha'_{\rm eff} \sim 0.25$. This leads, among other things, to the saturation of the particle multiplicity at low $p_t$, and to a slow growth of the total cross section. Indeed, the model predicts a  relatively low total cross section at the LHC -- $\,\sigma_{\rm tot}({\rm LHC})\simeq 90$ mb. On the other hand, the small-size component of the Pomeron is weakly screened, leading to an anticipated growth of the particle multiplicity at large $p_t~~(\sim 5$ GeV) at the LHC. Thus the model has the possibility to embody a smooth matching of the perturbative QCD Pomeron to the `soft' Pomeron.

\section{Long-range rapidity correlations}
We emphasize that each multi-Pomeron exchange diagram describes simultaneously a
few different processes. The famous AGK cutting rules \cite{AGK} gives the relation between the different subprocesses originating from the same  diagram. 

Note that the eikonal model predicts a
long-range correlation between the secondaries produced in different rapidity intervals. Indeed, we have possibility to cut any number of Pomerons. Cutting $n$ Pomerons we get an event with multiplicity $n$ times larger than that generated by one Pomeron. The probability to observe a particle from a diagram where $n$ Pomerons are cut is $n$ times larger than that from the diagram with only one cut Pomeron. The observation of a particle at rapidity $y_a$, say, has the effect of enlarging the relative contribution of diagrams with a larger number of cut Pomerons. For this reason the probability to observe another particle at quite a different rapidity $y_b$ becomes larger as well. This can be observed experimentally via the ratio of inclusive cross sections
\begin{equation}
R_2~=~\frac{\sigma_{\rm inel}d^2\sigma/dy_ady_b}{(d\sigma/dy_a)( d\sigma/dy_b)}-1~=~
\frac{d^2N/dy_ady_b}{(dN/dy_a)(dN/dy_b)}-1,
\label{eq:R2}
\end{equation}
where $dN/dy=(1/\sigma_{\rm inel})d\sigma/dy$ is the particle density.

Without multi-Pomeron effects the value of $R_2$ exceeds zero only when the
two particles are close to each other, that is when the separation 
$|y_a-y_b|\sim 1$ is not large. Such short-range correlations arise from resonance or jet production. However, multi-Pomeron exchange leads to a
long-range correlation, $R_2>0$, even for a large rapidity difference between the particles, $|y_a-y_b|\sim Y$.

\section{Rapidity gap survival \label{sec:survival}}

Now that we have a model of high-energy soft interactions, we can estimate the rapidity gap survival factors $S^2_{\rm eik}$ and $S^2_{\rm enh}$
of the process $pp \to p+H+p$ shown in Fig. \ref{fig:pAp}. We start with $S^2_{\rm eik}$.

\subsection{Eikonal rescattering}
The gap survival factor caused by {\it eikonal} rescattering of the
diffractive eigenstates \cite{GW}, for a fixed impact parameter ${\mathbf b}$, is
\begin{equation}  S^2_{\rm eik}({\mathbf b})~ = ~\frac{\left| {\displaystyle\sum_{i,k}} |a_{i}|^2~|a_{k}|^2~{\mathcal
M}_{ik}({\mathbf b})~\exp(-\Omega^{\rm tot}_{ik}(s,{\mathbf b})/2)\right|^2}{\left|{\displaystyle \sum_{i,k}}
|a_{i}|^2~|a_{k}|^2~{\mathcal M}_{ik}({\mathbf b}) \right |^2} \,.
\label{eq:c3pp}
\end{equation}
where $\Omega^{\rm tot}_{ik}(s,{\mathbf b})$ is the total opacity of the $ik$ interaction, and the $a_i$'s occur in the decomposition of the proton wave function  in terms of diffractive eigenstates $|p\rangle = \sum_i a_i |\phi_i \rangle$.  The total opacity has the form $\Omega^a_k(y)\Omega^a_i(y')$ integrated over the impact parameters ${\mathbf b_1},{\mathbf b_2}$ (keeping a fixed impact parameter separation ${\mathbf b}={\mathbf b_1}-{\mathbf b_2}$ between the incoming protons) and summed over the different Pomeron components $a$. Recall $y'=Y-y={\rm ln}s-y$, see Fig.~\ref{fig:2lad}.  The exact shape of the matrix element ${\cal M}_{ik}$ for the hard subprocess $gg \to H$ in ${\mathbf b}$ space and the relative couplings to the various diffractive eigenstates $i,k$ should be addressed further.

\begin{figure}
\begin{center}
\includegraphics[height=7cm]{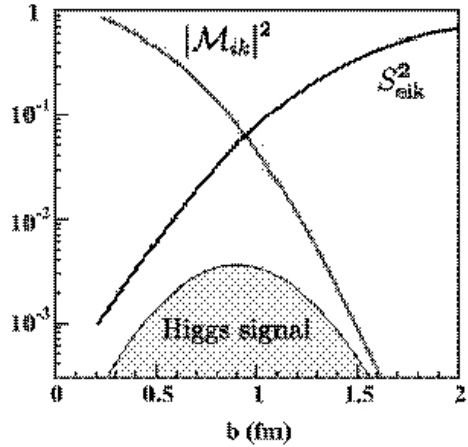}
\caption{A ``first look'' at the impact parameter dependence of the signal for 120 GeV Higgs production at the LHC after including an eikonal rescattering correction. }
\label{fig:higgs}
\end{center}
\end{figure}

One possibility is to say that the ${\mathbf b}$ dependence of ${\cal M}$
should be, more or less, the same as that observed for diffractive
$J/\psi$ electroproduction ($\gamma+p\to J/\psi+p$), and the coupling
to the $|\phi_i \rangle$ component of the proton should be proportional to the same factor 
$\gamma_i$ as in a soft interaction. This leads to
\begin{equation}
{\cal M}_{ik}\propto \gamma_i\gamma_k\exp(-b^2/4B)
\label{eq:m1}
\end{equation}
with $t$-slope $B\simeq 4$ GeV$^{-2}$.
The resulting ``first look'' predictions obtained using the `soft' model of \cite{KMRnns}, for the exclusive production of a scalar 120 GeV Higgs at the LHC, are shown in Fig.~\ref{fig:higgs}. After we integrate over $b$, we find that the survival probability of the rapidity gaps in $pp \to p+H+p$ to eikonal rescattering is $\langle S^2_{\rm eik}\rangle$=0.017, with the Higgs signal concentrated around impact parameter  $b=0.8$ fm.
Expressing the survival factors in this manner is too simplistic and even sometimes misleading, for the reasons we shall explain below; nevertheless these numbers are frequently used as a reference point. 

\subsection{Enhanced rescattering}

As indicated in Fig.~\ref{fig:pAp}, besides {\it eikonal} screening, $S_{\rm eik}$, caused by soft interactions between the protons, we must also consider so-called {\it `enhanced'} rescattering, $S_{\rm enh}$, which involves intermediate partons. Since we have to multiply the probabilities of absorption on each individual intermediate parton, the final effect is {\it enhanced} by the large multiplicity of intermediate partons. Unlike $S^2_{\rm eik}({\mathbf b})$, the enhanced survival factor $S^2_{\rm enh}({\mathbf b})$ cannot be considered simply as an overall multiplicative factor. The probability of interaction with a given intermediate parton depends on its position in configuration space; that is, on its impact parameter ${\mathbf b}$ and its momentum $k_t$. This means that $S_{\rm enh}$ simultaneously changes the distribution of the active partons which finally participate in the hard subprocess. It breaks the soft-hard factorization of (\ref{eq:M}).

Do we anticipate that $S_{\rm enh}$ will be important? Working at LO (of the collinear approximation) we would expect that effect may be neglected. Due to strong $k_t$-ordering the
transverse momenta of all the intermediate partons are very large
(i.e. the transverse size of the Pomeron is very small) and therefore the absorptive
effects are negligible. Nevertheless, this may be not true at a very low
$x$, say $x \sim 10^{-6}$, where the parton densities become close to saturation and the
small value of the absorptive cross section is compensated by the large
value of the parton density.
Indeed, the contribution of the first {\it enhanced} diagram, which
describes the absorption of an intermediate parton, was estimated in
the framework of the perturbative QCD in Ref.\cite{BBKM}. It turns out
that it could be quite large. On the other hand, such an effect does not reveal itself
experimentally. The absorptive corrections due to enhanced
screening must increase with energy. This is not observed in the
present data (see \cite{JHEP} for a more detailed discussion).
One reason is that the gap survival factor $S^2_{\rm eik}$
already absorbs almost the whole contribution
from the centre of the disk. The parton essentially only survives eikonal
rescattering on the periphery; that is, at large impact parameters $b$. On the other hand,
on the periphery, the parton density is rather small and the probability
of {\it enhanced} absorption is not large.

\subsection{Gap survival for exclusive Higgs production \label{sec:sigma}}

Now, the model of Ref.~\cite{KMRnns}, with its multi-component Pomeron, allows us to calculate the survival probability of the rapidity gaps, to {\it both} eikonal and enhanced rescattering. Recall that the evolution equations in rapidity (like (\ref{eq:evol1})) have a matrix form in $aa'$ space, where $a=1,2,3$ correspond to the large-, intermediate- and small-size components of the Pomeron.    We start the evolution from the large component $P_1$, and since the evolution equations allow for a transition from one component to another (corresponding to BFKL diffusion in ln$k_t$ space), we determine how the enhanced absorption will affect the high-$k_t$ distribution in the small-size component $P_3$, which contains the active gluon involved in forming the Higgs.
Moreover, at each step of the evolution the equations include absorptive factors of the form $e^{-\lambda(\Omega^a_k +\Omega^a_i)/2}$. By solving the equations with and without these suppression factors, we could quantify the effect of enhanced absorption. However, there are some subtle issues here. First, since we no longer have soft-hard factorization, we must first specify exactly what is included in the bare hard amplitude.

Another relevant observation is that the phenomenologically determined generalised gluon distributions are usually taken at $p_t=0$, and then the observed ``total'' cross section is calculated by integrating over $p_t$ of the recoil protons {\it assuming} an exponential behaviour $e^{-Bp_t^2}$; that is
\be
\sigma ~=~\int\frac{d\sigma}{dp_{1t}^2 dp_{2t}^2}dp_{1t}^2 dp_{2t}^2 ~=~\frac{1}{B^2}\left.\frac{d\sigma}{dp_{1t}^2 dp_{2t}^2}\right|_{p_{1t}=p_{2t}=0}~,
\ee
where
\be
\int dp^2_t~e^{-Bp_t^2}~=~1/B~=~\langle p_t^2\rangle.
\ee
However, the total soft absorptive effect changes the $p_t$ distribution in comparison to that for the bare cross section determined from perturbative QCD. Moreover, the correct $p_t$ dependence of the matrix element ${\cal M}$ of the hard $gg \to H$ subprocess does {\it not} have an exponential form. Thus the additional factor introduced by the soft interactions is not just the gap survival $S^2$, but rather $S^2\langle p^2_t \rangle^2$, where the square arises since we have to integrate over the $p_t$ distributions of {\it two} outgoing protons. Indeed in all the previous calculations the soft prefactor had the form $S^2/B^2$.  Note that, using the model of Ref.~\cite{KMRnns}, we no longer have to {\it assume} an exponential ${\mathbf b}$ behaviour of the matrix element. Now the ${\mathbf b}$ dependence of ${\cal M}({\mathbf b})$ is driven by the opacities, and so is known. Thus we present the final result in the form $S^2\langle p^2_t \rangle^2$. That is, we replace $S^2/B^2$ in (\ref{eq:M}) by $S^2\langle p^2_t \rangle^2$.  So if we wish to compare the improved treatment with previous predictions obtained assuming $B=4~{\rm GeV}^{-2}$ we need to introduce the ``renormalisation'' factor $(\langle p_t^2 \rangle B)^2$. The resulting (effective) value is denoted by $S^2_{\rm eff}$.

Before we do this, there is yet another effect that we must include. We have to allow for a threshold in rapidity
\cite{JHEP}.
 The evolution equation for $\Omega^a_k$, (\ref{eq:evol1}), and the analogous one for $\Omega^a_i$, are written in the leading ln$(1/x)$ approximation, without any rapidity threshold. The emitted parton, and correspondingly the next rescattering, is allowed to occur just after the previous step. On the other hand, it is known that a pure kinematical $t_{\rm min}$ effect suppresses the probability to produce two partons close to each other. Moreover, this $t_{\rm min}$ effect becomes especially important near the production vertex of the heavy object. It is, therefore, reasonable to introduce some threshold rapidity gap, $\Delta y$, and to compute $S^2_{\rm enh}$ only allowing for absorption outside this threshold interval, as indicated in Fig.~\ref{fig:pAp}. For exclusive Higgs boson production at the LHC, the model gives $S^2_{\rm eff}=0.004,\; 0.009$ and 0.015 
for $\Delta y=0,\; 1.5$ and 2.3 respectively \cite{RMK2}. For $\Delta y=2.3$ all the NLL BFKL
corrections may be reproduced by the threshold
effect.  

\vspace{0.2cm}
Furthermore, Ref.~\cite{RMK2} presents arguments that
\be
\langle S^2_{\rm eff} \rangle~=~0.015^{~+0.01}_{~-0.005}
\label{eq:s}
\ee
should be regarded as a {\it conservative} (lower) limit for the gap survival probability in the exclusive production of a SM Higgs boson of mass 120 GeV at the LHC energy of $\sqrt{s}=14$ TeV.  Recall that this effective value should be compared with $S^2$ obtained using the exponential slope $B=4~{\rm GeV}^{-2}$. The resulting value for the cross section is, conservatively,
\be
\sigma(pp \to p+H+p)~\simeq~2-3 ~{\rm fb},
\label{eq:x}
\ee
with an uncertainty\footnote{Besides the uncertainty arising from that on $ S^2_{\rm eff}$, the other main contribution to the error comes from that on the unintegrated gluon distributions, $f_g$, which enter to the fourth power.}
of a factor of 3 up or down.

\subsection{Comments on other estimates of $S^2$ \label{sec:other}}

A very small value $S^2_{\rm enh}=0.063$ is claimed in
\cite{GLMM}, which would translate into an {\it extremely} small value of
$S^2_{\rm eff}=0.0235 \times 0.063=0.0015$. There are many reasons why this estimate is invalid.
In this model the two-particle irreducible amplitude
depends on the impact parameter $b$ {\it only} through the form factors of the incoming
protons. The enhanced absorptive effects (which result
from the sum of the enhanced diagrams) are the same
at any value of $b$. Therefore, the enhanced screening effect does not
depend on the initial parton density at a particular impact parameter $b$,
and does not account for the fact that at the periphery of the proton, from where the main
contribution comes (after the $S_{\rm eik}$ suppression), the parton density is
much smaller than that in the centre. For
this reason the claimed value of $S^2_{\rm enh}$ is much too small. Besides
this lack of $k_t \leftrightarrow b$ correlation, the model has no diagrams
with odd powers of $g_{3P}$. For example, the lowest triple-Pomeron diagram is
{\it missing}. That is, the approach does not contain the first, and most important at the lower energies, absorptive correction. Next, recall that in a theory which contains the
triple-Pomeron coupling only, without the four-Pomeron term (and/or 
more complicated multi-Pomeron vertices), the total cross section
{\it decreases} at high energies. On the other hand, the
approximation used in \cite{GLMM} leads to 
saturation (that is, to a constant cross section) at very high energies. In other words, the approach is not valid at high energies\footnote{The model of Ref. \cite{GLMM} should contain a parameter which specifies the energy interval where the assumed approximation is valid. This has not been given.}. This means that such an approximation can only be justified in a limited energy interval; 
at very high energies it is inconsistent with asymptotics, while
at relatively low energies the first term, proportional to the first power of the
triple-Pomeron coupling $g_{3P}$, is missing.
Finally, the predictions of the model of \cite{GLMM} have not been compared to the CDF
exclusive data of Section \ref{sec:cdf}.

The values of $x\sim 10^{-6}$ relevant to the evaluation of $S^2_{\rm enh}$ at LHC energies are quite small. At leading order (LO) the gluon density increases with $1/x$. So, at first sight, it appears that this may lead to the Black Disk Regime where the enhanced absorptive corrections will cause the cross section for exclusive production to practically vanish. This is the basis of the low value of $S^2_{\rm enh}$ estimated in \cite{strik}. However, NLO global parton analyses show that at relatively low scales ($k^2_t=\mu^2\sim 2\ -\ 4$ GeV$^2$) the gluons are flat for $x<10^{-3}\, -\, 10^{-4}$ (or even decrease when $x\to 0$).
Recall that the contribution of enhanced diagrams from  larger scales
decrease as $1/k^2_t$ (see \cite{BBKM}). The anomalous dimension
$\gamma <1/2$ is not large and so the growth of the gluon density, $xg(x,\mu^2)\propto (\mu^2)^\gamma$, cannot compensate the factor $1/k^2_t$.
Therefore the whole enhanced diagram contribution should be evaluated at rather low scales where the NLO gluon is flat\footnote{Note that the flat $x$-behaviour of NLO gluons allow the justification of the inequalities in (\ref{eq:hier}) below.}
 in $x$. Thus there is no reason to expect a strong energy dependence of $S^2_{\rm enh}$.

Another observation against the extremely small gap survival factor, $S_{\rm enh}$, estimated in \cite{GLMM,strik} is that
the analysis of \cite{RMK2} shows the following hierarchy of the size of the  gap survival factor to enhanced rescattering 
\be
S^{\rm LHC}_{\rm enh}(M_H>120 ~{\rm GeV})~~>~~S^{\rm Tevatron}_{\rm enh}(\gamma\gamma;E_T>5~{\rm GeV})~~>~~S^{\rm Tevatron}_{\rm enh}(\chi_c),
\label{eq:hier}
\ee
which reflects the size of the various rapidity gaps of the different exclusive processes. The fact that $\gamma\gamma$ and $\chi_c$ events have been observed at the Tevatron confirms that there is no danger that enhanced absorption will strongly reduce the exclusive SM Higgs signal at the LHC energy.

\section{Exclusive processes at the Tevatron and the LHC \label{sec:cdf}}

In preparation for the studies of the exclusive process, $pp \to p+H+p$, at the LHC, it is important to know how reliable is the `Durham' prediction of the cross section (\ref{eq:x}), particularly as some other estimates are much lower, see Section \ref{sec:other}.

\begin{figure}
\begin{center}
\includegraphics[height=3cm]{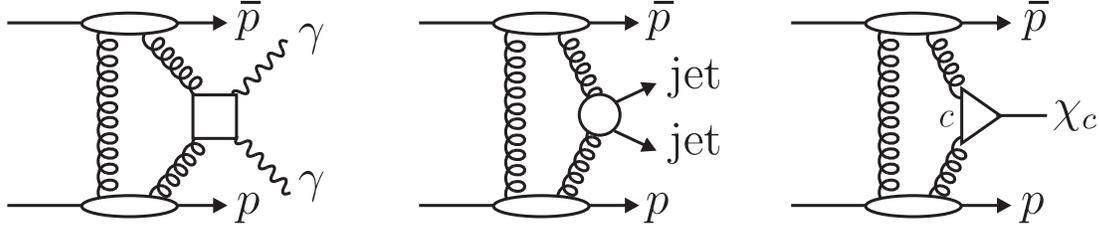}
\caption{The mechanism for the exclusive processes observed by CDF at the Tevatron. The survival probabilities of the rapidity gaps are not indicated in the sketches.  }
\label{fig:cdf}
\end{center}
\end{figure}  
Exclusive diffractive processes of the type $\bar{p}p \to \bar{p}+A+p$ have {\it already} been observed by CDF at the Tevatron, where $A=\gamma\gamma$ \cite{CDFgg} or dijet \cite{CDFdijet} or $\chi_c$ \cite{CDFchi}.   As the sketches in Fig.~\ref{fig:cdf} show, these processes are driven by the same mechanism as that for exclusive Higgs production, but have much larger cross sections. They therefore serve as ``standard candles''.

CDF observe three candidate events for $\bar{p}p \to \bar{p}+\gamma\gamma+p$ with $E_T^{\gamma}>5$ GeV and $|\eta^{\gamma}|<1$ \cite{CDFgg}. Two events clearly favour the $\gamma\gamma$ hypothesis and the third is likely to be of $\pi^0 \pi^0$ origin. The predicted number of events for these experimental cuts is $0.8^{+1.6}_{-0.5}$ \cite{KMRSgg}, giving support to the `Durham' approach used for the calculation of the cross sections for exclusive processes.

Especially important are the recent CDF data \cite{CDFdijet} on
 exclusive production
of a pair of high $E_T$ jets, $p\bar {p} \to p+jj+\bar {p}$.
Such measurements could provide
 an effective $gg^{PP}$ `luminosity
\begin{figure} [h]
\begin{center}
\includegraphics[height=7.5cm]{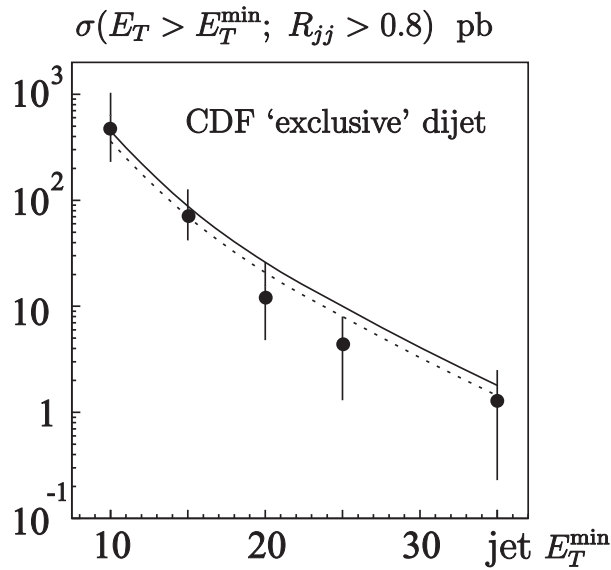}
\caption{The cross section  for `exclusive' dijet production
 at the Tevatron as a function $E_T^{\rm min}$ as measured by CDF \cite{CDFdijet}. 
  The data integrated over
the domain $R_{jj} \equiv M_{\rm dijet}/M_{PP} > 0.8$ and $E_T > E_T^{\rm min}$. A jet cone of $R<0.7$ is used.
The curves are the pure exclusive cross section
predicted by the `Durham' model using the CDF event selection. 
The solid curve is obtained \cite{KMR1} by rescaling the parton (gluon)  transverse momentum
$p_T$ to the measured jet
transverse energy $E_T$ by  $E_T=0.8 p_T$. The dashed curve assumes
$E_T=0.75 p_T$. The rescaling procedure effectively accounts for the hadronization
and radiative 
effects, and for jet energy losses outside the selected jet cone.}
\label{fig:sjj}
\end{center}
\end{figure}
monitor' for the kinematical region appropriate for Higgs
production. The corresponding cross section was evaluated to
be about 10$^4$ times larger than that for the exclusive production of a SM Higgs boson.
Since the exclusive dijet cross section is rather large, this  process appears
to be an ideal `standard candle'.
A comparison of the data with 
analytical predictions, obtained using the `Durham' model, is given in Fig.~$\ref{fig:sjj}$. 
It shows the  $E_T^{\rm min}$ dependence  
for the dijet events with $R_{jj} \equiv M_{{\rm dijet}}/M_{PP} > 0.8$,
where $M_{PP}$ is the invariant energy of the incoming
Pomeron-Pomeron system.  The agreement with the theoretical expectations lends credence to the predictions for the exclusive
Higgs production. 

Moreover, in the {\it early} data runs of the LHC it is possible to observe a range of diffractive processes which will illuminate the different components of the theoretical formalism. Some information is possible even without tagging the outgoing protons \cite{KMRearly}. For example, the observation of the rapidity, $y_A$, dependence of the ratio of diffractive (single gap) $A$ production to inclusive $A$ production will probe the effect of enhanced rescattering. The object $A$ may be an $\Upsilon$ or a $W$ boson or a dijet system. The ratio should avoid normalisation problems. Other informative examples are $W$ (or $Z$) + rapidity gaps events or central 3-jet production. The exclusive process $pp \to p+\Upsilon +p$ is interesting. For low $p_t$ of the outgoing proton, the process is mediated by photon exchange and probes directly the unintegrated gluon distribution. At larger $p_t$, the process is driven by odderon exchange and could be the first hint of the existence of the odderon.

\section{Conclusions}

We emphasized the value of installing near beam proton detectors to the ATLAS and CMS experiments in order to assist the study of the Higgs sector at the LHC via the exclusive process $pp \to p+H+p$. This is a unique chance to study the $H \to b\bar{b}$ decay, due to the large suppression of the QCD $b\bar{b}$ background, and to determine the spin, $C$ and $P$ values of the Higgs.  We described how the prediction of the $pp \to p+H+p$ cross section depends on a mixture of `soft' and `hard' physics.
We introduced a model of `soft' high-energy $pp$ interactions, which possessed all the requirements to give a reliable estimate of the survival probability of the rapidity gaps to both eikonal and enhanced `soft' rescattering effects. Finally, we noted that the rates of the exclusive processes already observed by CDF are in good agreement with the predictions of the Durham model. This lends valuable support to the exciting proposal to indeed install the proton taggers to explore the Higgs sector via exclusive production at the LHC.

%% file: marzaniUCLproc.tex
\begin{center}
{\Large \bf Gaps between jets and soft gluon resummation}

\vspace*{1cm}
Simone Marzani, Jeffrey Forshawm, James Keates
\vspace*{0.5cm} 
 
School of Physics \& Astronomy, University of Manchester,\\Oxford Road, Manchester, M13 9PL, U.K.

\vspace*{1cm}
\end{center}




\begin{abstract}
We study the effect of soft gluon resummation on the gaps-between-jets cross-section at the LHC. We review the theoretical framework that enables one to sum logarithms of the hard scale over the veto scale to all orders in perturbation theory. We then present a study of the phenomenological impact of Coulomb gluon contributions and super-leading logarithms on the gaps between jets cross-section at the LHC.
\end{abstract}


\setcounter{section}{0}
\setcounter{figure}{0}
\setcounter{table}{0}
\setcounter{equation}{0}

\section{Jet vetoing: gaps between jets}
We consider dijet production with transverse momentum $Q$ and a veto on the emission of additional radiation in the inter-jet rapidity region, $Y$, harder than $Q_0$.
 We shall refer generically to the ``gaps between jets'' process, although the veto scale is chosen to be large, $Q_0= 20$~GeV, so that we can rely on perturbation theory. Thus  a ``gap'' is simply a region of limited hadronic activity. 

Gaps between jets is a pure QCD process, hence the cross-section is large and studies can be performed with early LHC data. It is interesting because it allows one to investigate a remarkably diverse range of QCD phenomena. 
For instance, the limit of large rapidity separation corresponds to the limit of high partonic centre of mass energy and  BFKL effects are expected to become important~\cite{muellernavelet}.  On the other hand one can study the limit of emptier gaps, becoming more sensitive to wide-angle soft gluon radiation. Furthermore, if one wants to investigate both of these limits simultaneously, then  the non-forward BFKL equation enters the game~\cite{muellertang}.  In the following we discuss only wide-angle soft emissions.

Accurate studies of these effects are important also in relation to other processes, in particular the production of a Higgs boson in association with two jets. It is well known that this process can occur via gluon-gluon fusion and weak-boson fusion (WBF). QCD radiation in the inter-jet region is clearly  different in the two cases and, in order to enhance the WBF channel, one can put a cut on emission between the jets \cite{Barger:1994zq,Kauer:2000hi}. This situation is very closely related to gaps between jets since the Higgs carries no colour charge, and QCD soft logarithms can be resummed using the same technique~\cite{softgluonshiggs}. 

Given a hard scattering process, we can study how it is modified by the addition of soft radiation. If the observable is inclusive enough, then we  have no effects because soft contributions cancel when real and virtual corrections are added together, as a result of the Bloch-Nordsieck theorem. However, if we restrict the real radiation to a corner of the phase space, as happens for the gap cross-section, we encounter a miscancellation and are left with a logarithm of the ratio of the hard scale and veto scale, $Q/Q_0$.
The resummation of wide-angle soft radiation in the gaps between jets process was originally performed assuming that the real--virtual cancellation is perfect outside the gap, so that one needs only to consider virtual gluon corrections integrated over momenta for which real emissions are forbidden, i.e. over the ``in gap'' region of rapidity and with
$k_T$ above the veto scale $Q_0$~\cite{KOS,OS, Oderda}. We shall refer to these contributions as global logarithms. The resummed squared matrix element can be written as:
\bea \label{resummedpartonicxsec1}
|\m|^2 &=& \frac{1}{V_c} \langle m_0 | e^{- \xi \mathbf{\Gamma}^{\dagger}}e^{- \xi \mathbf{\Gamma}} |m_0 \rangle\,, \nonumber \\
 \xi  &=&\frac{2}{\pi} \int_{Q_0}^{Q} \frac{d k_T}{k_T} \as(k_T)\,,
\eea
where $V_c$ is an averaging factor for initial state colour.
The vector $|m_0 \rangle$ represents the Born amplitude and the operator $\mathbf{\Gamma}$ is the soft anomalous dimension:
\beq \label{gammaoperator}
\mathbf{\Gamma} = \frac{1}{2}Y \mathbf{t}_t^2+ i \pi \mathbf{t}_a\cdot \mathbf{t}_b +\frac{1}{4}\rho_{\rm jet}(Y, |\dy|)(\mathbf{t}_c^2+\mathbf{t}_d^2)\,,
\eeq
where $\mathbf{t}_i$ is the colour charge of parton $i$ and the function $\rho_{\rm jet}(Y,\dy)$ is related to the jet definition.
The operator $\mathbf{t}_t^2$ represents the colour exchanged in the $t$-channel:
\beq
\mathbf{t}_t^2=(\mathbf{t}_a+\mathbf{t}_c )^2 = \mathbf{t}_a^2+\mathbf{t}_c^2 + 2  \,\mathbf{t}_a\cdot \mathbf{t}_c\,.
\eeq
The imaginary part of Eq.~(\ref{gammaoperator}) is due to Coulomb gluon exchange. These contributions play an important role in the proof of QCD factorization and they are also responsible for super-leading logarithms~\cite{SLL1,SLLind}. We notice that for processes with less than four coloured particles, such as deep-inelastic scattering or Drell-Yan processes, the imaginary part of the anomalous dimension does not contribute to the cross-section. For instance, if we consider three coloured particles, then colour conservation implies that $ \mathbf{t}_a+\mathbf{t}_b + \mathbf{t}_c=0$, and consequently 
\beq
 i\pi \, \mathbf{t}_a \cdot \mathbf{t}_b =\frac{i\pi}{2} \left( \mathbf{t}_c^2 -\mathbf{t}_a^2-\mathbf{t}_b^2 \right)\,,
 \eeq
which contributes as a pure phase. Coulomb gluons do play a role in dijet production, but they are not implemented in angular-ordered parton showers. We shall evaluate the impact of these contributions on the cross-section in the next section.

It was later realised~\cite{DS} that the above procedure is not enough to capture the full leading logarithmic behaviour. Real gluons emitted outside of the gap are forbidden to re-emit back into the gap and this gives rise to a new tower of logarithms, formally as important as the primary emission corrections, known now as non-global logarithms.
 The leading logarithmic accuracy is therefore achieved by considering all $2 \to n $ processes, i.e. \mbox{$n-2$ out-of-gap gluons}, dressed with ``in-gap'' virtual corrections, and not only the virtual corrections to the $2\to 2$ scattering amplitudes.
The colour structure quickly becomes intractable and, to date, calculations have been performed only in the large $N_c$ limit~\cite{DS,appleby2,nonlinear}.

A different approach was taken in~\cite{SLL1,SLLind}, where the specific case of only one gluon emitted outside the gap, dressed to all orders with virtual gluons but keeping the full $N_c$ structure, was considered. That calculation had a very surprising outcome, namely the discovery of a new class of ``super-leading'' logarithms (SLL), formally more important than the ``leading'' single logarithms.
Their origin can be traced to a failure of the DGLAP ``plus-prescription'', when the out-of-gap gluon becomes collinear to one of the incoming partons. Real and virtual contributions do not cancel as one would expect and one is left with an extra logarithm.  This miscancellation  first appears at the fourth order relative to the Born cross-section and it is caused by the imaginary part of loop integrals, induced by Coulomb gluons. These SLL contributions  have been recently resummed to all orders in~\cite{jetvetoing}. The result takes the form:
\beq \label{master2}
|\m_1^{\rm SLL}|^2 =   - \frac{2 }{\pi} \int_{Q_0}^{Q} \frac{d k_T}{k_T}\as(k_T) \left(  \ln \frac{Q}{k_T} \right)  \left( \Omega^{\rm coll}_R + \Omega^{\rm coll}_V\right), 
\eeq  
where $ \Omega^{\rm coll}_{R(V)}$ is the resummed real (virtual) contribution in the limit where the out-of-gap gluon becomes collinear to one of the incoming partons. The presence of SLL has been also confirmed by a fixed order calculation in~\cite{SLLfixed};  in this approach SLL have been computed at $\ord (\as^5)$ relative to Born, i.e. going beyond the one out-of-gap gluon approximation.
 
\section{LHC phenomenology}
In this section we perform two different studies. Firstly we consider the resummation of global logarithms and we study the importance of Coulomb gluon contributions, comparing the resummed results to the ones obtained with a parton shower approach. We then turn our attention to SLL and we evaluate their phenomenological relevance. In both studies we consider $\sqrt{S}=14$~TeV, $Q_0=20$~GeV, jet radius $R=0.4$ and we use the MSTW 2008 LO parton distributions~\cite{mstw08}. 
\begin{figure*}
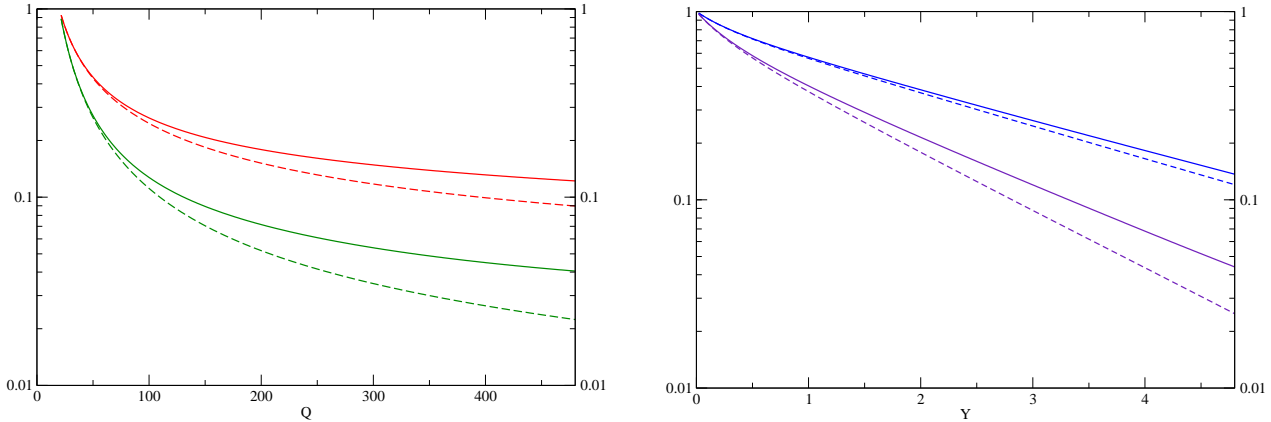

\begin{center}
\includegraphics[width=0.47\textwidth]{marzani_simoneGAP.fig1.eps} \hspace{.5cm}
\includegraphics[width=0.47\textwidth]{marzani_simoneGAP.fig2.eps}
\caption{On the left we plot the gap fraction for $Y=3$ (upper red curves) and $Y=5$ (lower green curves) as a function of $Q$ and on the right as a function of $Y$, for $Q=100$~GeV (upper blue curves) and $Q=500$~GeV (lower violet curves). The solid lines are the full resummation of global logarithms, while the dashed ones are obtained by omitting the $i \pi$ terms in the anomalous dimension.}\label{fig:kfact1}
\end{center}
\end{figure*}

\begin{figure*}
\begin{center}
\includegraphics[width=0.6\textwidth]{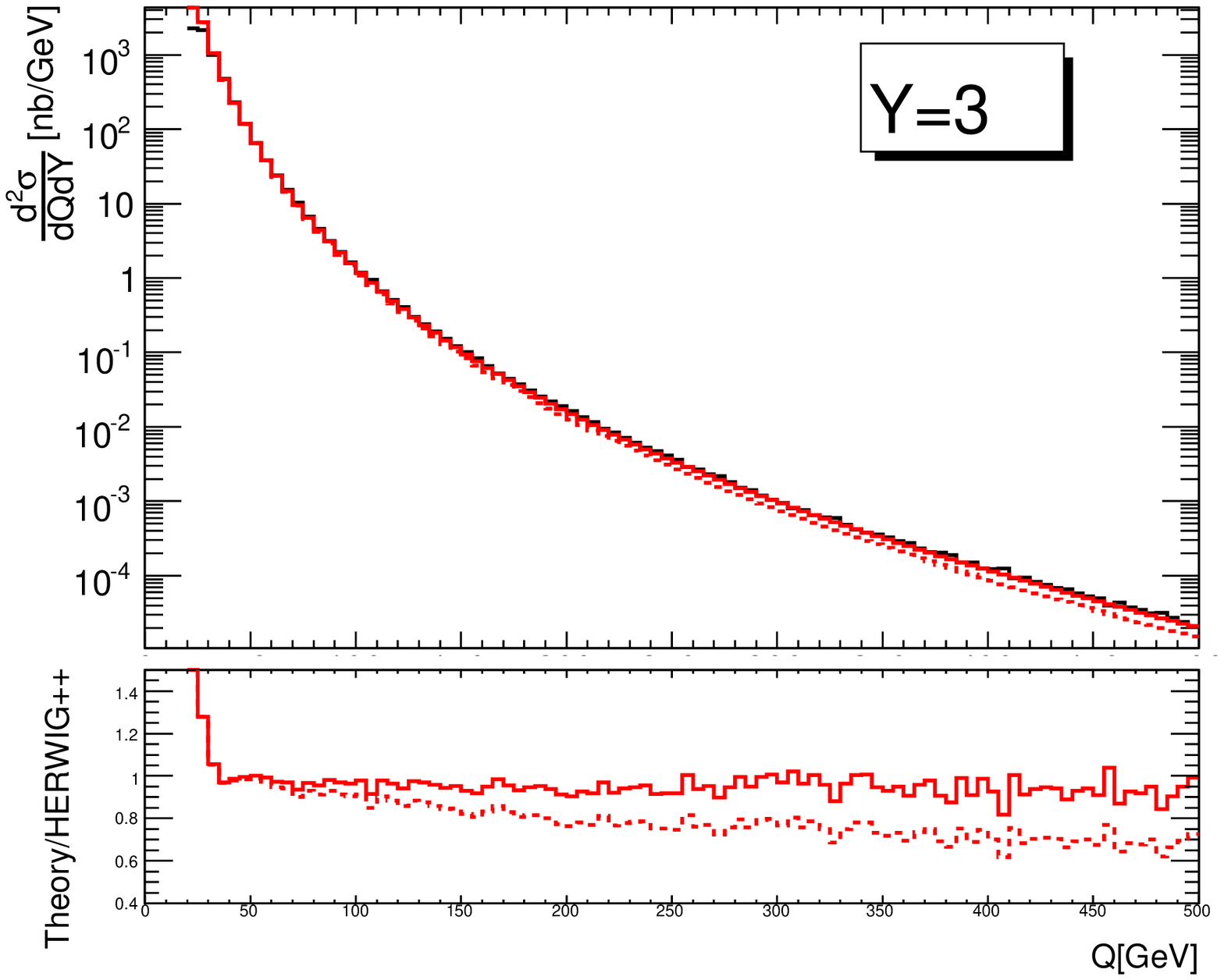}
\caption{The gap cross-section obtained using \textsc{Herwig++} (black histogram) is compared to the one from resummation (red curves). As before the solid line is the full result, while the dashed line is obtained by omitting the Coulomb gluon contributions.
At the bottom we plot the ratio between the results obtained from the resummation and the one from \textsc{Herwig++}.}\label{fig:gapx}
\end{center}
\end{figure*}

Soft logarithmic contributions are implemented in \textsc{Herwig++} via angular ordering of successive emissions. Such an approach cannot capture the contributions coming from the imaginary part of the loop integrals, due to Coulomb gluon exchange. We evaluate the importance of these contributions in Fig.~\ref{fig:kfact1}. On the left  we plot the gap cross-section, normalised to the Born cross-section (i.e. the gap fraction), as a function of $Q$ at two different values of $Y$ and, on the right, as a function of $Y$ at two different values of $Q$. The solid lines represent the results of the resummation of global logarithms; the dashed lines are obtained by omitting the $i \pi$ terms in the soft anomalous dimension matrices.  As a consequence, the gap fraction is reduced by $7\%$ at $Q=100$~GeV and $Y=3$ and by as much as $50\%$ at $Q=500$~GeV and $Y=5$. Large corrections from this source herald the breakdown of the parton shower approach.  In Fig.~\ref{fig:gapx} we compare the gap cross-section obtained after resummation  to that obtained using \textsc{Herwig++}~\cite{ThePEG,Bahr:2008pv,KLEISSCERN9808v3pp129,Gieseke:2003rz} after parton showering ($Q$ is taken to be the mean $p_T$ of the two leading jets). The broad agreement is encouraging and indicates that effects such as energy conservation, which is included in the Monte Carlo, are not too disruptive to the resummed calculation. Nevertheless, the histogram ought to be compared to the dotted curve rather than the solid one, because \textsc{Herwig++} does not include the Coulomb gluon contributions. 
The resummation approach and the parton shower differ in several aspects: some non-global logarithms are included in the Monte Carlo and the shower is performed in the large $N_c$ limit. Of course the resummation would benefit from matching to the NLO calculation and this should be done before comparing to data.  

\begin{figure*}
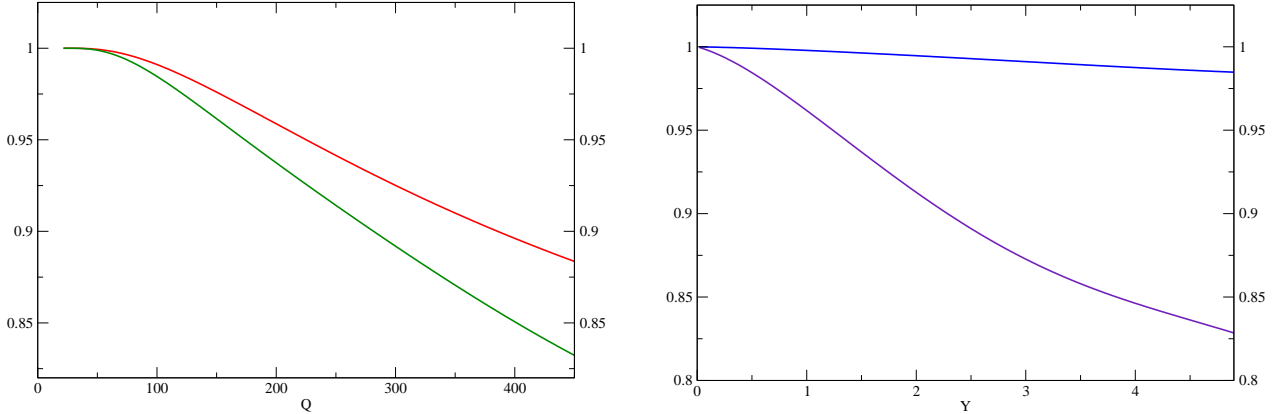

\begin{center}
\includegraphics[width=0.47\textwidth]{marzani_simoneGAP.fig4.eps} \hspace{.5cm}
\includegraphics[width=0.47\textwidth]{marzani_simoneGAP.fig5.eps}
\caption{On the left we plot the $K$-factor as defined in Eq.~(\ref{kSLL}) as a function of $Q$ for $Y=3$ (upper red curve) and $Y=5$ (lower green curve);  on the right we plot it as a function of $Y$, for $Q=100$~GeV (upper blue curve) and $Q=500$~GeV (lower violet curve) }\label{fig:kSLL}
\end{center}
\end{figure*}

Finally we want to study the relevance of the SLL contributions. In order to do that we define
\beq \label{kSLL}
K^{(1)}= \frac{\sigma^{(0)}+\sigma^{(1)}}{\sigma^{(0)}} \,,
\eeq
where $\sigma^{(0)}$ contains the resummed global logarithms and  $\sigma^{(1)}$ the resummed SLL contribution coming from the case where one gluon is emitted outside of the gap.
The results are shown in Fig.~\ref{fig:kSLL}.
Generally the effects of the SLL are modest, reaching as much as 15\% only for jets with \mbox{$Q > 500$~GeV} and rapidity separations $Y > 5$. The contribution coming from $n \ge 2$ out-of-gap gluons is thought to be less important~\cite{jetvetoing}.
Remember that we have fixed the value of the veto scale $Q_0=20$~GeV and that the impact will be more pronounced if the veto scale is lowered. 
\section{Conclusions and Outlook}
There is plenty of interesting QCD physics in ``gaps-between-jets'' and measurement can be performed with early LHC data.  
There are significant contributions arising from the exchange of Coulomb gluons, especially at large $Q/Q_0$ and/or large $Y$, which are not implemented in the parton shower Monte Carlos. However before comparing to data, there is a need to improve the resummed results by matching to the fixed order calculation. These observations will have an impact on jet vetoing in Higgs-plus-two-jet studies at the LHC.

We have studied the super-leading logarithms that occur because gluon emissions that are collinear to one of the incoming hard partons are forbidden from radiating back into the veto region. Even if their phenomenological relevance is generally modest, they deserve further study because they are deeply connected to the fundamental ideas behind QCD factorization.
 
\newpage

\begin{footnotesize}

\end{footnotesize}


%% file: london09_proc.tex
\begin{center}
{\Large \bf Jet-based approach for underlying-event characterization}

\vspace*{1cm}
Sebastian Sapeta\footnote{sapeta@lpthe.jussieu.fr}

\vspace*{0.5cm}
LPTHE, UPMC -- Paris 6, CNRS UMR 7589, Paris, France
\end{center}

\vspace*{1cm}

\begin{abstract}
We discuss which characteristics of the underlying event might be useful to measure for improving understanding of its properties and simulating it well in Monte Carlo generators. We use a method of selection of the underlying event which is based on jets analysis  and does not require any phase space cuts. We analyze the predictions of several tunes of simulation programs for selected quantities.
\end{abstract}

\section{Introduction}
\label{sec:intro}

In $pp$ collisions the hard interaction is always accompanied by soft background activity, called the \emph{underlying event} (UE), involving the spectator partons. 
 Underlying event has a largely non-perturbative nature and its separation from the hard process is to some extent a matter of definition. Therefore, our understanding of UE is rather incomplete and it is not clear how the UE should be modeled. Yet, the underlying event is very important and has a big impact on the results of data analysis since it adds a considerable amount of transverse momentum to the event. 

Improper estimation of UE may lead e.g. to modification of the inclusive jet spectrum  by up to 50\% (at moderate transverse momenta), incorrect kinematic reconstruction of mass peak position, degradation of the mass peak as well as lowering the efficiency of lepton isolation.
Therefore, it is important that the models implemented in Monte Carlo generators (MC) describe various aspects of UE as well as possible. This requirement leads to the question about the relevant properties of UE that could help to better tune the generators as well as further constrain the models.
Of particular value would be the characteristics for which the current MCs and tunes give different predictions when extrapolated to the LHC energy.

In the usual method of the underlying event analysis, which we call `topological', as a first step, one identifies the direction of the leading jet and defines the cone around it in the transverse plane. Then, the UE is measured using only those regions which are transverse with respect to the leading jet. This method was widely used in Tevatron data analysis \cite{Acosta:2004wqa,Albrow:2006rt, Moraes:2007rq}.

In these proceedings we present the study of the underlying event using an alternative technique based on the method for pileup and UE estimation proposed in \cite{Cacciari:2007fd}. 

\section{The method}

The method used in our analysis is jet-based, exploits the concept of jet areas \cite{Cacciari:2008gn} and can be carried out using the {\tt FastJet} package \cite{Cacciari:2005hq, FastJet}. It proceeds as follows: 

For each event, all true particles together with a large number of extremely soft particles (called `ghosts') are clustered with an infrared and collinear safe jet finding algorithm. This way, we obtain a list of jets with transverse momenta, $p_t$, ranging from the scale of hard collision to essentially zero (for the so called `pure ghost jets'). The susceptibility to the contamination from the soft radiation for each jet $j$ is given by the jet area $A_j$, proportional to the number of ghosts in the jet. Hence, the amount of $p_t$ added by this radiation to jet $j$ is $\rho A_j$, where $\rho$ is the level of transverse momentum per unit area characteristic for the UE.

In order to measure $\rho$, the median of the $p_{t, j}/A_j$ distribution is determined for the single event
\be
\label{eq:median}
\rho = {\rm median} \Big[\Big\{ \frac{p_{t, j}}{A_j}\Big\}\Big].
\ee
The main advantage of using the median is its very small sensitivity to perturbative radiation, which starts to be relevant only at the order $\alpha_s^{24}-\alpha_s^{47}$ (depending on jet definition and rapidity range) \cite{Cacciari:2007fd}. At the same time,  there is no need for cuts in $(y,\phi)$-plane. This is also important since such cuts can potentially introduce a bias (see \cite{inpreparation}).
Since the separation between UE and the hard part of the event is based solely on the value of $p_{t, j}/A_j$ we refer to this method as a `dynamical' selection.

One is also interested in the uncertainty of $\rho$ caused by fluctuations of UE from point to point within the event. This quantity, denoted by $\sigma$, is determined from the sorted list of $\{p_{t, j}/A_j\}$ and given by the value for which $15.86\% $ of jets have smaller $p_{t, j}/A_j$. With such definition, in the case of Gaussian distribution of UE,  $68.27\% $ of jets satisfy
$\rho - \sigma/\sqrt{A_j} < p_{t, j}/A_j < \rho + \sigma/\sqrt{A_j}$.

\section{The results}

\begin{figure}[t]
\includegraphics[width=6cm, angle=-90]{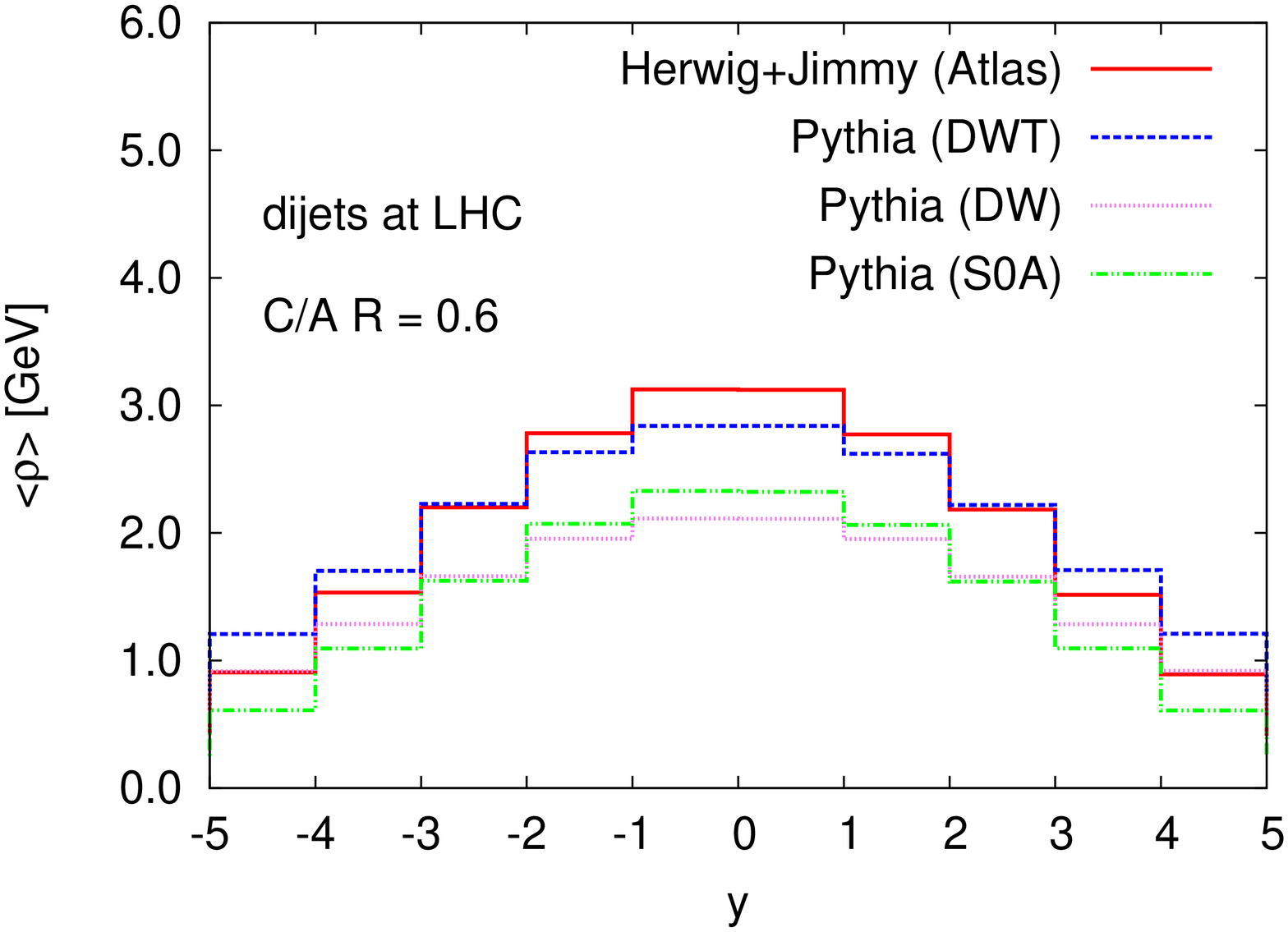}
\includegraphics[width=6cm, angle=-90]{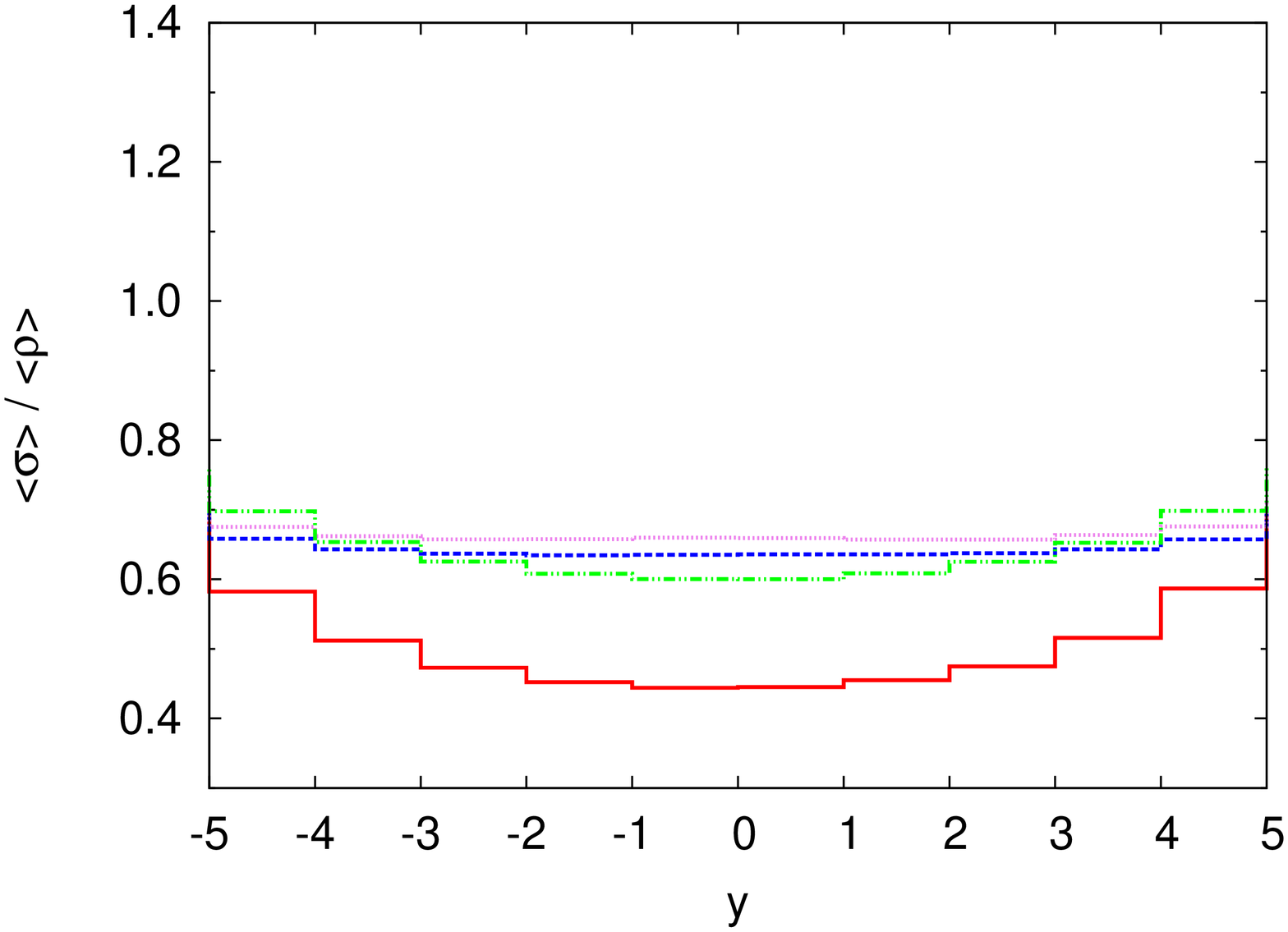}
\caption{
Left: Rapidity dependence of the average level of UE for various generators/tunes.
Right: Average level of point-to-point fluctuations of UE normalized to $\langle \rho \rangle$ as a function of rapidity.
}
\label{fig:av_rho_sigma}
\end{figure}
\begin{figure}[t]
\includegraphics[bb=70 0 600 700, width=6.5cm, angle=-90]{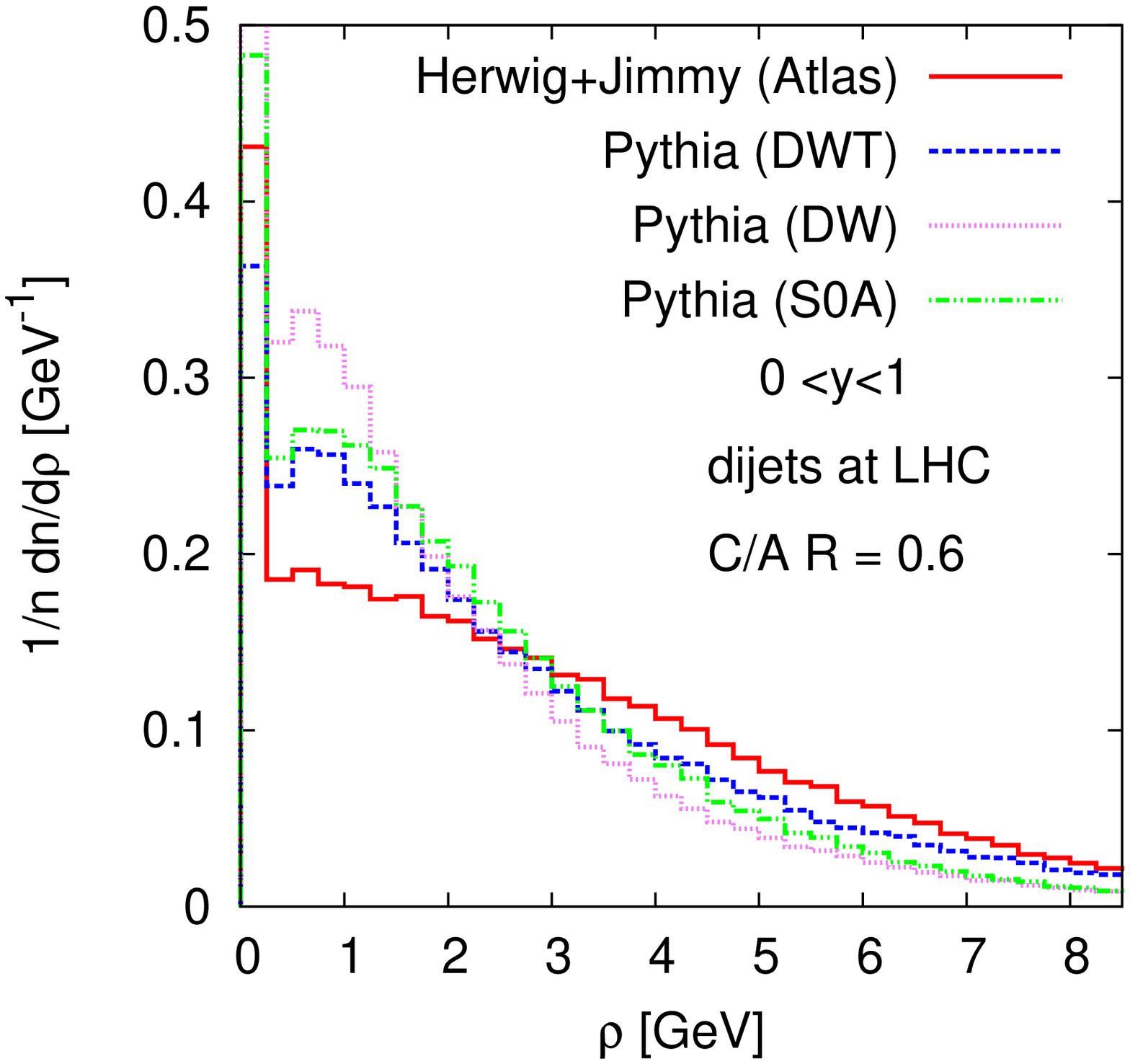}
\includegraphics[width=6cm, angle=-90]{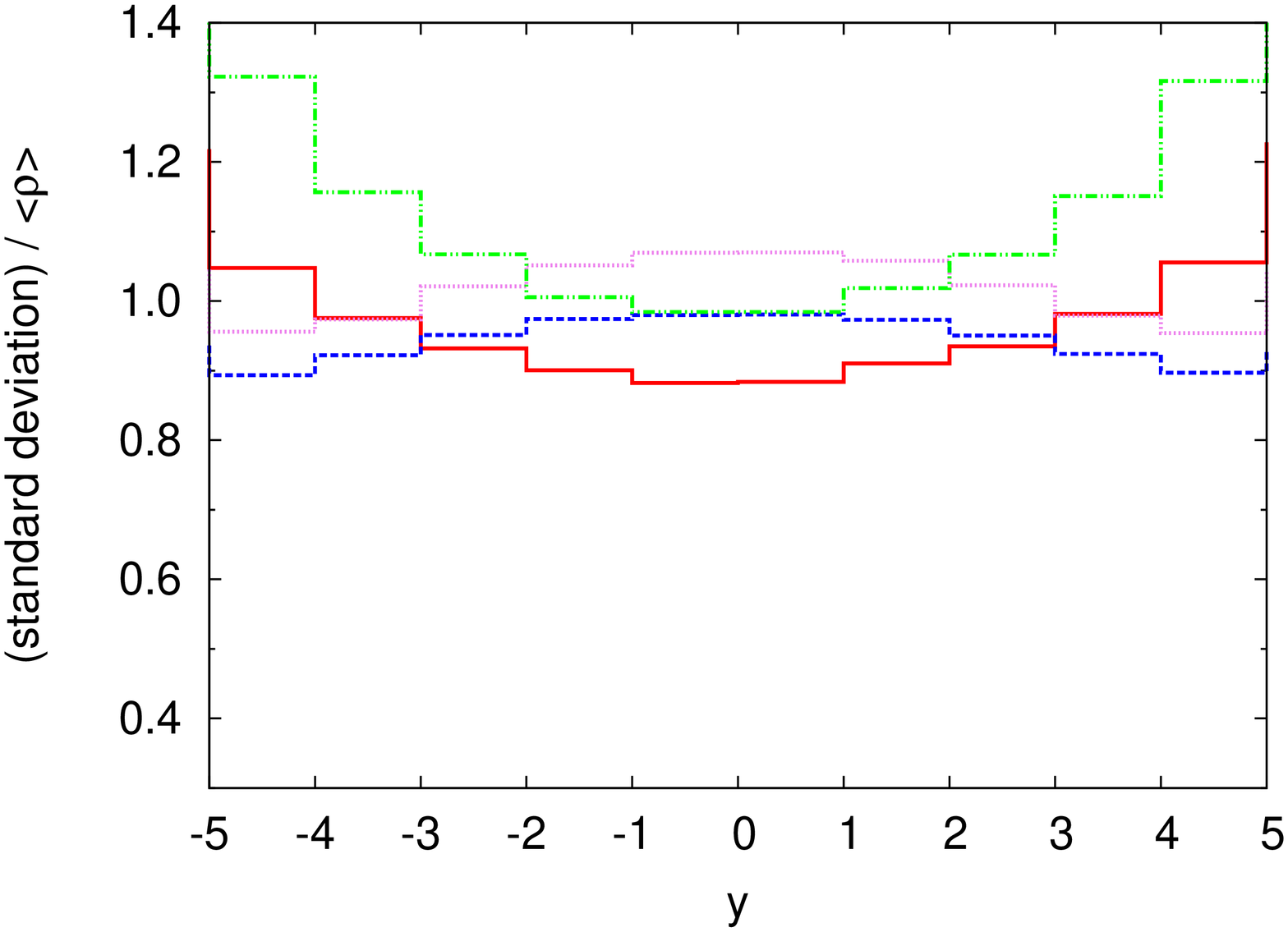}
\caption{
Left: Distribution of $\rho$ determined in rapidity range $0<y<1$ from $2\cdot 10^5$ generated events for various generators/tunes.
Right: Standard deviation of $\rho$ distributions like the one from left hand side, normalized to $\langle \rho \rangle$, as a function of rapidity.
}
\label{fig:rho_dist}
\end{figure}
In this study we are interested in the UE itself. We address the question which properties of UE are relevant and what are the predictions for these properties from different Monte Carlo tunes. One important extension with respect to \cite{Cacciari:2007fd,Cacciari:2008gn, Cacciari:2008gd} is that we go differential in rapidity. We note that the dynamical method works well only if the number of jets  used to calculate median is large enough. This sets some limit on the smallest possible rapidity range that can be used for $\rho$ determination.

We study the UE characteristics for a series of Monte Carlo generators/tunes:
DW and DWT tunes of Pythia~6.4~\cite{Sjostrand:2006za} by R.~Field~\cite{Albrow:2006rt}  that come with the `old' shower and S0A tune~\cite{Skands:2007zg,Buttar:2006zd,Sjostrand:2004ef} by P.~Skands with the `new' shower, Herwig~6.5~\cite{Corcella:2000bw,Corcella:2002jc} + Jimmy 4.3~\cite{Butterworth:1996zw} in an Atlas tune by A.Moraes \cite{Albrow:2006rt} as well as the Herwig 6.5 default UE model.  All models of UE, except for Herwig default case, are based on multiple parton interactions.

We consider dijet events at the LHC, $\sqrt{s} = 14$ TeV, generated with $p_{t,{\rm min}} = 50$ GeV. The analysis is performed with {\tt FastJet 2.4.1} \cite{FastJet}. Except where stated,  we cluster particles using the Cambridge/Aachen algorithm with R = 0.6 and active area with explicit ghosts. 
The results discussed in these proceedings were obtained without further rapidity-based selection. 
We have established that for such a case one can safely go with size of the rapidity strip down to $\Delta y = 1$.
For the analysis of the impact of further rapidity selection we refer to \cite{inpreparation}. 
Here, we note only that the number of jets in a given range of rapidity of size of one unit (which is typically around 15) may not be sufficient if one requires e.g. the two hardest jets to lie also in this range. In such case, one can improve the quality of $\rho$ determination either by removing the hardest jets or by doubling the size of the rapidity strip.

In the left panel of Fig.~\ref{fig:av_rho_sigma} we show the average value of $\rho$ as a function of $y$, determined from $2\cdot 10^5$ events for different tunes. We observe a significant dependence of the level of UE on rapidity. The predictions for the UE at the LHC differ also for different tunes. They are very close for Pythia DW and S0A since both tunes have the same $\sqrt{s}$ scaling of $p_{t, {\rm min}}$. This scaling is faster than in Pythia DWT tune hence the latter gives larger level of UE, which is similar to the predictions of Herwig+Jimmy in Atlas tune.

In the right panel of Fig.~\ref{fig:av_rho_sigma} we show the level of fluctuations of the UE within an event from different tunes. It is given in terms of the average $\sigma$ normalized to the average $\rho$. We see that these fluctuations are large and that predictions from Pythia and Herwig+Jimmy differ significantly.

But UE fluctuates also from event to event. To illustrate this, we plot in Fig.~\ref{fig:rho_dist} the distribution of $\rho$ determined in rapidity strip $0<y<1$. It is this distribution (and analogous ones for different rapidity ranges) that we used to compute $\langle \rho \rangle$ depicted in the left hand side of Fig.~\ref{fig:av_rho_sigma}. 
The leftmost bin with the high yield contains very quiet events with $\rho < 0.25$  GeV per unit area.
We notice sizable differences in shape for various generators/tunes. All distributions, however, are fairly broad, which indicates high fluctuations of UE from one event to another. This constitutes 
an argument in favor of an event-by-event analysis of the underlying event. 
In the right hand side of Fig.~\ref{fig:rho_dist} we show the standard deviation (normalized to the average $\rho$) for the distribution from the left hand side of this figure and similar ones for different rapidity bins. We see, in particular, that the level of these fluctuations is considerably larger than the average level of fluctuations within an event shown in Fig.~\ref{fig:av_rho_sigma}.

\begin{figure}[t]
\includegraphics[width=8.5cm, angle=-90]{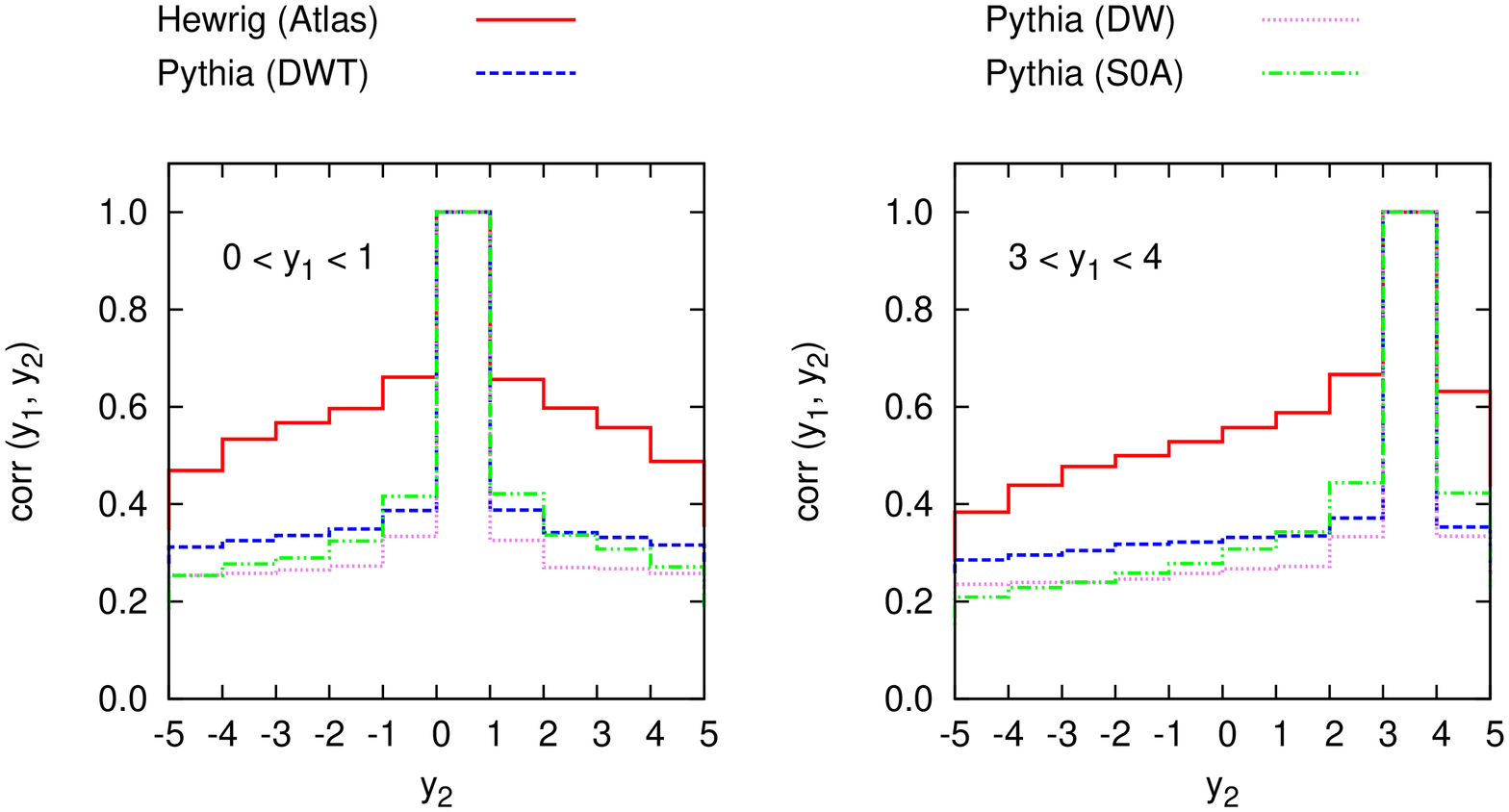}
\caption{
Correlations of the underlying event between different rapidity ranges. The coefficient from Eq.~(\ref{eq:corrcoef}) is plotted as function of $y_2$ for two $y_1$ bins and for a range of generators/tunes. 
}
\label{fig:correlations}
\end{figure}

Correlations between values of $\rho$ in different rapidity strips is another interesting characteristic of the underlying event. To quantify it, we use the standard definition of the correlation coefficient
\be
\label{eq:corrcoef}
{\rm corr}(y_1,y_2) =
\frac{\big\langle \rho(y_1)\rho(y_2)\big\rangle -
      \big\langle \rho(y_1)\big\rangle \langle\rho(y_2)\big\rangle}
     {\sqrt{\big\langle \rho(y_1)^2\big\rangle- \big\langle \rho(y_1)\big\rangle^2}
     \sqrt{\big\langle \rho(y_2)^2\big\rangle- \big\langle \rho(y_2)\big\rangle^2}},
\ee
where $y_1$, $y_2$ are the rapidity bins of width $\Delta y = 1$ and the $\langle \dots \rangle$ denotes the average over  many events ($2\cdot 10^5$, in the case of this study).
In Fig.~\ref{fig:correlations} we show the coefficient from Eq.~(\ref{eq:corrcoef}) as a function of $y_2$ for two bins of $y_1$: $0<y_1<1$~(left) and $3<y_1<4$ (right). We see that the predictions for the amount of correlations of the underlying event at the LHC are very different for Pythia and Herwig+Jimmy. 
They are fairly similar for various Pythia tunes is spite of (in some cases, like DWT or DW vs. S0A) important differences in UE models that come with them.
We notice that the difference between Pythia and Herwig+Jimmy is the most pronounced discrepancy among all quantities that we have studied, which makes the correlation coefficient a particularly valuable quantity from the point of view of improving the fits as well as the model of UE. We also remark that the observed higher correlations for Herwig+Jimmy with respect to Pythia is qualitatively consistent with the lower level of average point to point fluctuations in the event discussed previously in the context of Fig.~\ref{fig:av_rho_sigma}.

\begin{figure}[t]
\includegraphics[width=6.5cm, angle=-90]{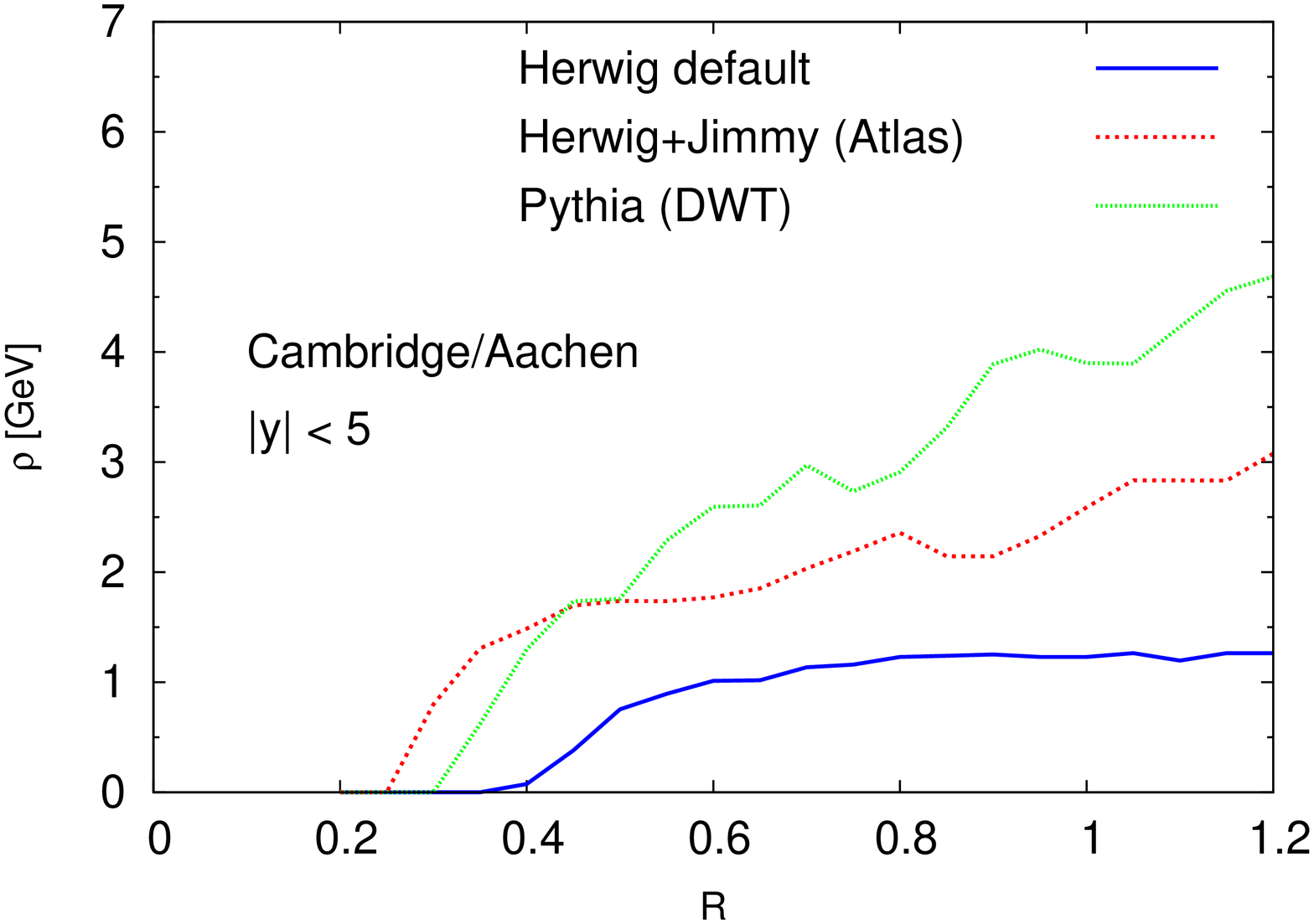}
\caption{Dependence of the level of the underlying event, $\rho$, on the jet radius $R$ for a single, typical event from different generators/tunes.}
\label{fig:rdep}
\end{figure}
\begin{figure}[t]
\includegraphics[width=5.5cm, angle=-90]{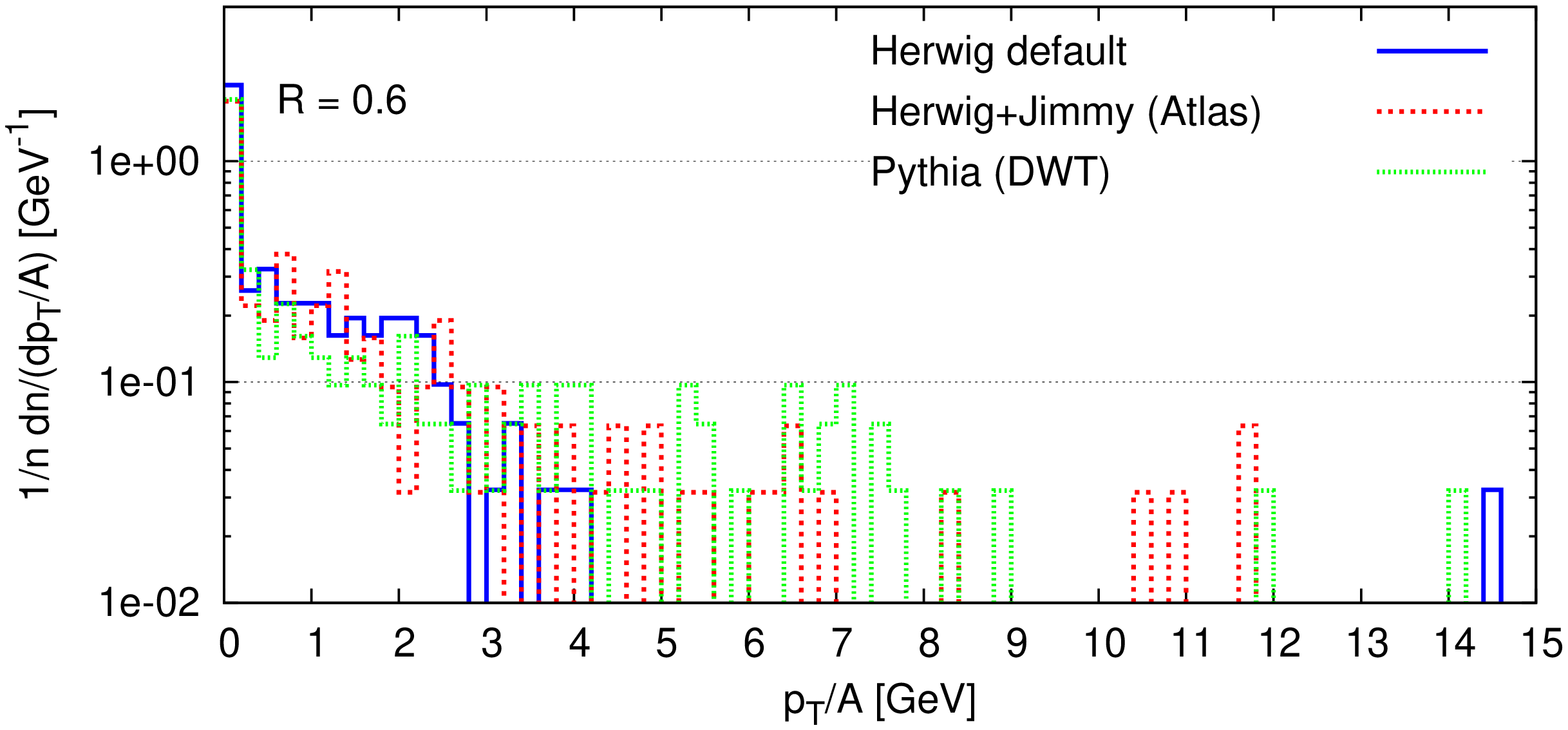}
\caption{
Histograms of $p_t/A$ corresponding to the events shown in Fig.~\ref{fig:rdep}.
}
\label{fig:ptarhist}
\end{figure}

The results presented up to this point were obtained with $R= 0.6$, a choice motivated by the findings from \cite{Cacciari:2007fd}.
We conclude our discussion of selected characteristics of UE that are of potential interest by examining the dependence of the level of UE,  $\rho$, on the radius $R$ in the jet definition. Let us first consider, for the purpose of illustration, a simple two-component model of UE in which $p_t/A$ of soft particles is given by the Gaussian distribution with mean $\rho_0$. The second component is just $n_h$ hard (with respect to the scale of the Gaussian component) particles.  It can be shown~\cite{inpreparation} that for the case of this model 
\be
\label{eq:2comp}
\rho = \Theta(R-R_0) (\rho_0 + {\rm const}\cdot  \, \sigma \, n_h\, R ).
\ee
The value of $R$ has to be larger than some critical value $R_0$ so that the number of jets that contain physical particle is larger than the number of ghost jets and the median formula~(\ref{eq:median}) gives the value considerably larger than zero. Further increasing of R leads to linear growth of estimated $\rho$ with the coefficient which is proportional to the number of hard jets $n_h$ as well as to the width of the Gaussian distribution $\sigma/\sqrt{\langle A\rangle}$ (this behavior is observed also in a toy model discussed in \cite{Cacciari:2009fi}). The qualitative features of Eq.~(\ref{eq:2comp}) turn out to be reproduced by the realistic simulation of dijet production at the LHC, as depicted in Fig.~\ref{fig:rdep}, which shows $\rho$, determined from the wide rapidity range $|y| < 5$, for a typical, single event and for three different Monte Carlo models of UE.  

For better understanding, in Fig.~\ref{fig:ptarhist} we show the corresponding histograms of $p_t/A$. 
We see that in the case of Herwig default there are practically no hard jets except those from  the dijet system and the whole noise is concentrated around a single soft scale. That is why the corresponding curve in Fig.~\ref{fig:rdep} is almost flat for $R$ greater than some threshold value, in agreement with Eq.~(\ref{eq:2comp}) for very small $n_h$. The distribution of $p_t/A$ for the case of Herwig+Jimmy acquires some tail due to multiple parton interactions present in this model. Hence, 
 the event contains larger number of jets with hard or semi-hard scale. This is consistent with a non-zero slope of $\rho(R)$ for Herwig+Jimmy in Fig.~\ref{fig:rdep}. The slope for Pythia, in the same figure, is even larger which again is qualitatively consistent with a longer tail (more hard jets) in Fig.~\ref{fig:ptarhist}.

\section{Conclusions}
\label{sec:conclusions}

We presented the analysis of the underlying event predictions for the LHC. We used the jet-based method of dynamical selection of UE introduced in \cite{Cacciari:2007fd}. In this approach the underlying event is analyzed on an event-by-event basis and one does not impose cuts in $(y,\phi)$-plane, hence the whole event is used. One can, however, analyze the UE locally in limited ranges of rapidity (or/and azimuthal angle) to study how it depends on the region of phase space.

We discussed several quantities which we find useful as the UE characteristics: transverse momentum density per unit area, $\rho$,  point-to-point fluctuations, $\sigma$, event-to-event fluctuations, $\rho$ distributions over many events, point-to-point correlations. We analyzed how these quantities depend on rapidity and jet radius for various Monte Carlo models and tunes.

We found that the existing MC models predict for the LHC the amount of 1-3 GeV of transverse momentum per unit area coming from the underlying event. This value depends on rapidity, varies from one event to another, and changes with jet radius $R$ used as a parameter of the jet finding algorithm.  The amount of the UE predicted for the LHC differs also between generators/tunes. 

We observe large fluctuations of UE in the simulated data both within a single event (40--70$\%$) and from event to event (90--130$\%$). Here, the differences between MCs are even more pronounced than for the case of $\rho$. But the quantity which shows the largest discrepancy between Herwig+Jimmy and Pythia is the correlation coefficient. 

We believe that all these makes the discussed characteristics very promising for further MC tunes as well as improving the models of the underlying event.

\section{acknowledgements}
I would like to thank Gavin P. Salam and Matteo Cacciari, with whom the original results presented here have been obtained, for collaboration and comments on the manuscript.
I am also grateful to the organizers of the `London workshop on Standard Model discoveries with early LHC data' for the opportunity to give this talk.


%% file: watt_smlhc.tex
\begin{center}
{\Large \bf Exclusive photoproduction of vector mesons and $Z/\gamma^*$}\\
\vspace*{1cm}
G. Watt\\
\vspace*{0.5cm}
Institute for Particle Physics Phenomenology, University of Durham, DH1 3LE, UK
\vspace*{1cm}
\end{center}

\begin{abstract}
  We review selected aspects of exclusive diffractive photoproduction at the Tevatron and discuss the prospects for the early LHC running.  This talk~\cite{url} is based on the results presented in Ref.~\cite{Motyka:2008ac}, with some updates due to recent experimental results.
\end{abstract}

\setcounter{section}{0}
\setcounter{figure}{0}
\setcounter{table}{0}
\setcounter{equation}{0}

\section{Introduction}
\begin{figure}[h]
  \includegraphics[width=0.4\textwidth]{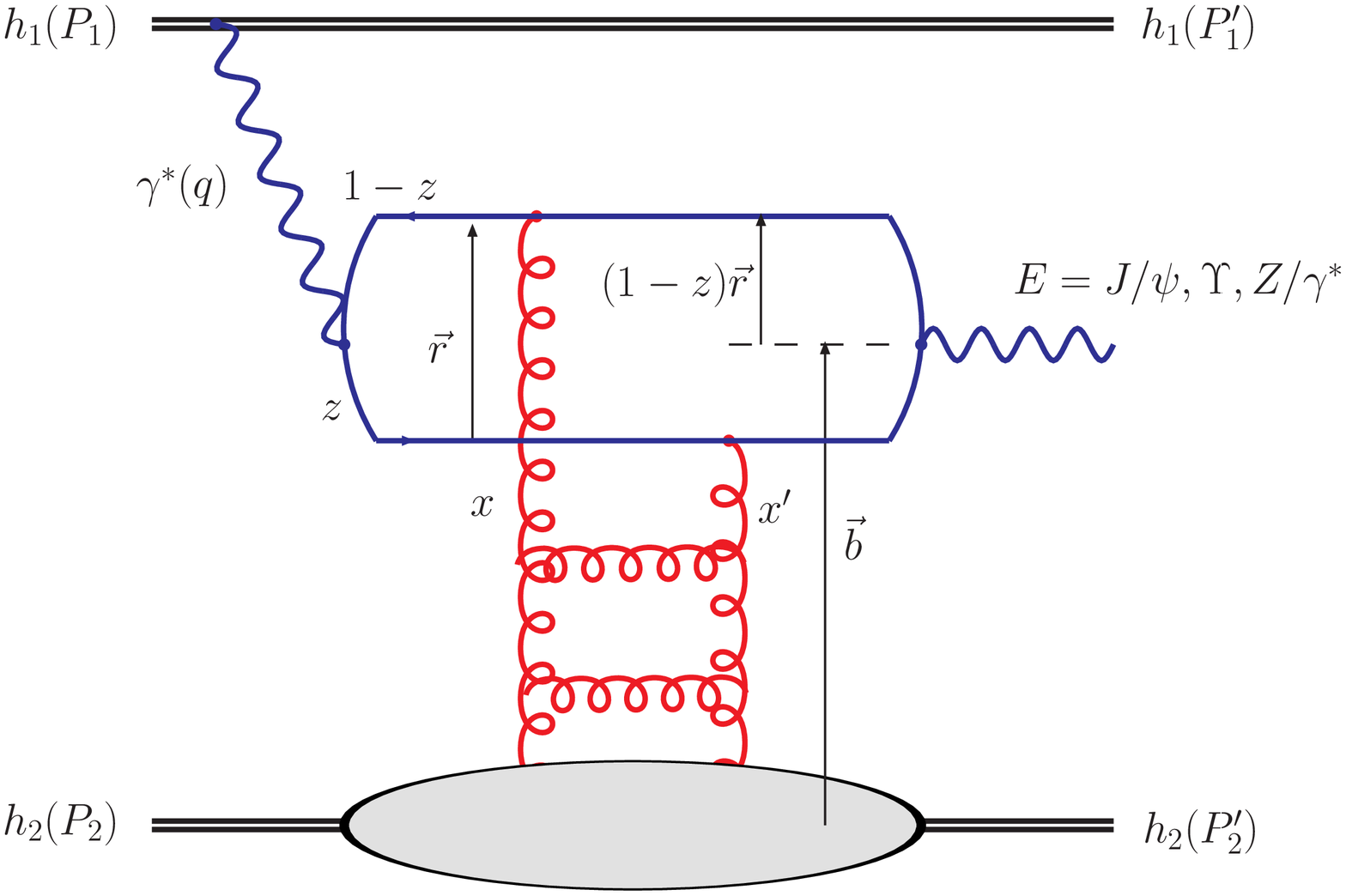}
  \caption{\label{fig:diagram}Exclusive photoproduction of a massive final state $E=J/\psi,\Upsilon,Z/\gamma^*$ in hadron--hadron collisions.}
\end{figure}

Exclusive diffractive Higgs boson production at the LHC, $pp\to p+H+p$, has attracted increasing attention in recent years as an alternative way to explore the Higgs sector~\cite{Martin:2009}.  However, the theoretical uncertainties are comparatively large relative to the inclusive production.  In particular, the generalised (or skewed) unintegrated gluon distribution, which can be written in terms of the usual gluon distribution, enters to the fourth power, and is required in the region of small $x$ and low scales $\sim 1$--$2$ GeV where the gluon distributions obtained from global fits have large uncertainties.  While current Tevatron and early LHC data will prove vital in checking the predictions, it is also important to look for complementary processes to constrain the various ingredients of the calculation.  One such process, where the gap survival factor is expected to be much closer to 1, is exclusive photoproduction, $\gamma p\to E+p$, where the photon is radiated from one of the two incoming hadrons; see Fig.~\ref{fig:diagram}.

\section{\label{sec:phpatlhc}Exclusive photoproduction}
Exclusive diffractive vector meson production, $\gamma^{(*)}p\to V+p$, and deeply virtual Compton scattering (DVCS), $\gamma^*p\to \gamma+p$, have been extensively studied at HERA.  These processes provide a valuable probe of the gluon density at small $x$~\cite{Ryskin:1992ui,Martin:1999wb,Martin:2007sb}.

To obtain the hadron--hadron cross section for exclusive production of a massive final state $E$ with rapidity $y$, we need to multiply the photon--hadron cross section by the flux $\mathrm{d}n/\mathrm{d}k$ of quasi-real photons with energy\footnote{The photon--hadron and hadron--hadron centre-of-mass energies are labelled $W$ and $\sqrt{s}$, respectively.} $k\simeq (M_E/2)\exp(y)\simeq W^2/(2\sqrt{s})$~\cite{Klein:2003vd}:
\begin{equation} \label{eq:rapdist}
  \frac{\mathrm{d}\sigma}{\mathrm{d}y}(h_1h_2\to h_1+E+h_2) = k\frac{\mathrm{d}n}{\mathrm{d}k}\sigma(\gamma p\to E+p),
\end{equation}
together with a second term with $y\to -y$ to account for the contribution from the interchange of the photon emitter and the target, neglecting interference.  We also neglect absorptive corrections, and only present cross sections integrated over final-state momenta, then these effects are expected to be largely washed out, with a rapidity gap survival factor $S^2\sim 0.7$--$0.9$.  Alternative calculations including a detailed treatment of these effects can be found in Refs.~\cite{Khoze:2002dc,Schafer:2007mm,Rybarska:2008pk,Cisek:2009hp}.

The photon energy spectrum $\mathrm{d}n/\mathrm{d}k$ in Eq.~\ref{eq:rapdist} is given by a modified equivalent-photon (Weizs\"acker--Williams) approximation~\cite{Klein:2003vd,Drees:1988pp}:
\begin{equation} \label{eq:photonflux}
  \frac{\mathrm{d}n}{\mathrm{d}k} = \frac{\alpha_{\rm em}}{2\pi k}\left[1+\left(1-\frac{2k}{\sqrt{s}}\right)^2\right]
\times\left(\ln A-\frac{11}{6}+\frac{3}{A}-\frac{3}{2A^2}+\frac{1}{3A^3}\right),
\end{equation}
where $A=1+(0.71\,{\rm GeV}^2)/Q_{\rm min}^2$, $Q_{\rm min}^2\simeq k^2/\gamma_L^2$ and $\gamma_L=\sqrt{s}/(2m_p)$ is the Lorentz factor of a single beam.\footnote{The results of Ref.~\cite{Motyka:2008ac} were erroneously obtained with $A^2$ instead of $A^3$ in the last term of Eq.~\ref{eq:photonflux}.  The results presented here have been corrected, although the numerical impact is completely negligible.}

If the photon-induced contribution to exclusive vector meson production is known precisely enough, there is potential for \emph{odderon} discovery.  The odderon is the $C$-odd partner of the Pomeron, and in perturbative QCD is modelled by exchange of three gluons in a colour-singlet state.  Cross sections were calculated using $k_T$-factorisation in Ref.~\cite{Bzdak:2007cz} for different scenarios.  However, no attempt was made to tune the photoproduction contribution to HERA data, hence the odderon-to-photon ratios are more reliable than the absolute cross sections.  The odderon-to-photon ratios for $\mathrm{d}\sigma/\mathrm{d}y|_{y=0}$ at the Tevatron are 0.3--0.6 ($J/\psi$) and 0.8--1.7 ($\Upsilon$), while the LHC ratios are smaller at 0.06--0.15 ($J/\psi$) and 0.16--0.38 ($\Upsilon$)~\cite{Bzdak:2007cz}.  Moreover, odderon exchange leads to a different meson $p_T$ distribution than the photon-induced contribution.  In the following, we restrict attention to photoproduction.

To calculate the exclusive $\gamma p$ cross section in Eq.~\ref{eq:rapdist} we use the dipole model approach for exclusive diffractive processes~\cite{Bartels:2003yj}, where the amplitude factorises into the light-cone wave functions of the incoming and outgoing particle and a dipole cross section describing the interaction of the $q\bar{q}$ dipole with the proton.  The parameterisation of the dipole cross section is given in terms of a DGLAP-evolved gluon density, fitted to $F_2$ data, with Gaussian impact parameter ($b$) dependence, denoted ``b-Sat'' model~\cite{Kowalski:2003hm,Kowalski:2006hc}, which has already been shown to give a good (parameter-free) description of a wide variety of HERA data on exclusive $\gamma^{*}p\to V+p$ ($V=\rho,\phi,J/\psi,\gamma$) and inclusive $F_2^{c\bar{c}}$, $F_2^{b\bar{b}}$, $F_L$, $F_2^{\rm D}$.  Note that although the ``b-Sat'' model incorporates saturation effects via the eikonalisation of the gluon density, these saturation effects are expected to be only moderate for $J/\psi$ production and negligible for $\Upsilon$ and $Z^0$ production.  We use a ``boosted Gaussian'' vector meson wave function~\cite{Forshaw:2003ki,Kowalski:2006hc}.

\section{Results for Tevatron and LHC}
\subsection{Exclusive $J/\psi$ production}
\begin{figure*}
  \begin{minipage}{0.5\textwidth}
    (a)\hfill$\,$\\
    \includegraphics[width=\textwidth]{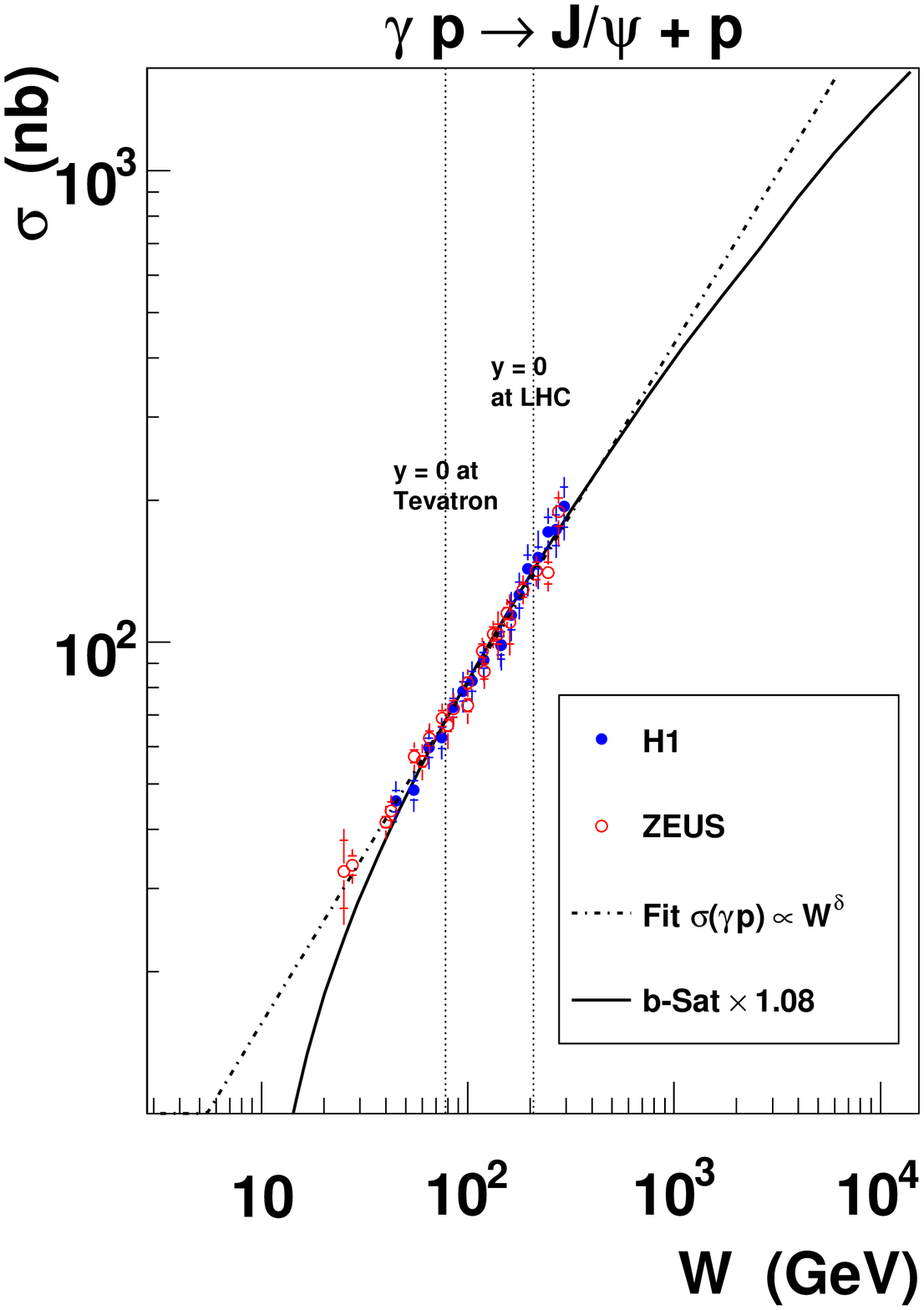}
  \end{minipage}%
  \begin{minipage}{0.5\textwidth}
    (b)\hfill$\,$\\
    \includegraphics[width=\textwidth]{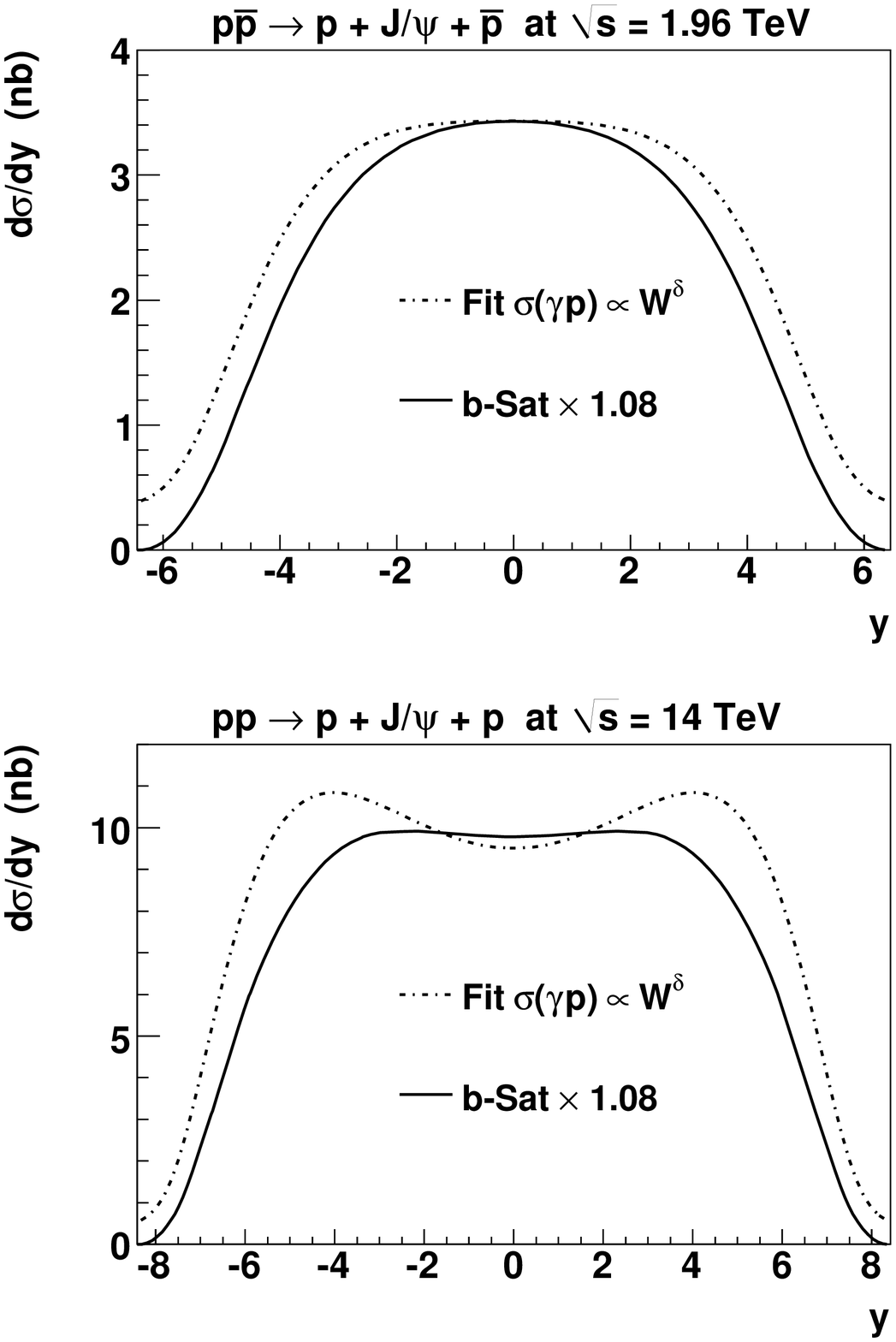}
  \end{minipage}
  \caption{\label{fig:jpsi}Exclusive $J/\psi$ photoproduction in hadron--hadron collisions.  No absorptive corrections are included.}
\end{figure*}
In Fig.~\ref{fig:jpsi}(a) we show the $\gamma p$ cross section as a function of $W$, where we indicate the $W$ values corresponding to central production ($y=0$) at the Tevatron and LHC.  The ``b-Sat'' model predictions are normalised to best fit the HERA $J/\psi$ data~\cite{Chekanov:2002xi,Aktas:2005xu} by a factor 1.08.  Also shown are the results of a direct power-law fit to the HERA data, which gives $\sigma(\gamma p\to J/\psi+p) = (2.96\,\mathrm{nb})(W/\mathrm{GeV})^{0.721}$.  In Fig.~\ref{fig:jpsi}(b) we show the rapidity distributions at the Tevatron and LHC given by Eq.~\ref{eq:rapdist}.  The fact that $y=0$ corresponds to $W$ values where precise HERA data are available means that the uncertainties in the predictions are small.  CDF have recently measured $\left.\mathrm{d}\sigma/\mathrm{d}y\right\rvert_{y=0} = (3.92\pm 0.62)$~nb~\cite{Aaltonen:2009kg}.  The ``b-Sat'' model prediction of $3.4$ nb multiplied by $S^2\simeq 0.9$~\cite{Schafer:2007mm} and an estimated odderon contribution of a factor 1.3--1.6~\cite{Bzdak:2007cz} gives a total theory prediction of (4.0--4.9) nb, in agreement with the Tevatron data.  At the LHC, measurement of exclusive $J/\psi$ production is unlikely to be possible by ATLAS or CMS due to lack of a low-$p_T$ trigger on leptons, but the measurement should be possible by ALICE~\cite{Schicker:2009} and by LHCb.

\subsection{Exclusive $\Upsilon$ production}
\begin{figure*}
  \begin{minipage}{0.5\textwidth}
    (a)\hfill$\,$\\
    \includegraphics[width=\textwidth]{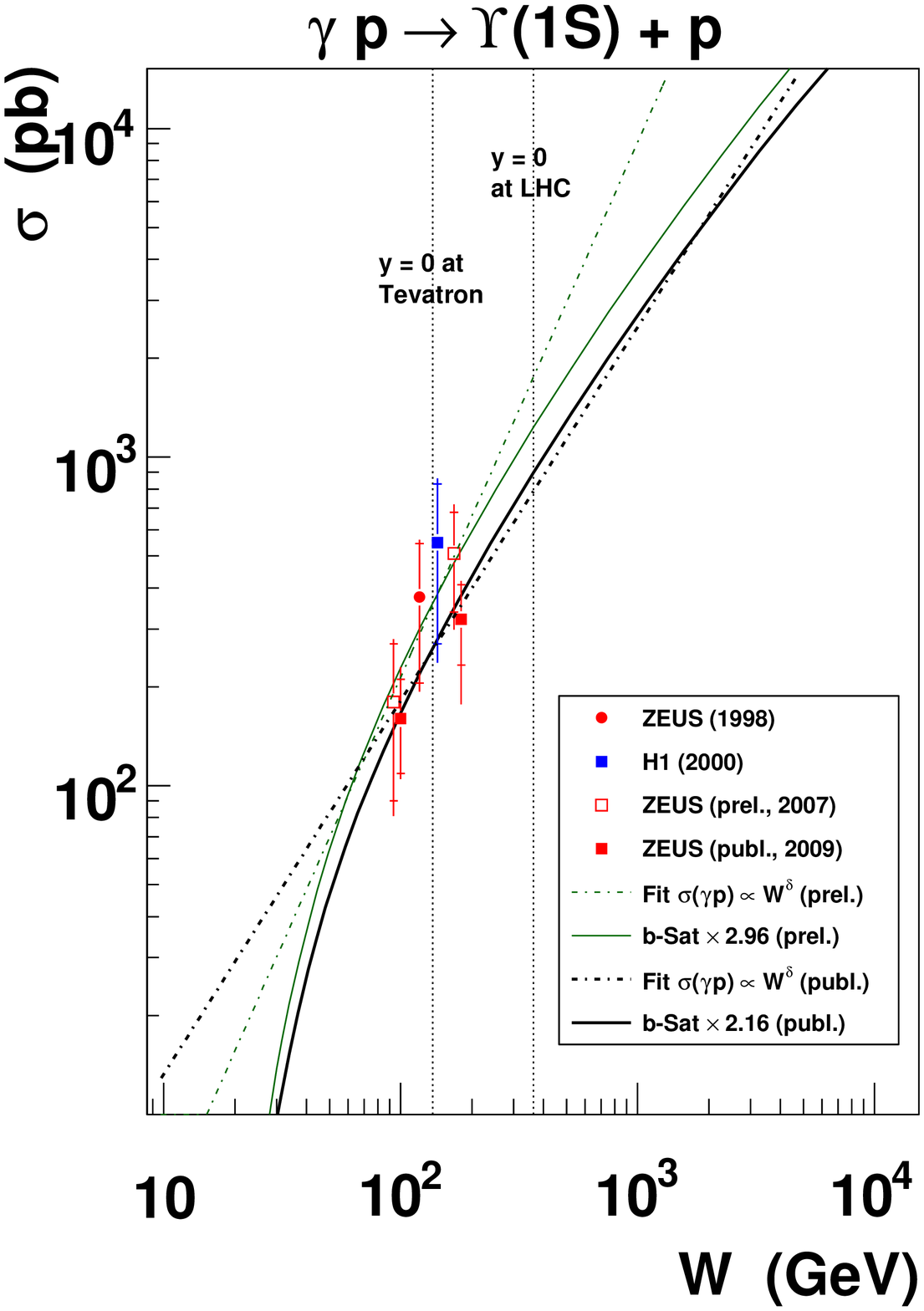}
  \end{minipage}%
  \begin{minipage}{0.5\textwidth}
    (b)\hfill$\,$\\
    \includegraphics[width=\textwidth]{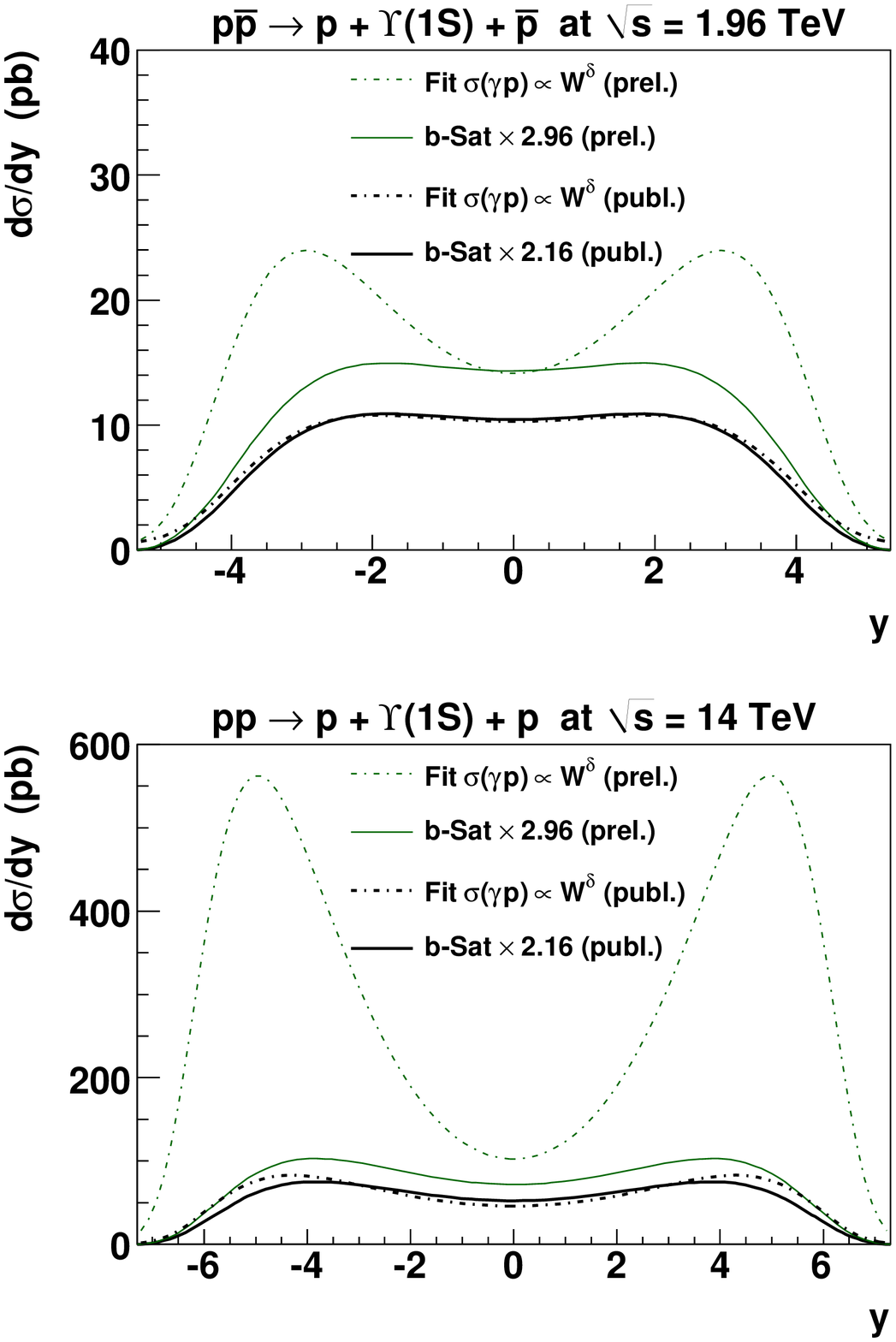}
  \end{minipage}
  \caption{\label{fig:upsilon}Exclusive $\Upsilon$ photoproduction in hadron--hadron collisions.  No absorptive corrections are included.}
\end{figure*}
Only very sparse data are available on exclusive $\Upsilon$ photoproduction at HERA, and consequently, there is a much larger uncertainty in the theory predictions extrapolated to the Tevatron and LHC.  Indeed, until very recently, only two data points with very large errors were published~\cite{Breitweg:1998ki,Adloff:2000vm}, together with a further two preliminary ZEUS data points~\cite{ZEUSupsilon}.  A power-law fit made in Ref.~\cite{Motyka:2008ac} to these four data points, shown in Fig.~\ref{fig:upsilon}(a), gave $\sigma(\gamma p\to \Upsilon+p) = (0.119\,\mathrm{nb})(W/\mathrm{GeV})^{1.63}$.  The two preliminary ZEUS data points, especially the point at highest $W$, moved down slightly in the final measurement~\cite{Chekanov:2009xb}, with a reduction in the size of the uncertainty, so that a revised power-law fit to the four published data points gives $\sigma(\gamma p\to \Upsilon+p) = (0.968\,\mathrm{nb})(W/\mathrm{GeV})^{1.14}$, also shown in Fig.~\ref{fig:upsilon}(a).  The ``b-Sat'' model predictions have a very similar $W$ dependence to this revised power-law fit, but lie more than a factor two below the data for the default choice of $m_b=4.5$ GeV and the ``boosted Gaussian'' $\Upsilon$ wave function.  Given that there are uncertainties in these two choices, which are expected to mainly affect the normalisation but not the $W$ dependence, we have simply rescaled the ``b-Sat'' predictions by a factor 2.16 to best fit the final HERA $\Upsilon$ data.  In Fig.~\ref{fig:upsilon}(b) we show the rapidity distributions at the Tevatron and LHC given by Eq.~\ref{eq:rapdist}.  Note that the results of Ref.~\cite{Motyka:2008ac} for the LHC rapidity distribution, using the preliminary ZEUS $\Upsilon$ data, indicated a large difference between the rescaled ``b-Sat'' predictions and the HERA power-law fit.  But with the new published ZEUS data, the two predictions are in much better agreement.  Of course, the fact that the central value of the power-law fit changes so much when two of the data points shift within their experimental errors, see Fig.~\ref{fig:upsilon}(a), means that the uncertainty on the power-law parameterisation is large.  Ideally, the errors on the two parameters in the power-law fit (given by the experimental errors on the 4 HERA data points), would be propagated through to the Tevatron and LHC rapidity distributions.  An early attempt in this direction, but only for the error in the normalisation and not in the power, was made in Ref.~\cite{Klein:2003vd}.

Candidate exclusive $\Upsilon$ events have been found by CDF and the cross section measurements are eagerly awaited.  Note that the odderon contribution to exclusive $\Upsilon$ production at the Tevatron is predicted to be about the same or even greater than the photon-induced contribution~\cite{Bzdak:2007cz}.  A feasibility study has been carried out by CMS for 100 pb$^{-1}$ of integrated luminosity~\cite{Piotrzkowski:2009}.

The effect of the decay lepton acceptance of the Tevatron and LHC experiments has been studied in Ref.~\cite{Cox:2009ag}, where the results of the power-law fit from Ref.~\cite{Motyka:2008ac} (using the preliminary ZEUS data) were compared with predictions obtained using an alternative parameterisation of the dipole cross section.  Note that LHCb has the potential to measure exclusive $\Upsilon$ production for more forward lepton pseudorapidities than ATLAS or CMS.

\subsection{Exclusive $Z/\gamma^*$ production}
DVCS at HERA, $\gamma^* p\to \gamma+p$, is theoretically cleaner than exclusive vector meson production since there is no uncertainty from the wave function.  The existing data are well described by the ``b-Sat'' dipole model~\cite{Kowalski:2006hc,Watt:2007nr}, although they are not as precise as the HERA data on exclusive $J/\psi$ production.  The DVCS process in $ep$ scattering interferes with the purely electromagnetic Bethe--Heitler (BH) process where the real photon is instead emitted from either the incoming or outgoing electron.  The BH process is precisely calculable in QED and is therefore subtracted in existing DVCS measurements at HERA.  Analogous processes to DVCS at hadron--hadron colliders are exclusive $Z^0$ photoproduction, $\gamma p\to (Z^0\to\ell^+\ell^-)+p$, and timelike Compton scattering (TCS), $\gamma p\to (\gamma^*\to\ell^+\ell^-)+p$.  Similarly to the DVCS case, there is interference of TCS with the pure QED subprocess ($\gamma\gamma\to\ell^+\ell^-$) which is precisely calculable and can be reduced with suitable cuts~\cite{Pire:2008ea}.

Wave functions for an outgoing $Z/\gamma^*$ with \emph{timelike} $q^2=M^2>0$ were derived in Ref.~\cite{Motyka:2008ac}.  Differences were found with respect to the usual \emph{spacelike} case, $q^2=-Q^2<0$, such that the amplitude for $\gamma p\to Z^0 + p$ is \emph{not} simply the DVCS amplitude at $Q^2 = M_Z^2$ with a different coupling.  In particular, we pick up a \emph{real} contribution to the amplitude related to the contribution of an on-shell $q\bar{q}$ pair in addition to the usual imaginary part.  In the dipole picture, direct numerical integration over the dipole size $r$ proved to be difficult due to a wildly oscillatory integrand.  This problem was solved by taking the analytic continuation to complex $r$, then choosing an appropriate integration contour~\cite{Motyka:2008ac}.  Alternatively, there are no such problems if working in transverse momentum space and using $k_T$-factorisation~\cite{Cisek:2009hp}.  
\begin{figure*}
  \begin{minipage}{0.5\textwidth}
    (a)\hfill$\,$\\
    \includegraphics[width=\textwidth]{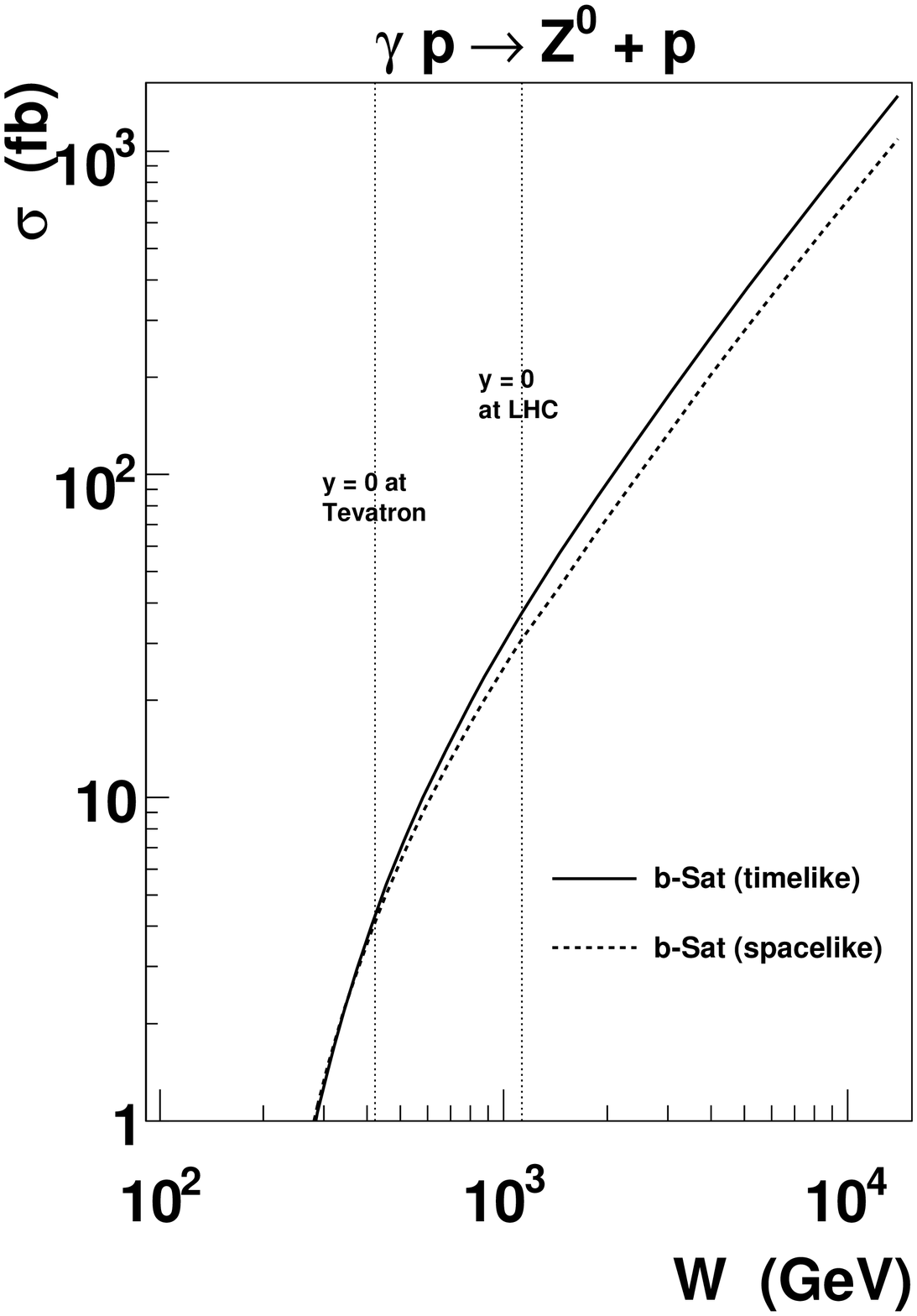}
  \end{minipage}%
  \begin{minipage}{0.5\textwidth}
    (b)\hfill$\,$\\
    \includegraphics[width=\textwidth]{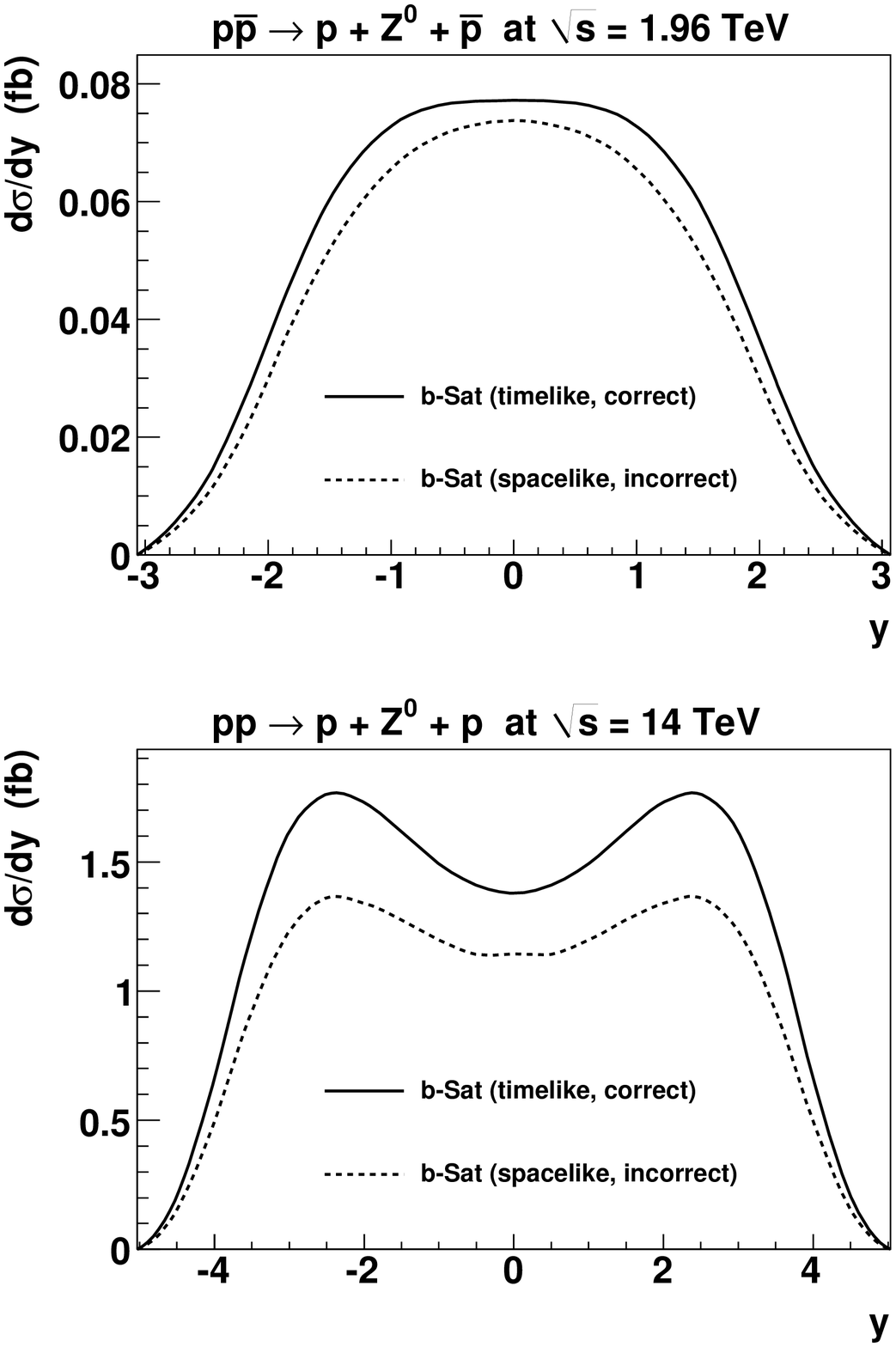}
  \end{minipage}
  \caption{\label{fig:z}Exclusive $Z^0$ photoproduction in hadron--hadron collisions.  No absorptive corrections are included.}
\end{figure*}
We show the results in Fig.~\ref{fig:z}, where it can be seen that the cross sections at $y=0$ are enhanced by 5\% at the Tevatron and 21\% at the LHC in the (correct) timelike case compared to the (incorrect) spacelike case.  (The odderon contribution to exclusive $Z^0$ production is expected to be strongly suppressed~\cite{Motyka:2008ac}.)  The recent calculations of Ref.~\cite{Cisek:2009hp} found cross sections larger by a factor $\sim$3, presumably due mainly to differences in the gluon density at the relevant $x\sim M_Z/\sqrt{s}$, while absorptive corrections were found to lower the cross section by a factor 1.5--2.

CDF have made a search for exclusive $Z^0$ production at the Tevatron~\cite{Aaltonen:2009cj}.  Eight candidate events were found in 2.20 (2.03) fb$^{-1}$ of data in the electron (muon) channel with $M_{\ell\ell}>40$ GeV and $\eta_\ell<4$, consistent with the QED prediction for $\gamma\gamma\to \ell^+\ell^-$.  No candidate events were found in the $Z^0$ mass window, allowing an upper limit to be placed for the exclusive $Z^0$ cross section of $\sigma<0.96$ pb at 95\% confidence-level, compared to the theory prediction of 0.3 fb~\cite{Motyka:2008ac}, i.e.~3000 times lower than the experimental limit.  The theory prediction of $\sigma = 13$~fb~\cite{Motyka:2008ac} at the LHC looks slightly more promising.

\section{Summary}
A summary table of predictions is given in Table \ref{tab:results}, where the event rates (but not the cross sections) include the appropriate leptonic branching ratio.  The $\Upsilon$ cross sections have been revised with respect to Ref.~\cite{Motyka:2008ac}.  Note that the Tevatron and LHC design luminosities are assumed in all cases as an illustrative comparison of the event rates for different processes, but these are not realistic values for early LHC running.  In particular, exclusive $J/\psi$ production is likely to be measured only at ALICE where the nominal luminosity (and so the event rate) is smaller by a factor of 2000 and at LHCb where the corresponding reduction factor is 50.
\begin{table}
    \begin{tabular}{c|c|c|c}
    $J/\psi$ & $\mathrm{d}\sigma/\mathrm{d}y|_{y=0}$ (nb) & $\sigma$ (nb) & Event rate (s$^{-1}$) \\ \hline
    Tevatron & 3.4 & 28 & 0.33 \\
    LHC & 9.8 & 120 & 71 \\
    \hline\hline\multicolumn{4}{c}{}\\\hline\hline
    $\Upsilon(1S)$ & $\mathrm{d}\sigma/\mathrm{d}y|_{y=0}$ (pb) & $\sigma$ (pb) & Event rate (hr$^{-1}$) \\ \hline
    Tevatron & 10 & 83 & 1.5 \\
    LHC & 53 & 771 & 688 \\
    \hline\hline\multicolumn{4}{c}{}\\\hline\hline
    $Z^0$ & $\mathrm{d}\sigma/\mathrm{d}y|_{y=0}$ (fb) & $\sigma$ (fb) & Event rate (yr$^{-1}$) \\ \hline
    Tevatron & 0.077 & 0.30 & 0.065 \\
    LHC & 1.4 & 13 & 134 \\
  \end{tabular}
  \caption{Event rates include leptonic branching ratio and assume a luminosity $\mathcal{L} = 2\times 10^{32}\,{\rm cm}^{-2}\,{\rm s}^{-1}$ (Tevatron) and $\mathcal{L} = 10^{34}\,{\rm cm}^{-2}\,{\rm s}^{-1}$ (LHC).  No gap survival factor included.}
  \label{tab:results}
\end{table}

%% file: LamiLondon09.tex
\begin{center}
{\Large \bf The Totem experiment at the LHC}

\vspace*{1cm}    
S. Lami (On behalf of the TOTEM Collaboration)\\

\vspace*{0.5cm}        
INFN Pisa,    Largo Pontecorvo 3 - 56127 Pisa, Italy 
\end{center}                                                    
                                                    
\vspace*{1cm}   

\begin{abstract}
The TOTEM experiment at the LHC is dedicated to the measurement of the  
total proton-proton cross-section and to the study of the elastic scattering and
diffractive dissociation processes. The main features of the TOTEM experimental 
apparatus and of its physics programme are here 
presented, together with some prospects for early measurements in the first year
of the LHC.

\end{abstract}
\section{Introduction}\label{sec:intro}

TOTEM~\cite{Totem_TDR} foresees specific measurements and experimental techniques which are
very different from the other `general purpose' experiments at LHC.
A precise `luminosity independent' measurement of $\sigma_{TOT}^{pp}$, based on
the Optical Theorem,  will
be achievable in special beam optics runs by simultaneously measuring: 1) the 
elastic scattering rate at low transfer momentum, possibly as small as $|t|=10^{-3}$ GeV$^2$, 
and 2) the inelastic scattering rate with the largest possible coverage to
reduce losses to few percents.
The first goal requires detectors located into units mounted into the vacuum chamber of
the accelerator, called Roman Pots, as the scattered protons are emitted at angles
of the order of 10~$\mu$rad, therefore without leaving the beam--pipe.
The latter requires the measurement of all the inelastically produced particles
in the very forward direction with respect to the $pp$ collision point; this can
be achieved by using tracking detector telescopes with a complete azimuthal
coverage around the beam--pipe.
A flexible trigger provided by TOTEM detectors will allow to take data under all 
LHC running scenarios.
The combination of the CMS and TOTEM experiments
will also allow the study of a wide range of diffractive
processes with an unprecedented coverage in rapidity. For this purpose the 
TOTEM trigger and data acquisition (DAQ) systems are designed to be compatible with the CMS 
ones, in order to allow common data taking periods foreseen at a later stage~\cite{CMS_TOTEM_TDR}.  
Finally, the aim of the TOTEM experiment to obtain accurate information on the
basic properties of proton-proton collisions should also provide
a significant contribution to the understanding of very high energy
cosmic ray physics.
In the following, after a general overview of the experimental apparatus, the main 
features of the TOTEM physics programme will be described.

\section{Experimental Apparatus}\label{sec:detec}
\begin{figure}
\vspace*{-.3in}
\includegraphics[scale=0.6]{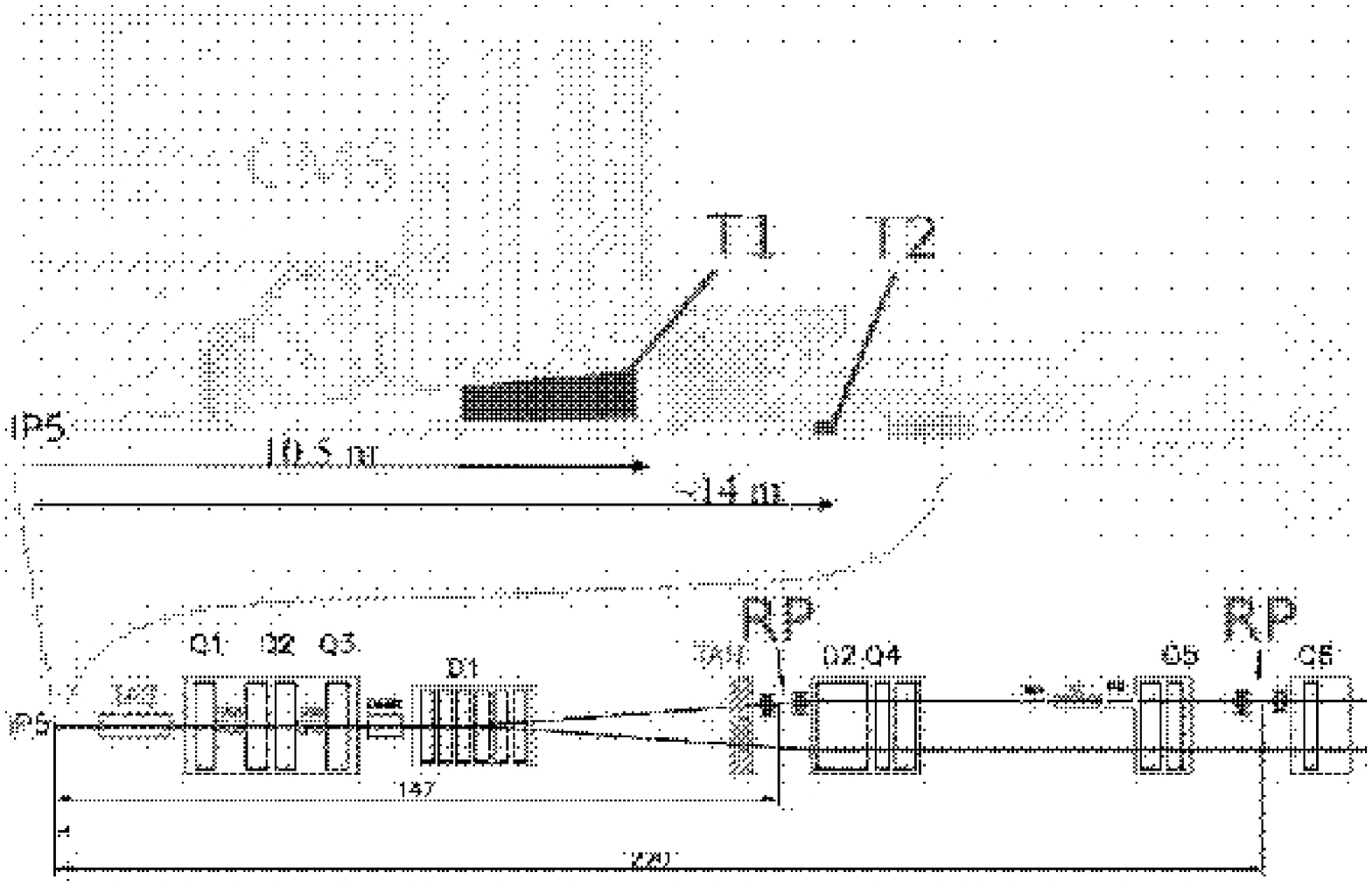}
\caption{Top: The T1 and T2 telescopes embedded into the forward 
region of the CMS detector. Bottom: Roman Pots location along the LHC beam-line. All TOTEM detectors are located on both sides of IP5.}
\label{fig:totem_exp}
\end{figure}
Located on both sides of the interaction point IP5 (shared with the CMS experiment), 
the TOTEM experimental apparatus comprises ``Roman Pot'' (RP) detectors and the T1 and T2 
inelastic telescopes (Fig.~\ref{fig:totem_exp}).
The RPs, placed on the beam-pipe of the outgoing beam in two stations at about 147 m and 220 
m from IP5, are special movable beam-pipe insertions designed 
to detect ``leading'' protons with a scattering angle down to few $\mu$rad. 
T1 and T2, embedded inside the forward region of CMS, provide charged track reconstruction 
for 3.1 $<$ $|\eta|$ $<$ 6.5 ($\eta = -ln(tan\frac{\theta}{2})$) with a 2$\pi$ coverage and 
with a very good efficiency. These detectors will provide a full inclusive trigger 
for all inelastic and diffractive events, minimizing 
losses to a few percent, and will be also used for the reconstruction of the event 
interaction vertex, so to reject background events~\cite{Totem_JNST}. 
The read-out of all TOTEM sub-detectors is based on the digital VFAT chip~\cite{Totem_JNST}, 
specifically designed for TOTEM and characterized by trigger capabilities. 

\begin{figure}
\includegraphics[height=3.9cm]{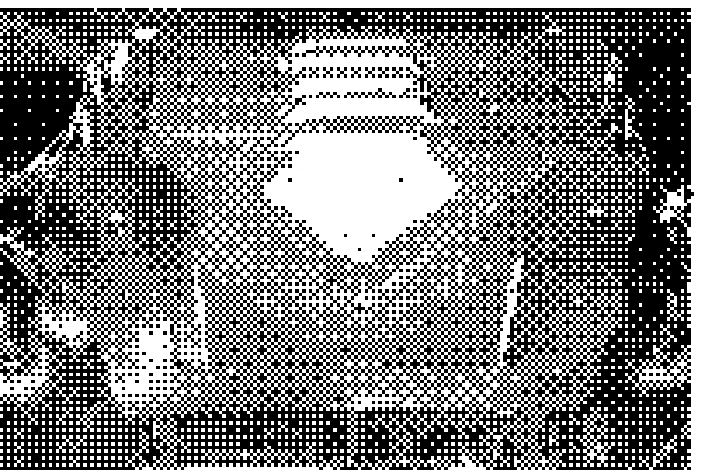}
\includegraphics[height=3.9cm]{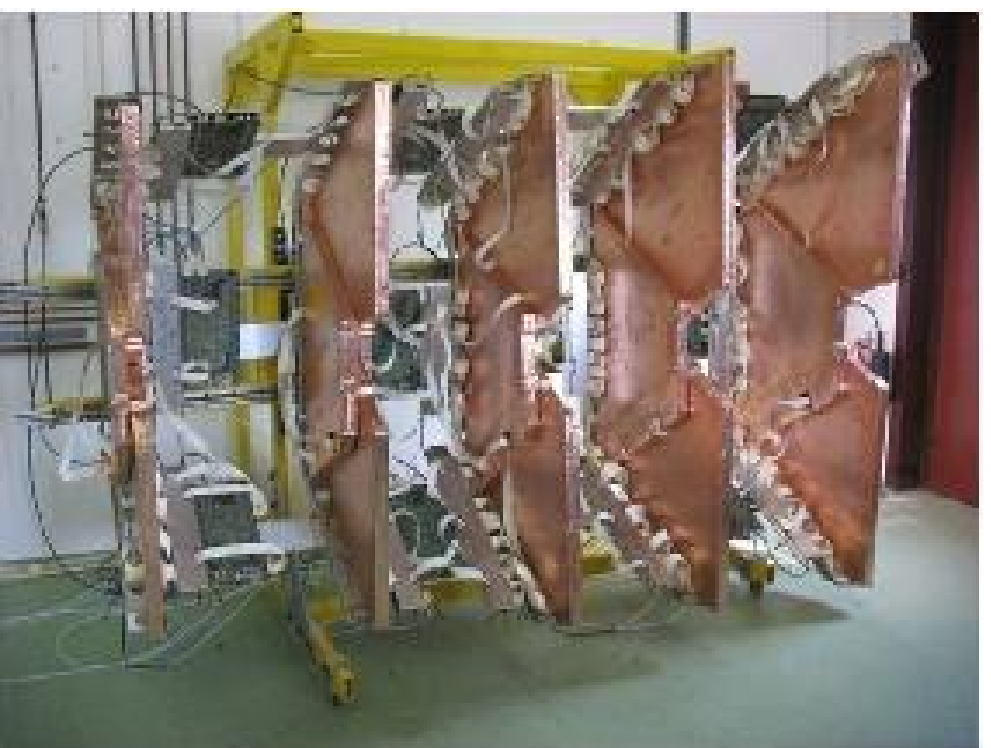}
\includegraphics[height=3.9cm]{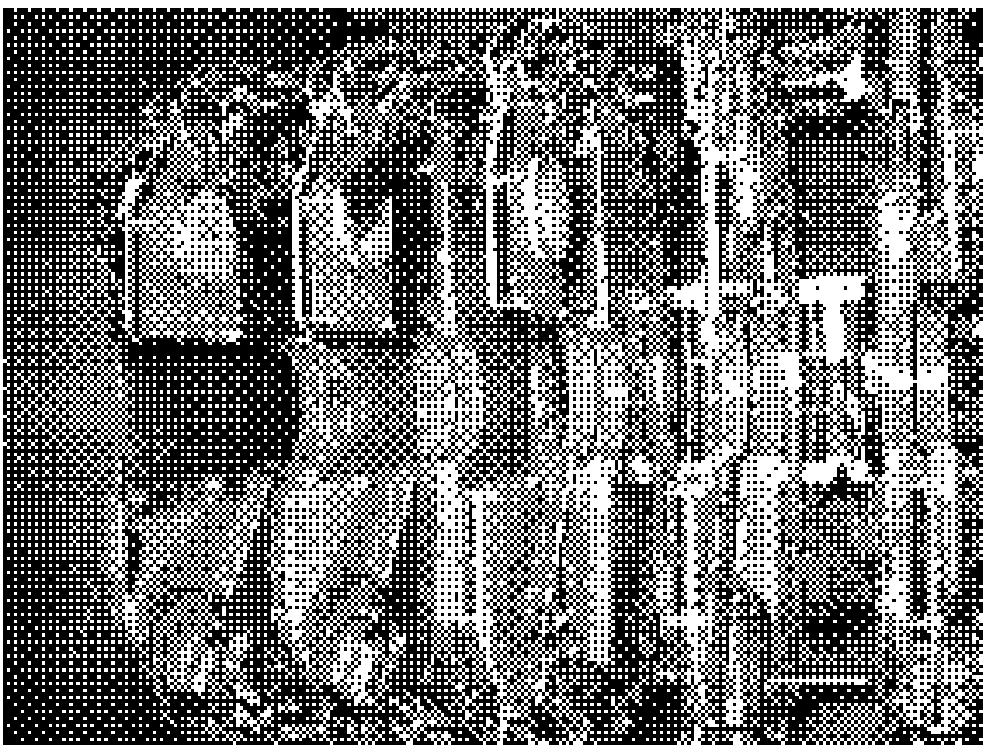}
\caption{Left: Silicon detectors hosted in one pot. Center: 
    One half-arm of the T1 telescope. Right: The installation of one half-arm of the T2 telescope.}
\label{fig:totem_det}
\end{figure}

The RPs host silicon detectors which are moved very close to the beam when it is 
in stable conditions. 
Each RP station is composed of two units in order to have 
a lever arm for local track reconstruction and trigger selections by track angle. 
Each unit consists of three pots, two vertical and one horizontal completing the 
acceptance for diffractively scattered protons. 
Each pot contains a stack of 10 planes of silicon strip detectors 
(Fig.~\ref{fig:totem_det}, left). Each plane has 512 strips (pitch of 
66 $\mu$m) allowing a single hit resolution of about 20 $\mu$m. 
As the detection of protons elastically scattered at angles down to few 
$\mu$rads requires a detector active area as close to the beam as $\sim$ 1 mm, 
a novel ``edgeless planar silicon'' detector technology has been developed 
for TOTEM RPs in order to minimize an edge dead zone to only about 
50 $\mu$m~\cite{RP_Silicon}. 

Each T1 telescope arm, covering the range 3.1 $<$ $|\eta|$ $<$ 4.7, consists of five planes 
formed by six trapezoidal ``Cathode Strip Chambers'' (CSC)~\cite{Totem_TDR} 
(Fig.~\ref{fig:totem_det}, center). These CSCs, with 10 mm thick gas gap and a gas mixture 
of Ar/CO$_2$/CF$_4$ ($40\%/50\%/10\%$), provide three measurements of the charged particle 
coordinates with a spatial resolution of about 1 mm. The anode wires (pitch of 3 mm) 
will also provide 
level-1 trigger information; the 
cathode strips (pitch of 5 mm) are rotated by $\pm$ $60^o$ with respect to the wires.

The T2 telescope~\cite{T2article}, based on ``Gas Electron Multiplier'' (GEM) technology~\cite{GEM}, 
extends charged track reconstruction to the range 5.3 $<$ $|\eta|$ $<$ 6.5. 
Ten aligned detector planes, with a semicircular shape, are combined to form one 
of the half-arms located at $\sim$ 13.5 m from IP5 (Fig.~\ref{fig:totem_det}, right). 
This novel gas detector technology is ideal for the T2 telescope 
thanks to its good spatial resolution, excellent rate capability and good 
resistance to radiation. The T2 GEMs are characterized by 
a triple-GEM structure and a gas mixture of Ar/CO$_2$ (70$\%$/30$\%$)~\cite{T2article}. 
The read-out board has two separate layers with different patterns: 256x2 concentric 
circular strips (80\,$\mu$m wide, pitch of 400\,$\mu$m), allow the track radial coordinate
reconstruction with a resolution of about 100\,$\mu$m; while a matrix of 24x65 pads 
(from 2x2\,mm$^2$ to 7x7\,mm$^2$ in size) provide level-1 trigger information and 
track azimuthal coordinate reconstruction.

\section{Physics Programme}\label{sec:phys}

The physics goals of the TOTEM experiment are the measurement of the total $pp$ cross section 
($\sigma_{tot}$) with an ultimate precision of 1$\div$2\,$\%$, the study of the nuclear elastic 
$pp$ differential cross section ($d{\sigma}_{el} / dt$) over a wide range of 
$|t|$ ($\sim 10^{-3} < |t| < 10\,{\rm GeV}^{2}$) and the study of the inelastic interactions
by measuring the cross section of soft diffractive processes and the forward charged particle flow. 
TOTEM can perform this programme operating in stand-alone mode, while the cooperation
with CMS will make possible the study of hard diffraction, low-x physics, central exclusive
diffractive production and the combination of particle multiplicity and energy flow in the
forward region~\cite{CMS_TOTEM_TDR}. 
TOTEM measurements will allow to distinguish among different models of soft proton 
interactions, giving a deeper understanding of the proton structure.
%
\begin{figure}
\includegraphics[height=5.9cm]{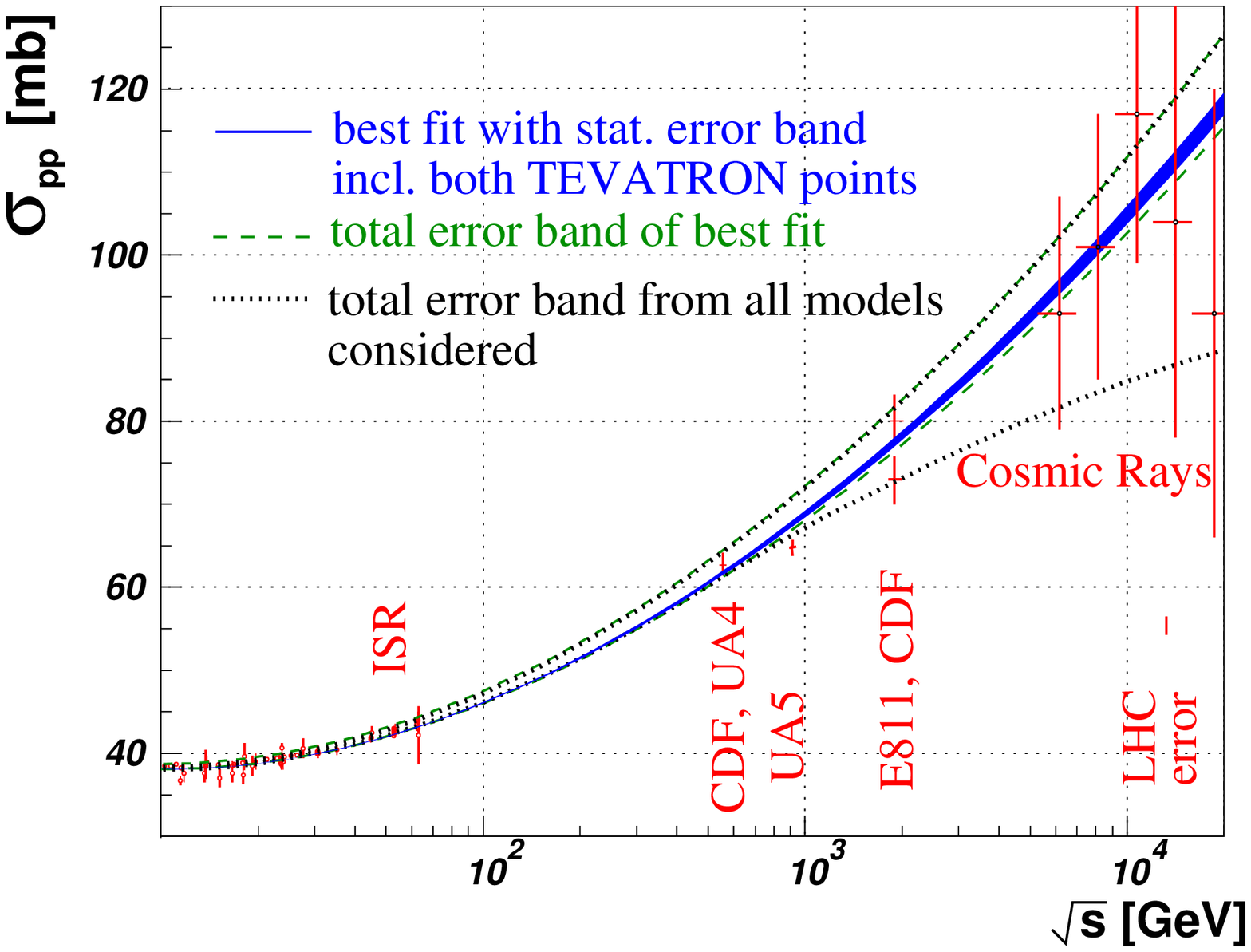}
\hspace*{1.cm}\includegraphics[height=6.2cm]{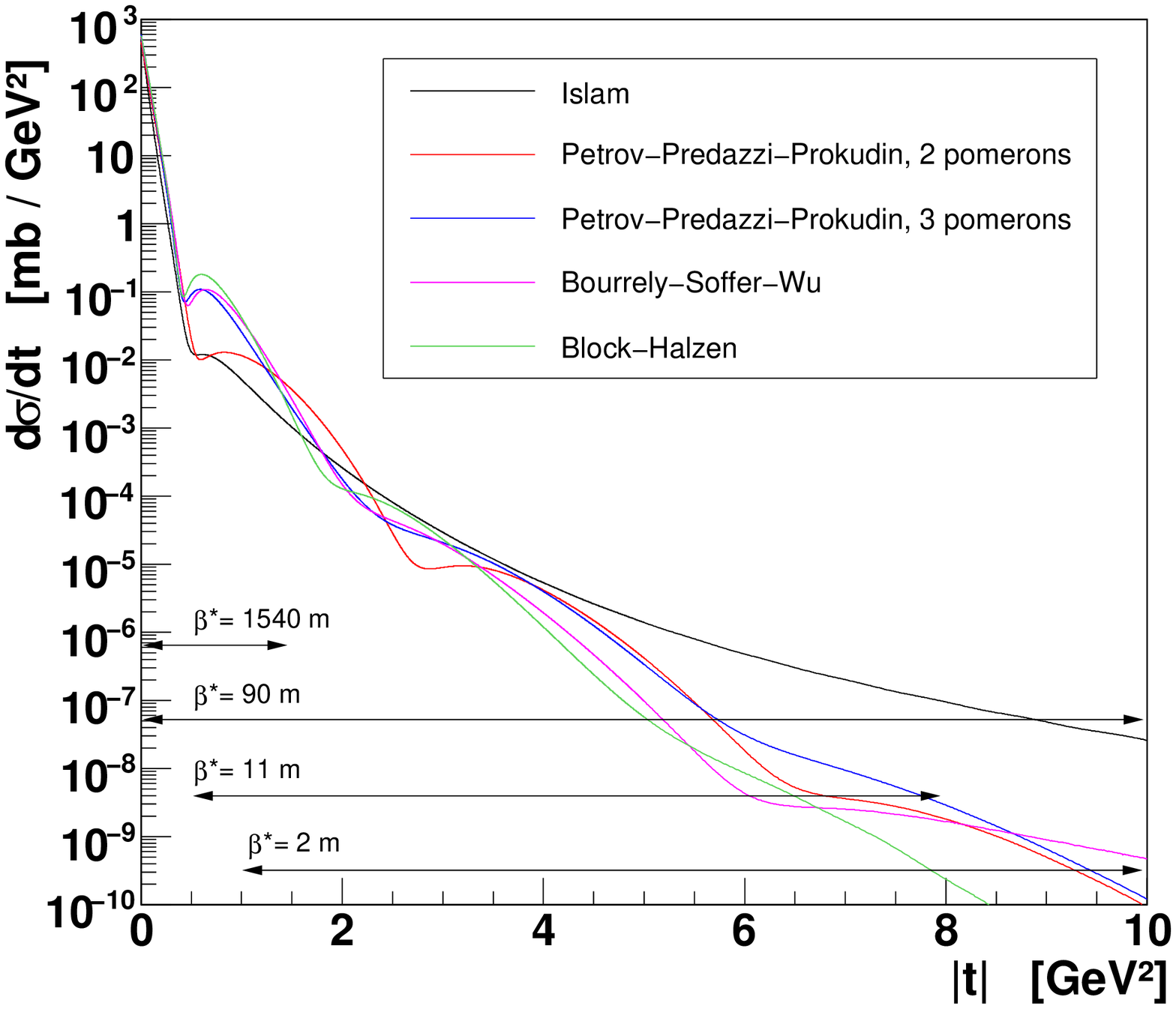}
\vspace*{-0.2cm}
\caption{Left: Fits from the COMPETE Collaboration to available $pp$ and $p\bar p$ scattering 
and cosmic ray data. Right: $d{\sigma}_{el} / dt$ at the LHC as predicted by 
different models; t-acceptance ranges for different machine optics are also shown.}
\label{fig:sigma_pp_el}
\end{figure}

Fig.~\ref{fig:sigma_pp_el} (left), summarizing the existing measurements
on $pp$ and $p\bar p$ scattering, shows predictions for $\sigma_{tot}$ from the COMPETE 
Collaboration based on fits according to different models~\cite{COMPETE}. 
The best fit predicts $\sigma_{tot} = 111.5 \pm 1.2 ^{+4.1}_{-2.1}$ mb for the LHC energy 
($\sqrt{s}$ = 14 TeV). The large uncertainties on available high energy data give a 
big error (90$\div$130 mb), depending on the model used for the extrapolation. 
TOTEM aims to measure $\sigma_{tot}$ with a precision down to 1$\div$2\,$\%$, 
therefore allowing to discriminate among the different models. The measurement will 
be based on the ``luminosity independent method'' which, combining the optical 
theorem with the total rate, gives $\sigma_{tot}$ and the machine 
luminosity ($\mathcal{L}$) as a function of measurable rates: 
\begin{equation}
   \sigma_{tot} = \frac{16 \pi}{1 + \rho^{2}} \cdot
   \frac{dN_{el}/dt |_{t=0}}{N_{el} + N_{inel}}
 \hskip 2.0cm
   \mathcal{L} = \frac{1 + \rho^{2}}{16 \pi} \cdot
   \frac{(N_{el} + N_{inel})^{2}}{dN_{el}/dt |_{t=0}} 
  \label{eq:stot_lumi}
\end{equation}
where $N_{el}$ and $N_{inel}$ are respectively the elastic and inelastic 
rate and $\rho$ is the ratio of real to imaginary part of the forward 
nuclear elastic scattering amplitude (given by theoretical predictions). 
Therefore, TOTEM will also provide an absolute measurement of $\mathcal{L}$ 
to be used for calibration purposes.
The uncertainty on the extrapolation of $dN_{el}/dt$ to $t = 0$ (optical point) depends on 
the acceptance for protons scattered at small $|t|$ values, hence 
at small angles. This requires a small beam angular divergence at the IP, which can be 
achieved in special runs with high $\beta^*$ machine optics and typically low $\mathcal{L}$. 
An approved optics with $\beta^* =$ 1540 m (and $\mathcal{L} \sim$ 10$^{28}$ cm$^{-2}$s$^{-1}$) 
will give a $\sigma_{tot}$ ($\mathcal{L}$) measurement at the level of 1$\div$2\,$\%$ 
(2\%). Another approved optics with $\beta^* =$ 90 m (and $\mathcal{L} \sim$
10$^{30}$ cm$^{-2}$s$^{-1}$), achievable without modifying the standard LHC injection optics, 
is expected to allow a preliminary $\sigma_{tot}$ ($\mathcal{L}$) measurement at the level 
of $\sim$ 5\% (7\%) in the first period of the LHC running. 
The experimental systematic error for 
the measurement with $\beta^* =$ 90 m will be dominated by the evaluation 
of $dN_{el}/dt |_{t=0}$, while with $\beta^{*} = 1540\,$m it will be 
dominated by the uncertainty on the corrections to trigger losses in Single 
Diffraction events for masses below $\sim$10\,GeV/c$^2$ ~\cite{Totem_JNST}. 
Given the high rates involved, the statistical error on $\sigma_{tot}$
will be negligible after few hours of data taking even at low $\mathcal{L}$.
The theoretical uncertainty related to the estimate of the $\rho$ parameter is expected 
to give a contribution on the relative uncertainty of less than 1.2\% (considering for 
instance the full error band on $\rho$ extrapolation as derived in ref~\cite{COMPETE}).
 
\begin{figure}
\includegraphics[height=6.cm]{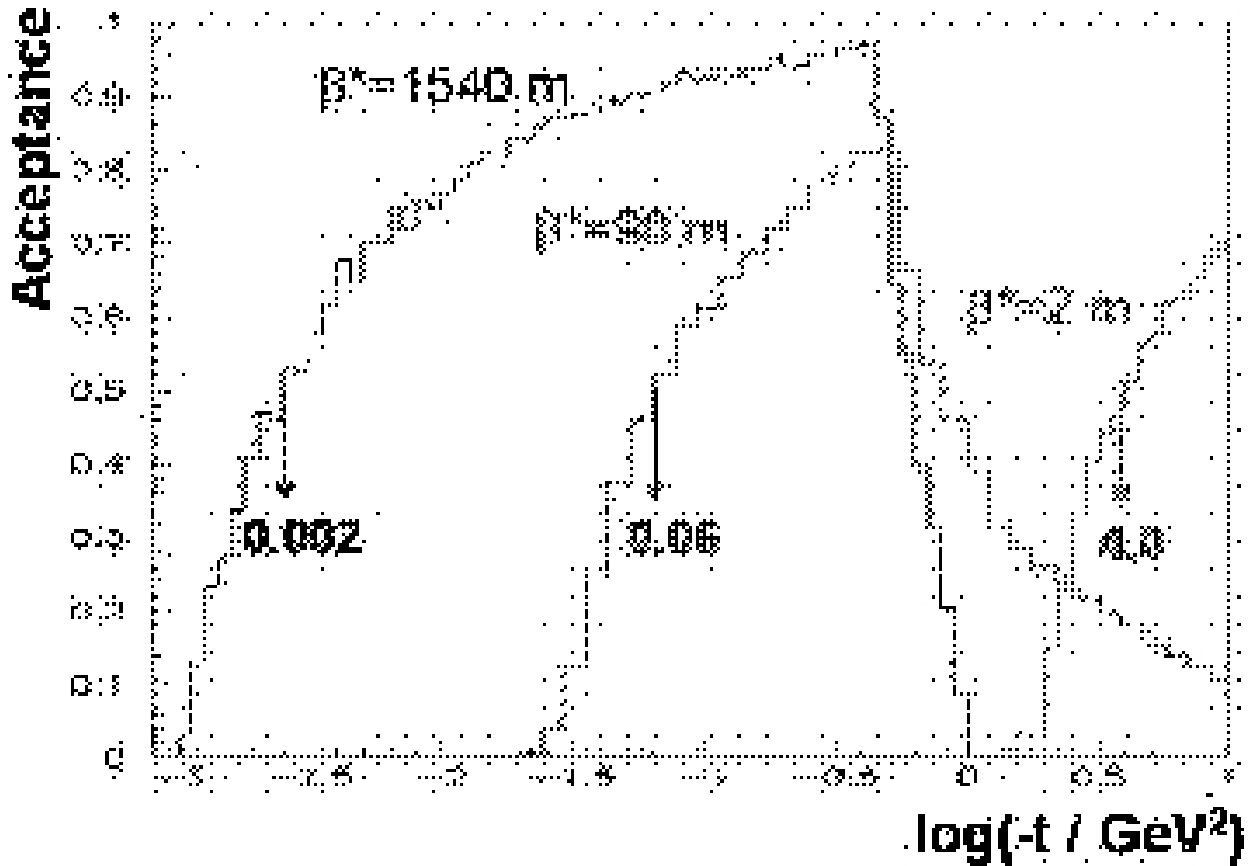}
\hspace*{.1cm}\includegraphics[height=6.cm]{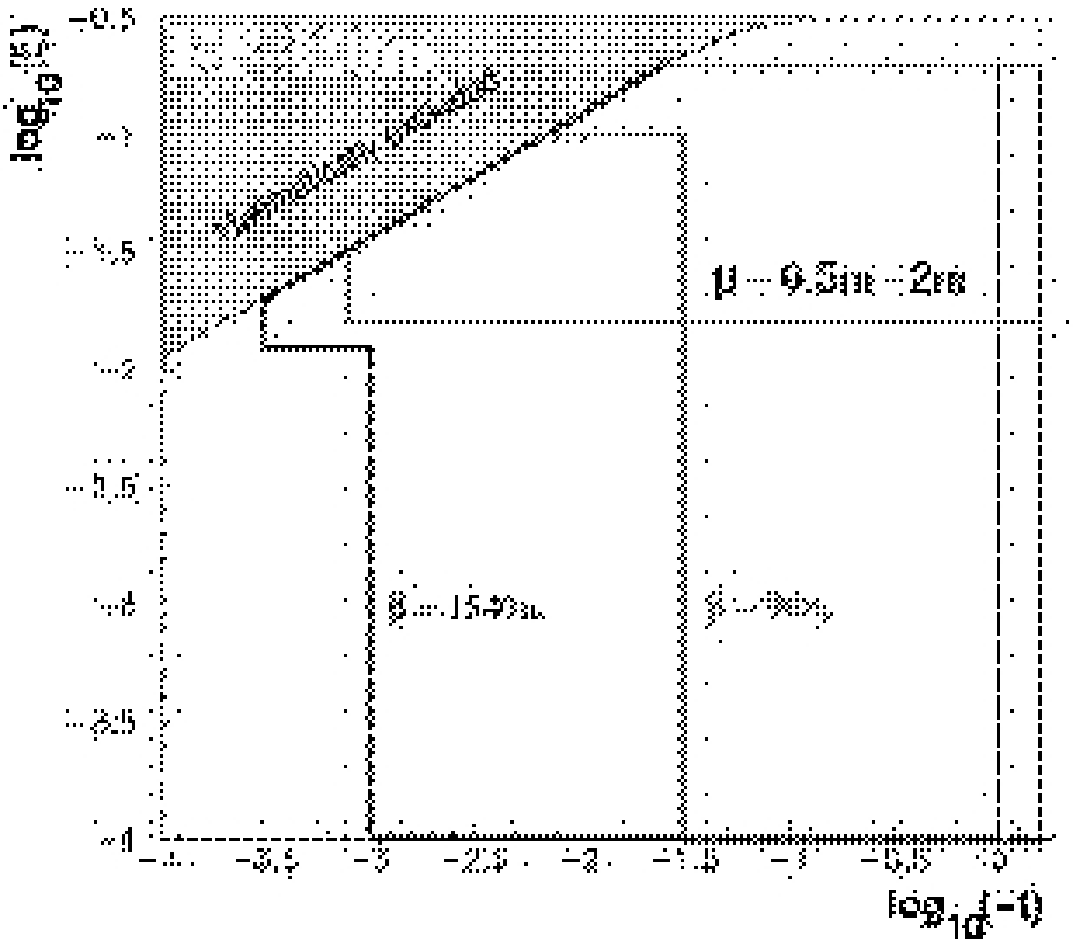}
\caption{Left: Proton acceptance  for the RPs at 220 m 
as a function of $t$ for three different running scenarios, where the 50$\%$ acceptance
is marked for each ${\beta}^*$ value.
 Right: Diffractive proton acceptance for the RPs at 220 m for different
optics as a function of $\xi$ and $t$.}
\label{fig:diff_acc}
\end{figure}

Fig.~\ref{fig:sigma_pp_el} (right) shows the distributions of $d{\sigma}_{el} / dt$
at $\sqrt{s}$ = 14 TeV as predicted by different models in the whole $|t|$-range 
accessible by TOTEM according to the different LHC optics settings~\cite{Totem_JNST}. 
Several regions are found at increasing $|t|$: for $|t| < 6.5 \times 10^{-4}$ GeV$^2$ 
(the Coulomb region)  
$d{\sigma}_{el} / dt \sim 1 / |t|^{2}$ is dominated by photon exchange;
for $|t|$ up to $\sim 10^{-3}$ GeV$^2$, the hadronic and Coulomb scattering 
interfere; for $10^{-3} < |t| < 0.5$ GeV$^2$ there is the hadronic region, 
e.g. described by ``single-Pomeron exchange'', characterized by an approximately 
exponential fall ($d{\sigma}_{el} / dt \sim {\rm e}^{-B\,|t|}$); 
the diffractive structure of the proton is then expected for $0.5 < |t| < 1$ GeV$^2$; 
for $|t| >1$ GeV$^2$ elastic collisions are described by pQCD, 
e.g. in terms of triple-gluon exchange with $d{\sigma}_{el} / dt \sim |t|^{-8}$. 
TOTEM will allow to discriminate among different models with a precise
measurement of $d{\sigma}_{el} / dt$ over all the accessible $t$-region, where 
$d{\sigma}_{el} / dt$ spans over 11 orders of magnitude.
In the hadronic region, important for the extrapolation of $d{\sigma}_{el} / dt$ to $t=0$,  
a fit on $B(|t|)$ is typically performed in the $|t|_{min} < |t| < 0.25$ GeV$^2$ range,  
$|t|_{min}$ depending on the acceptance for protons scattered at small angles, hence on 
the beam optics. 
Fig.~\ref{fig:diff_acc} (left) shows the proton acceptance  for the RPs at 220 m 
as a function of $t$ for three different running scenarios, where the 50$\%$ acceptance
is marked for each ${\beta}^*$ value. For safety reasons, the minimum distance
between RP detectors and the LHC beam is 10$\sigma_{beam}$; by adding an extra 0.5 mm
to take in account the distance from the edge of the sensitive detector area to the
bottom of the RP window, this distance is about 1.3 mm for RP220 with  ${\beta}^*=$ 1540 m,
corresponding to a minimum angle of about 5 $\mu$rad or equivalently 10$^{-3}$ GeV$^2$ of
squared 4-momentum transfer.

Diffractive (due to colour singlet exchange) and non-diffractive 
(due to colour exchange) inelastic interactions represent a big 
fraction (around $70\div 75$ $\%$) of $\sigma_{tot}$.
Nevertheless many details of these processes, with close ties to proton structure 
and low-energy QCD, are still poorly understood. 
The majority of diffractive events exhibits intact (``leading'') protons 
characterized by their $t$ and fractional momentum loss $\xi \equiv \Delta p/p$. 
TOTEM will be able to measure $\xi$-, $t$- and mass-distributions with acceptances 
depending on the beam optics. 
The diffractive proton acceptance for the RPs at 220 m has been computed for different
optics as a function of $\xi$ and $t$ (Fig.~\ref{fig:diff_acc}, right). 
For ${\beta}^*=$ 1540~m or 90 m most of the protons are seen independently of
their momentum loss. For low ${\beta}^*$ values, elastic scattering is detectable only
for $|t| > 2$ GeV$^2$, while diffracted protons are seen for $\xi >$ 2$\%$ for all $|t|$ values.

When leading protons
are detected on both sides of IP5, as in the case of Double
Pomeron Exchange (DPE), the central mass  $M$ acceptance is shown in Fig.~\ref{fig:early_plots}, left.
By combining data from low, intermediate and high $\beta^*$ runs, the differential
cross-section as a function of $M$ can be measured with good precision over the full
mass range.


\section{Early Physics Programme}\label{sec:early}

\begin{figure}
\includegraphics[height=6.cm]{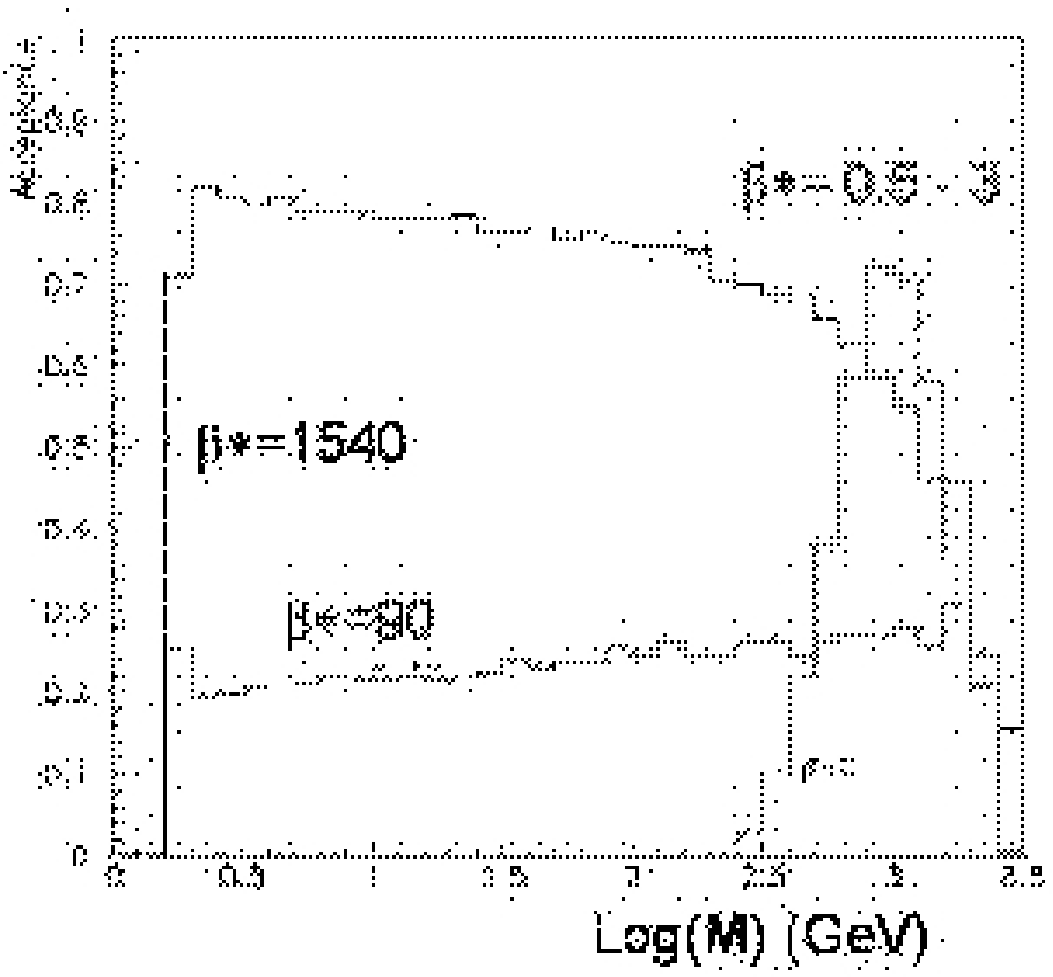}
\hspace*{.3cm}\includegraphics[height=6.2cm]{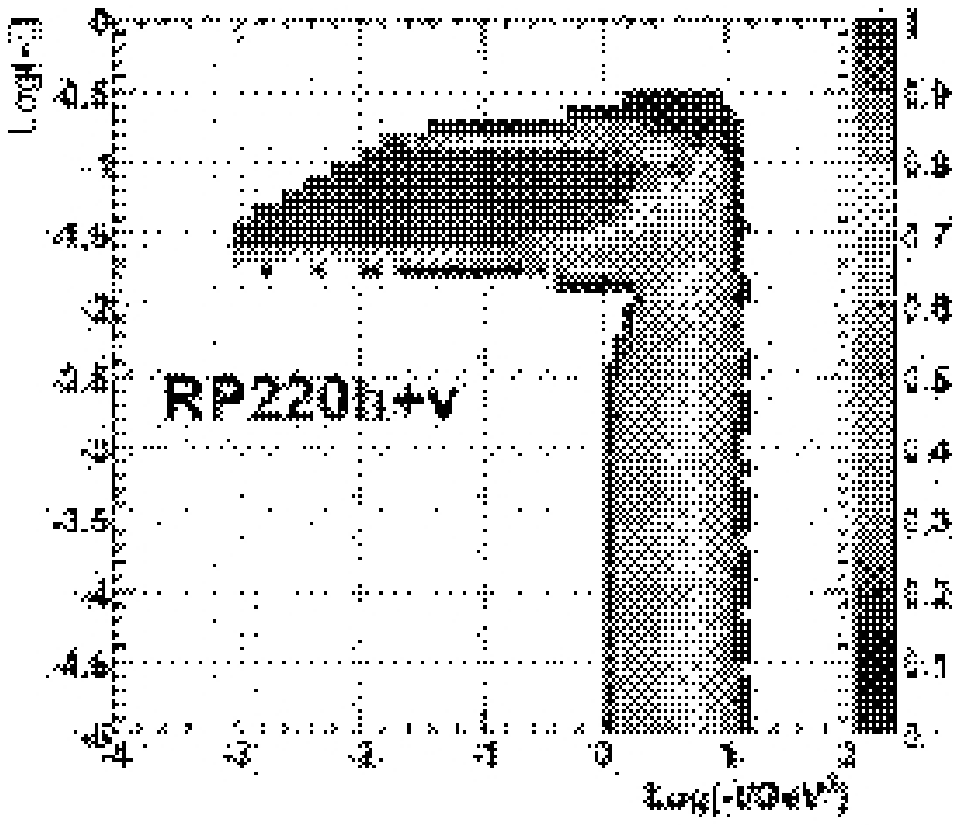}
\caption{Left: The DPE central mass $M$ distribution for three $\beta^*$ values. 
Right: Proton acceptance for early runs at the 220 m RP stations.}
\label{fig:early_plots}
\end{figure}

LHC will soon restart with a beginning center of mass energy of 7 TeV and a low
${\beta}^*$ optics. We are studying the TOTEM performance under these conditions, however
we are also asking for ${\beta}^* =$90~m for short runs in 2010. The acceptance for protons
at the 220 m RP stations is shown in Fig.~\ref{fig:early_plots} (right); they are mainly
seen for a momentum loss $\xi >$ 2$\%$, with a resolution on $\xi$ slightly less than
10$^{-2}$. These results are for a low ${\beta}^*$ optics, they depend very strongly on the
beam optics and little on the energy, so that they are similar to those obtained for 14 TeV.
For DPE,
the predicted central diffractive mass distribution would start at around 250 GeV, a
significant difference from the very low mass of few GeV reachable with a high
${\beta}^*$ optics (see Fig.~\ref{fig:early_plots}, left).
Therefore the very first measurements with a low ${\beta}^*$ will consist in: 1) by using the
horizontal RPs, in Single Diffractive and DPE for high masses of the
diffractive system; 2) by using the vertical RPs for the measurement of the elastic
scattering over a range of 1$< |t| <$ 10 GeV$^2$, which would not allow a measurement of
 $\sigma_{tot}$ but, however, it will already allow to discriminate among different models for
the prediction of $d{\sigma}_{el} / dt$.

The T1 and T2 telescopes will provide a measurement of charged particle multiplicity
for different processes as well as the identification and measurement of rapidity gaps.
Primary cosmic rays in the PeV (10$^{15}$ eV) energy range and above are a challenging
issue in astrophysics.
The LHC center of mass energy corresponds to a 100 PeV energy for a fixed target
collision in the air, at the same time providing a high event rate relative to
the very low rate of cosmic particles in this energy domain.
A primary cosmic ray entering the upper atmosphere experiences a nuclear interaction,
with the production of nuclear fragments and $\pi$ mesons, starting an air shower
with hadronic, electromagnetic and muon components. The real challenge is to
determine the nature of the primary interaction and the energy and composition
of the incident particle from the measurement of the shower.
Several high energy hadronic interaction models are nowadays available, which predict
energy flow, multiplicity and other quantities of such showers.
There are large differences between the predictions of currently available
models, with significant inconsistencies in the forward region.
Among the several quantities that can be measured by TOTEM and compared with
model predictions, the charged particle
multiplicity in the T1 and T2 acceptance region shows significant differences
in the predictions obtained by different available hadronic interaction
models for cosmic ray showers, once the events are passed through the
simulation of T1 and T2. Therefore this measurement can already
be used in early runs
to validate/tune the generators~\cite{CMS_TOTEM_TDR}.
  
\section{Summary and Conclusions}

The TOTEM experiment will be ready for data taking at the LHC restart. 
Running under all beam conditions, it will perform an important 
and exciting physics programme involving the measurement of $\sigma_{tot}$ and 
$d{\sigma}_{el} / dt$ in $pp$ interactions as well as studies on diffractive 
processes and on forward charged particle production. 
Special high ${\beta}^*$ runs will be required in order to measure 
$\sigma_{tot}$ at the level of $\sim$ 5 $\%$ (early measurement 
with ${\beta}^*$ = 90 m) and $\sim 1\div 2$ $\%$ (${\beta}^*$ = 1540 m). 
$d{\sigma}_{el} / dt$ will be studied in the range 
$\sim 10^{-3} < |t| < 10\,{\rm GeV}^{2}$ allowing to distinguish 
among several theoretical predictions. A common physics programme 
with CMS on soft and hard diffraction as well as on forward particle 
flow studies is also foreseen in a later stage.



%% file: Pilkington-Andrew-LondonSM2009.tex
\begin{center}
{\Large \bf Forward Physics with early ATLAS data}
\vspace*{1cm}

Andrew Pilkington (on behalf of the ATLAS Collaboration)
\vspace*{0.5cm}

School of Physics and Astronomy, University of Manchester, Oxford Road, Manchester, UK, M13 9PL.
\vspace*{1cm}
\end{center}

\begin{abstract}
The ATLAS forward detector system is presented. Following this, the forward physics measurements that are expected to be carried out with early ATLAS data are introduced and the relevant trigger and analysis strategies discussed.
\end{abstract}

\setcounter{section}{0}
\setcounter{figure}{0}
\setcounter{table}{0}
\setcounter{equation}{0}

\section{\label{sec:detectors} The ATLAS forward detector system}

ATLAS is a general multipurpose detector \cite{atlas} designed to be able to measure a wide variety of physics processes at the LHC. ATLAS has a standard layout of sub-detectors; an Inner Detector ($|\eta|<2.5$) for tracking purposes, electromagnetic calorimeters ($|\eta|<3.2$) for measuring the energy of electrons and photons, hadronic calorimeters ($|\eta|<4.9$) to measure the energy of mesons and baryons, and a muon spectrometer ($|\eta|<2.7$). In addition, there are a number of other sub-detectors designed to measure (forward) particle production; these are the MBTS, LUCID, ZDC and ALFA.

The Minimum Bias Trigger Scintillators (MBTS) are located at $|z|=3.6$~m and cover $2.1<|\eta|<3.8$ \cite{mbts}. The MBTS is designed to trigger so-called {\it minimum bias events}, identifying particle production from inelastic collisions. Each side of the MBTS (positive and negative) are divided into 16 segments: two $\eta$-rings and eight $\phi$-sections. The multiplicity of hit segments on each side will be available for the Level 1 (L1) trigger. The MBTS detector will only be available during the low luminosity phase of LHC operation, as it was not designed to have high radiation tolerance.

The LUCID detectors \cite{lucid} are located 17~m from the interaction point, one on each side of 
ATLAS, and provide coverage of $5.6 < |\eta|< 6.0$ for charged particles. Each LUCID detector 
is a symmetric array of polished aluminium tubes that surround the beam-pipe. Each tube 
is 15~mm in diameter and filled with C$_4$ F$_{10}$ gas, which results in Cerenkov emission from 
charged particles crossing the tube. The Cerenkov light is read out by photo-multiplier tubes. 
The initial LUCID detector, available for early data taking, has L1 trigger capabilities but only limited azimuthal coverage. An upgrade to provide full azimuthal coverage is under study.

The Zero Degree Calorimeter (ZDC) \cite{zdc} is located 140~m from the interaction point in 
the TAN region (target absorber for neutrals), where the single beam-pipe 
splits into two, and provides coverage of $|\eta| > 8.3$ for neutral particles. The ZDC consists 
of one electromagnetic and three hadronic tungsten/quartz calorimeters. Vertical quartz strips 
provide the energy measurements and horizontal quartz rods are used for coordinate readout. 
At LHC startup, when there are few bunches in the beam, the electromagnetic calorimeter 
is not installed and the space it would occupy is used by the LHCf experiment. After initial 
running, LHCf will be removed and the full ZDC installed. 

The ALFA roman pot (RP) spectrometers are located 240~m from the interaction point \cite{alfa}. Unlike other detectors, the RP spectrometers are not fixed relative to the beam. At 
injection, the ALFA detectors are in a withdrawn position far from the beam. After the protons have been injected and the beam 
has been stabilized, the detectors are moved to within 1.5~mm of the beam. Elastic and diffractive 
protons that have been deflected outside the beam envelope pass through arrays of scintillating fibre trackers (20$\times$64 fibres in each array), which measure the position of the protons with respect to the beam. This allows the momentum and angle of each proton to be reconstructed by using the known LHC lattice. ALFA will only be used 
during special LHC runs at low luminosities with high $\beta^{*}$ optics. 

\section{\label{sec:soft} Soft diffraction}

Single diffraction is a low $t$-process in which a colour singlet  (i.e. pomeron) is exchanged between the two protons and one of the protons breaks up into a dissociative system. There is a large rapidity gap between the outgoing proton and the dissociative system as a colour singlet object was exchanged. The cross section for soft single diffraction (SD) is expected to be of the order 10~mb at the LHC. There are two approaches that will be used to measure soft-SD at ATLAS.

The first approach will identify the dissociated system using the inner detector, calorimeters, LUCID and possibly the ZDC. The variable of interest is the fractional longitudinal momentum loss, $\xi$, given by 
\begin{equation}
\xi = \frac{M_{{\rm X}}^2}{s}
\end{equation}
where $M_{{\rm X}}$ is the invariant mass of the dissociative system and $s$ is the centre-of-mass energy of the $pp$ collision. It follows that the dissociated system for events with low-$\xi$ will be contained only in the forward detectors, whereas high-$\xi$ events will have activity in many areas of the central detector as well. The MBTS, LUCID and the ZDC will be required to trigger events across the full kinematic range. It has been estimated that a sample of one million events can be collected in two weeks given a luminosity of 10$^{31}$~cm$^{-2}$~s$^{-1}$.

The second approach is to tag the outgoing proton and measure $\xi$ directly using 
\begin{equation}
\xi = 1 - \frac{|p_{{\rm z}}^{\prime}|}{|p_{{\rm z}}|}
\end{equation}
where $p_{{\rm z}}$ and $p_{{\rm z}}^{\prime}$ are the longitudinal momenta of the incoming and outgoing protons respectively. This approach requires the ALFA RP detectors, which can only be used in special runs with high-$\beta^{*}$ optics at a luminosity of 10$^{27}$~cm$^{-2}$~s$^{-1}$. ALFA will be able to measure the fractional momentum loss, with an accuracy between 8\% (for $\xi \sim 0.01$) and 2\% (for $\xi\sim 0.1$). The trigger will be provided by ALFA and it is expected that 1.2-1.8 million events will be retained with just  100~hrs of data taking \cite{alfa}.  The analysis will also make use of LUCID and the ZDC, which are used to tag the dissociative system to separate the events from elastic scattering. 

In addition to soft single diffraction, it is expected that ATLAS will also be able to measure soft double diffraction, in which both protons form dissociative systems.

\section{\label{sec:dijet} Diffractive di-jet production}

\begin{figure*}[ht]
\centering
\mbox{
\subfigure[]{\includegraphics[width=.3\textwidth, height=4cm]{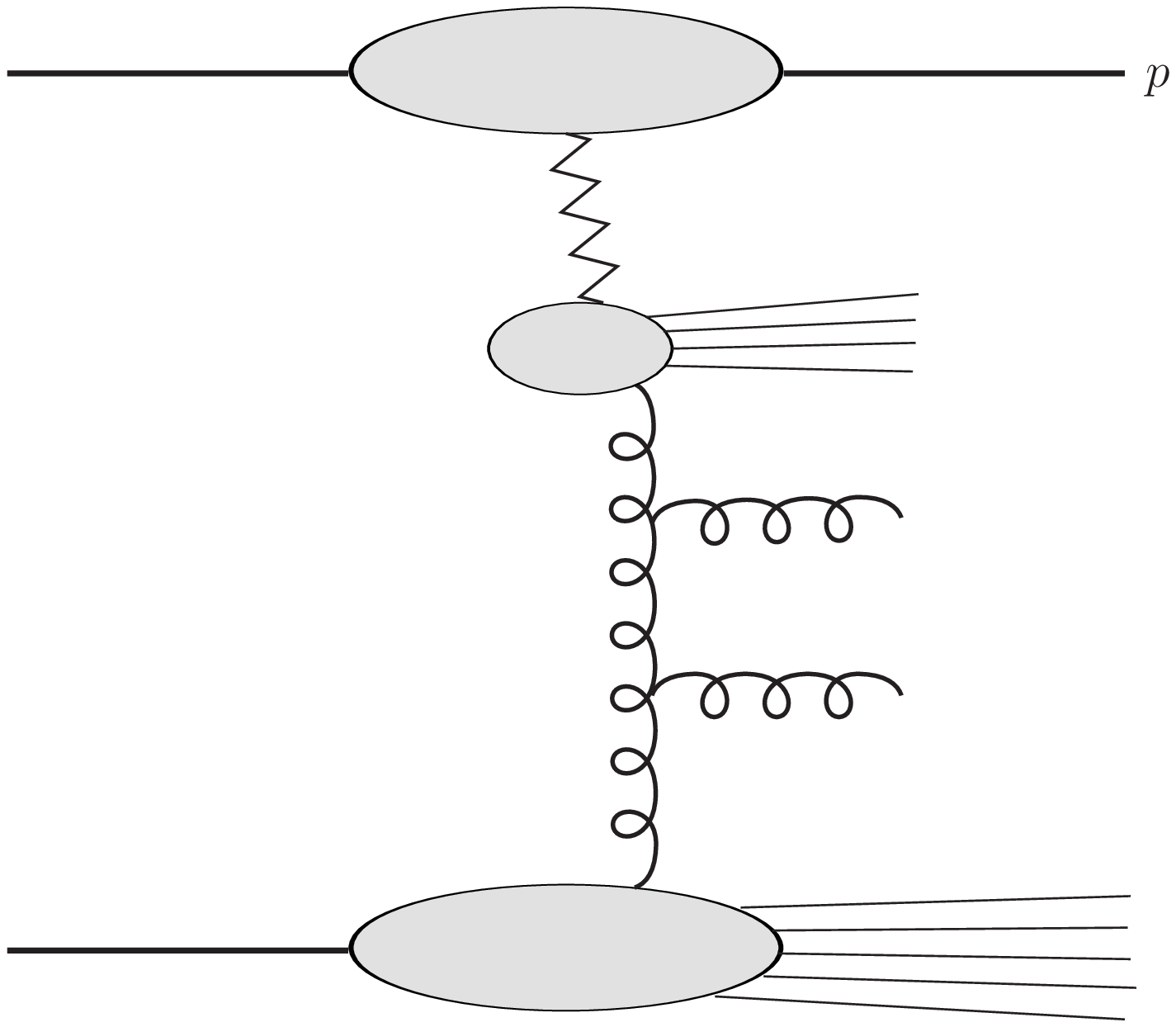}}\qquad
\subfigure[]{\includegraphics[width=.3\textwidth, height=4cm]{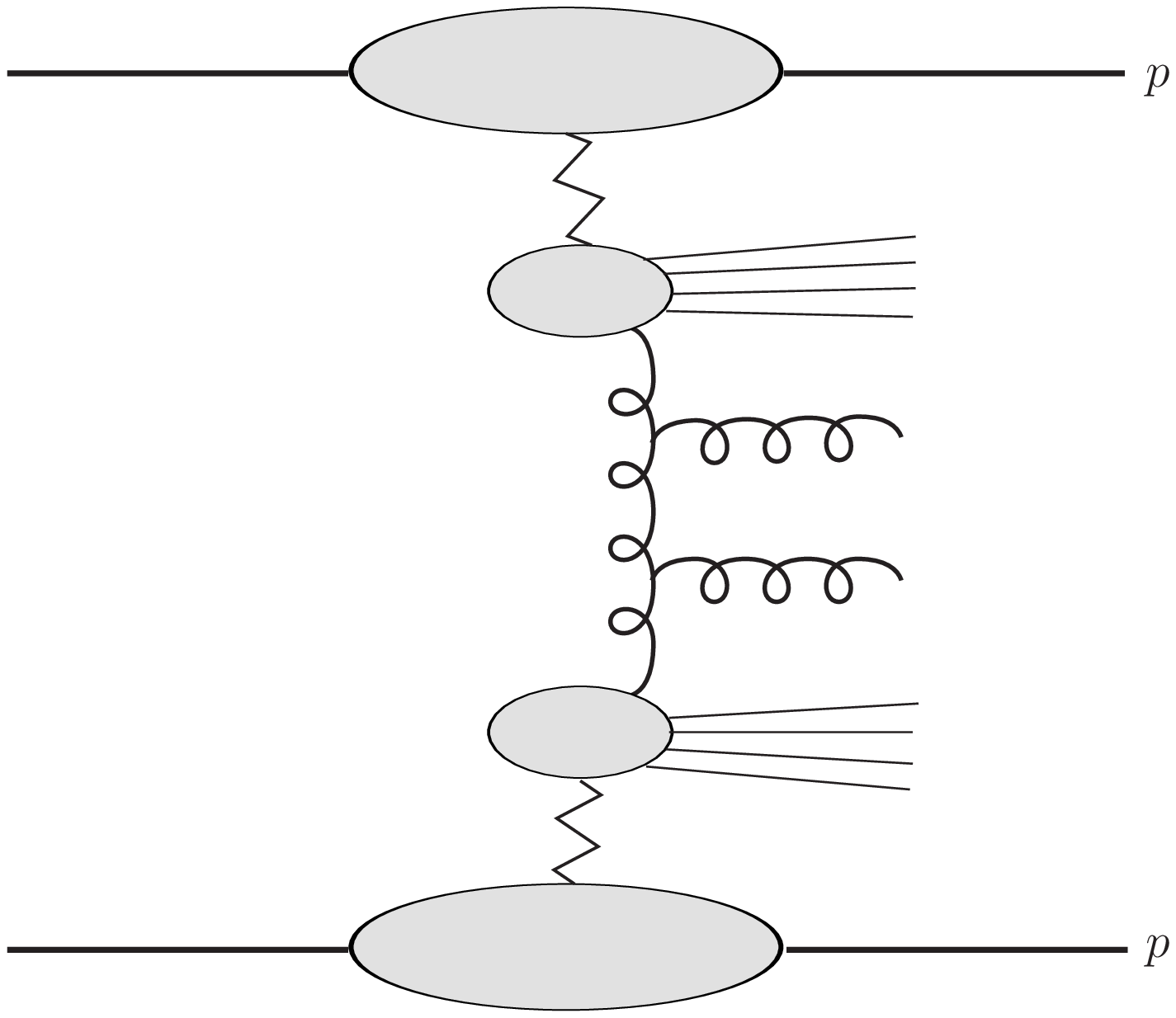}}\qquad
\subfigure[]{\includegraphics[width=.3\textwidth, height=4cm]{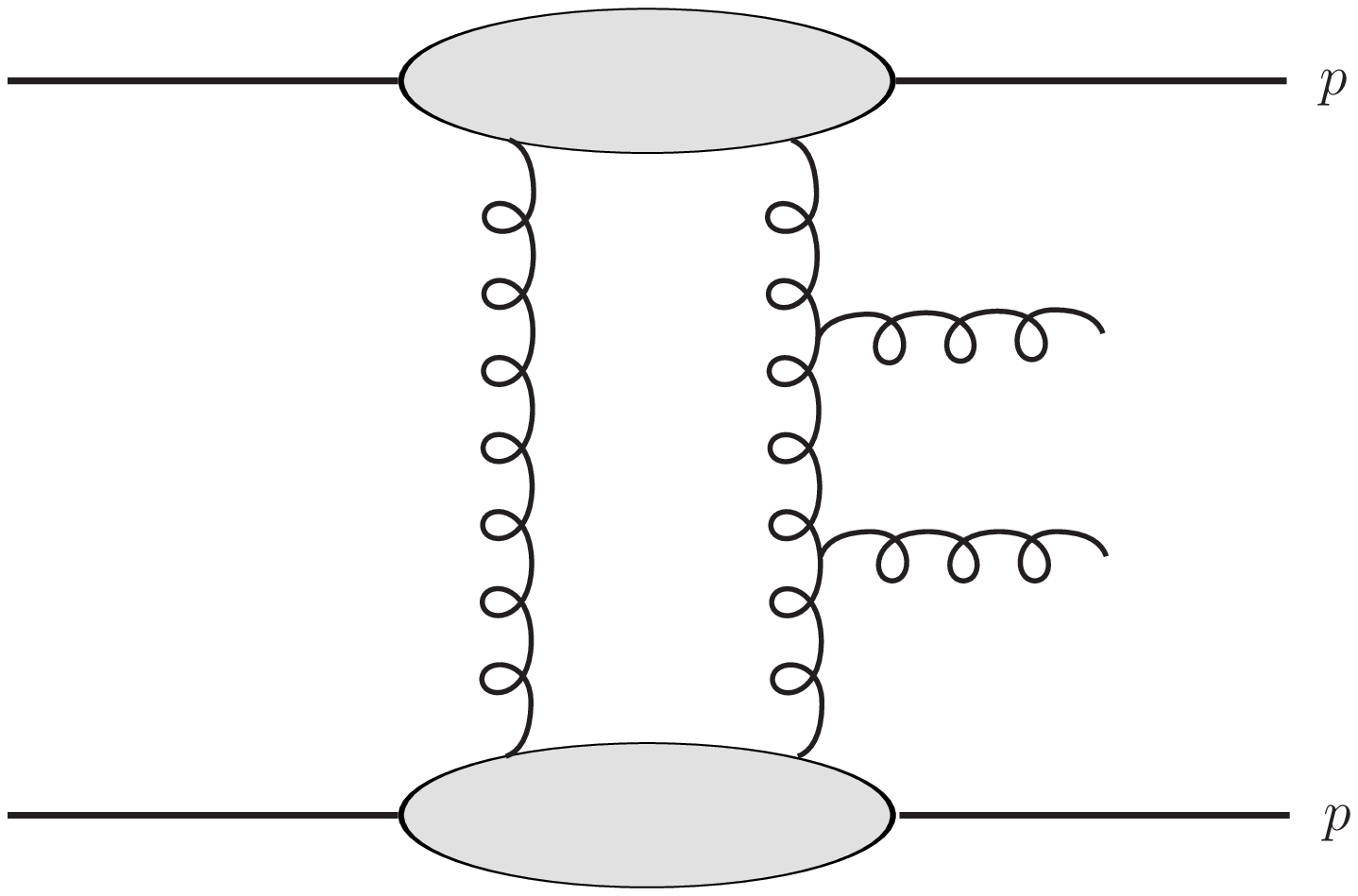}}
}
\caption{Single diffractive di-jet production (a), di-jet production via double pomeron exchange (b) and central exclusive di-jet production (c). In (a) and (b), the `zig-zag' line represents colour singlet (pomeron) exchange and there will be rapidity gap between the intact proton and the pomeron remnants. In (c), the di-jets are produced {\it exclusively}, implying no hadronic activity outside of the di-jet system. \label{fig:proc}}
\end{figure*}

ATLAS will be able to measure single diffractive di-jet production and di-jet production via double pomeron exchange. These processes, shown in Fig.~\ref{fig:proc} (a) and (b), will allow the study of diffractive parton density functions (dPDFs) and factorization breaking in diffractive events. Factorization breaking is the observation that the dPDFs obtained at HERA do not predict the correct cross section for diffractive events at hadron colliders \cite{sd} and is attributed to secondary scattering between spectator partons in the protons causing the proton to break up and the rapidity gap to be destroyed. Factorization breaking will be studied by measuring the ratio of single diffractive to non-diffractive di-jet events, R(SD/ND), and the ratio of double-pomeron exchange to single diffractive di-jet events, R(DPE/SD). 

It is expected that a few thousand SD di-jet events, with jet transverse energy, $E_{{\rm T}}>20$~GeV, will be available for analysis given 100~pb$^{-1}$ of data. The number of events is restricted by the large prescale applied in the Level 1 (L1) trigger for low transverse energy jets and the current focus is to develop new trigger strategies capable of retaining more events. The new triggers may include information from LUCID and/or the MBTS (discussed futher in Sec.~\ref{sec:cep}).

\section{\label{sec:cep} Central exclusive di-jet production}

Central exclusive di-jet production (CEP) is the process $pp \rightarrow p + jj + p$, where the `$+$' denotes a large rapidity gap from the outgoing protons, and is shown in Fig.~\ref{fig:proc} (c). CEP has received a great deal of attention in recent years due to the possibility of tagging the outgoing protons, using new forward proton detectors, in order to measure properties of the Higgs boson \cite{fp420}. Although the CDF measurements of exclusive di-jets, photons and charmonium are in good agreement with the theoretical predications \cite{cdf}, there remains a factor of three uncertainty in the calculation of the Higgs cross section at 14~TeV. Exclusive di-jet production offers the opportunity to constrain this uncertainty at the LHC with early ATLAS data. 

The analysis strategy is to define the exclusivity of an event using the di-jet mass fraction,
\begin{equation}
R_{\rm{jj}} =\frac{M_{{\rm jj}}}{M_{{\rm calo}}}
\end{equation}
where $M_{{\rm jj}}$ is the invariant mass of the di-jets and $M_{{\rm calo}}$ in the mass of all energy deposits in the calorimeter. Typically, an exclusive event will have $R_{\rm{jj}}\sim1$ and inclusive/diffractive events will have $R_{\rm{jj}}\ll 1$. However, the tail of the background distribution extends up to large $R_{\rm{jj}}$ values and the extraction of the exclusive signal will require a very good understanding of the background distributions.


Due to the large QCD cross sections, the standard low-$E_{\rm T}$ jet triggers will be heavily prescaled and will provide insufficient statistics to perform CEP measurements. Consequently, a new trigger was developed that, in addition to a jet with $E_{\rm T}>18$~GeV, required an absence of hit segments on at least one side of the MBTS. Simulations show that this trigger has a 60\% efficiency for signal events (with respect to the jet trigger alone) and provides a 10$^{4}$ rejection factor for standard QCD events. It is expected that a few hundred exclusive di-jet events will survive the final selection criteria for every 10~pb$^{-1}$ of data.

In addition to exclusive studies, this trigger strategy can also be used to retain single diffractive di-jet events. Although the overall trigger efficiency for single diffractive events is approximately 10$^{-2}$ (for this trigger), the majority of the events that are removed are at high-$\xi$. Thus, the trigger may allow a sample of low-$\xi$ SD di-jet events to be analyzed, which could then be used to constrain the dPDFs in an interesting kinematic region.

\section{\label{sec:forwardjets} Forward jets}

There are a number of signatures involving widely separated jets that can be measured with early ATLAS data. The processes of interest are $2\rightarrow 2$ partonic scatters mediated by $t$-channel colour octet or colour singlet exchange.

An inclusive measurement of interest is the azimuthal de-correlation of di-jets, the size of which is dependent on the pseudo-rapidity separation of the jets . The theoretical predictions of the azimuthal de-correlation is dependent on the framework used to make the prediction, namely fixed order (NLO), parton shower and BFKL. It is expected that LHC measurements will be sensitive to BFKL effects\cite{decorr}.

It is also useful to split the di-jet samples into gap and non-gap components. This is achieved by vetoing on radiation between the jets. Experimentally, we introduce a veto-scale which defines the maximum transverse energy that can be deposited between the jets in a gap event. It is then useful to study the fraction of gap-events as a function of the pseudo-rapidity separation of the jets, $\Delta \eta$. For large separation, $\Delta \eta \sim 6$, the gap events are predominantly due to colour singlet exchange. 
A prediction of BFKL \cite{mt} is that the fraction of events with little activity between the jets should rise with the 
separation of the jets, $\Delta \eta$, and was extensively searched for at the Tevatron and HERA \cite{gbj1}. The rise of the gap-fraction was not observed at the Tevatron, 
for example, because the centre-of-mass energy was too small; it was shown in \cite{cox1} that the 
rapidly falling PDFs at high $x$ tempered the rise and meant that a large enough sample 
of events with large jet separations could not be obtained (because $\Delta \eta = \rm{ln}(\hat{s}/\hat{t}$)). An improved measurement should be possible at the LHC due to the increased centre-of-mass 
energy. In principle, ATLAS should be able to measure the gap-fraction up to $\Delta \eta \sim 9, 9.5$ with approximately 10~pb$^{-1}$ of data. The events are retained for analysis by using a forward di-jet ($|\eta_{\rm jet}|>3.2$) trigger with $E_{{\rm T}}>18$~GeV.

Finally, for smaller pseudo-rapidity separations ($\Delta \eta \sim 4$), it is interesting to study the fraction of gap events as a function of $\rm{ln}(Q/Q_{0}$), where $Q$ is the transverse energy of the jets. In this region, the gap events also arise from single gluon (colour octet)  exchange. This observable is sensitive to QCD effects such as wide angle soft gluon radiation \cite{gbj2,gbj3}. Preliminary studies indicate that these measurements should be achievable with less than 100~pb$^{-1}$ of data using standard jet triggers.

%

%% file: white_chris.tex
\begin{center}
{\Large \bf Higgs Boson Production with Multiple Hard Jets}

\vspace*{1cm}
Chris D. White
\vspace*{0.5cm}

Nikhef, Science Park 105, 1098 XG Amsterdam, The Netherlands

\vspace*{1cm}
\end{center}            

\begin{abstract}
We describe a new framework for calculating multiple hard jet emission in Higgs boson production via gluon-gluon fusion. It is based on the well-known FKL factorisation formula. This formally applies only in a certain high energy limit, but we describe how to modify the description so as to work well at LHC energies. Approximate matrix elements thus obtained are validated by comparing with known tree level results, and the framework is implemented in a Monte Carlo event generator.
\end{abstract}

\setcounter{section}{0}
\setcounter{figure}{0}
\setcounter{table}{0}
\setcounter{equation}{0}

\section{Introduction}
Scattering events at the LHC will be dominated by multijet final states, ultimately arising from QCD radiation. The scattering amplitudes for such events involve Feynman diagrams with many external quark and gluon legs, which cannot be feasibly calculated using present computing power. Instead, approximate methods are usually used to estimate multijet final states. A common practice is to interface tree level matrix elements with parton shower algorithms. These simulate the effect of QCD radiation on a given hard matrix element, where the extra partons are approximated well in the soft and collinear limit i.e. when the emitted radiation has a low transverse momentum with respect to the external legs of the hard interaction. Such an approach, whilst undoubtedly useful, is known to be inadequate in many processes. The question then naturally arises of whether it is possible to estimate scattering amplitudes with many outgoing hard partons directly (i.e. with no restriction on the transverse momentum).

\begin{figure}[h]
\scalebox{0.6}{\includegraphics{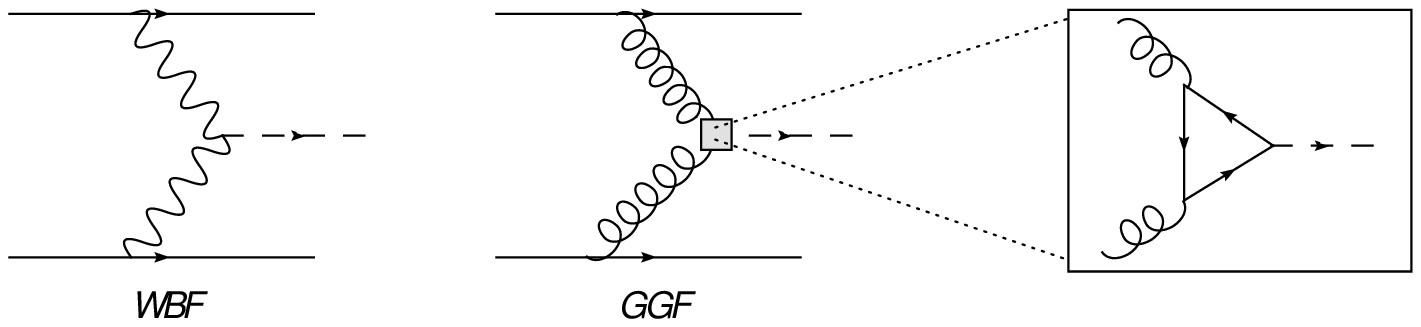}}
\caption{The two main Higgs production modes at the LHC.}\label{modes}
\end{figure}

One such process in which multiple hard jet production is relevant is Higgs boson production. There are two main production modes at the LHC, known as $W$ boson fusion (WBF) and gluon-gluon fusion (GGF) and depicted in figure~\ref{modes}.
Both of these can be used as a discovery channel, and each forms a background to the other. In WBF there is no colour exchange in the $t$-channel, and any jet activity is mostly confined to the forward and backward detector regions. In GGF however, in which the Higgs couples to gluons via a top quark loop, there is a colour octet exchange. One thus expects jet activity over the complete range in rapidity, and the nature of the radiation (e.g. whether it is hard or soft) has a significant impact on what one can measure about the Higgs boson, as we will see. Thus, a means for accurately estimating hard jet radiation in GGF is extremely well-motivated. 

\section{FKL factorisation}
\begin{figure}
\scalebox{0.6}{\includegraphics{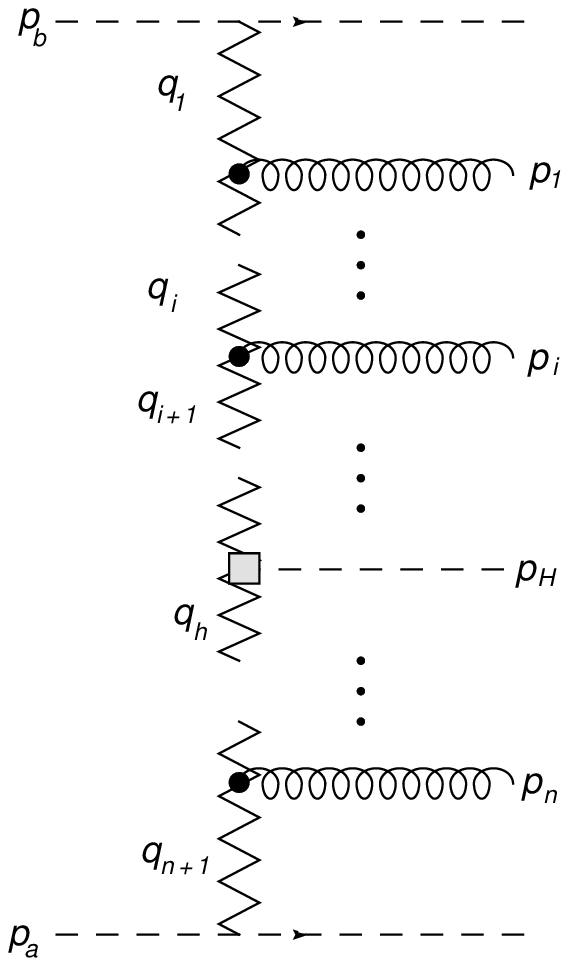}}
\caption{Diagrams contributing to the GGF scattering amplitude in the MRK limit, where the gluons couple with Lipatov vertices, and the virtual gluons are Reggeised (i.e. have resummed leading virtual corrections).}\label{FKLfig}
\end{figure}

Our aim is to estimate scattering amplitudes with many external hard parton legs, and our starting point is the FKL factorisation formula of~\cite{Fadin:1975cb}. This states that in the limit of multi-Regge kinematics (MRK) in which the rapidity gap between final state particles is large (i.e. large centre of mass energy but fixed momentum transfer), the amplitude for pure multiparton scattering (from initial state partons $\alpha$ and $\beta$) is dominated by the process
\begin{equation}
\alpha+\beta\rightarrow\alpha+\beta+ng,
\label{proc}
\end{equation}
for a given number $(n+2)$ of final state partons. Furthermore, the sum of these diagrams gives a factorised expression for the scattering amplitude, which (including also a Higgs boson in the final state) has the form

\begin{align}
&i{\cal M}^{ab\rightarrow ab j_1\ldots j_{n}}_{\mu_1\ldots\mu_n}=2s(g_s)^{n+2}\left(\prod_{i=1}^{n_1+1}\frac{1}{q_i^2}\exp[\hat{\alpha}(q_i^2)(y_{i-1}-y_i)]\right)\notag\\
&\times\left(\prod_{i=1}^{n_1}C_{\mu_i}(q_{i-1},q_i)\right)C_H(q_{n_1+1},q_{n_1+2})\notag\\
&\times\left(\prod_{i=n_1+2}^{n+1}\frac{1}{q_i^2}\exp[\hat{\alpha}(q_i^2)(y_{i-1}-y_i)]\right)\notag\\
&\times\left(\prod_{i=n_1+2}^nC_{\mu_i}(q_{i-1},q_i)\right).
\label{FKL}
\end{align}
In this formula, $\{q_i\}$ are the 4-momenta of the virtual gluons, $C_H$ is an effective vertex coupling the Higgs to gluons~\cite{DelDuca:2003ba}, $C^\mu$ the {\it Lipatov effective vertex} for gluon emission, and $\hat{\alpha}(q^2)$ a function which includes leading virtual corrections in the MRK limit (see~\cite{Andersen:2008ue,Andersen:2008gc} for more details). The form of this formula is shown schematically in figure~\ref{FKLfig}. 

The chief advantages of this formula are firstly that there is no restriction on the ordering of final state transverse momenta. Secondly, the factorised form means that scattering amplitudes with any number of partons can be efficiently calculated, thus this is indeed a good starting point for trying to describe hard final state radiation. Furthermore, the virtual corrections built into the Reggeised gluon propagators cancel the collinear singularities arising from real gluon emission. However, a significant disadvantage of eq.~(\ref{FKL}) is that it formally applies only in the MRK limit, which is not well-approximated by the phase space probed at the LHC or Tevatron. The solution is to modify the application of the FKL formula so as to build in known features of perturbation theory.
\section{Modified approach}
Our approach to calculating Higgs boson plus multiparton scattering amplitudes can be summarised as follows. We define the scattering amplitude with $(n+2)$ final state partons by eq.~(\ref{FKL}) implemented according to the following prescription:
\begin{enumerate}
\item Impose 4-momentum conservation at the emission vertices. 
\item Use full virtual 4-momenta $q_i$ instead of transverse components only.
\item Require gauge invariance of the Lipatov vertex ($k\cdot C$=0) over all of phase space.
\end{enumerate}
The first requirement is an obvious piece of physics, but is formally subleading in the centre of mass energy so is not included in most previous analytic studies of eq.~(\ref{FKL}). It implements corrections to the matrix element, as well as the phase space of the emitted gluons. The replacement of virtual 4-momenta by their transverse components (as is usually done in the BFKL implementation of eq.~(\ref{FKL})~\cite{Fadin:1975cb,Kuraev:1976ge,Kuraev:1977fs}) is allowable in the MRK limit, but leads to significant differences away from this corner of phase space. Using the full 4-momenta amounts to keeping known singularities of the scattering amplitude in the same place outside the MRK limit, rather than shifting them. The final requirement has the consequence that the squared Lipatov vertex is positive, thus enforcing local positivity of the matrix element (see~\cite{Andersen:2008ue,Andersen:2008gc} for details). The gauge invariance constraint overlaps with, but is not the same as, the {\em kinematic constraint} of~\cite{Catani:1989yc}.

\begin{figure}
\scalebox{0.4}{\includegraphics{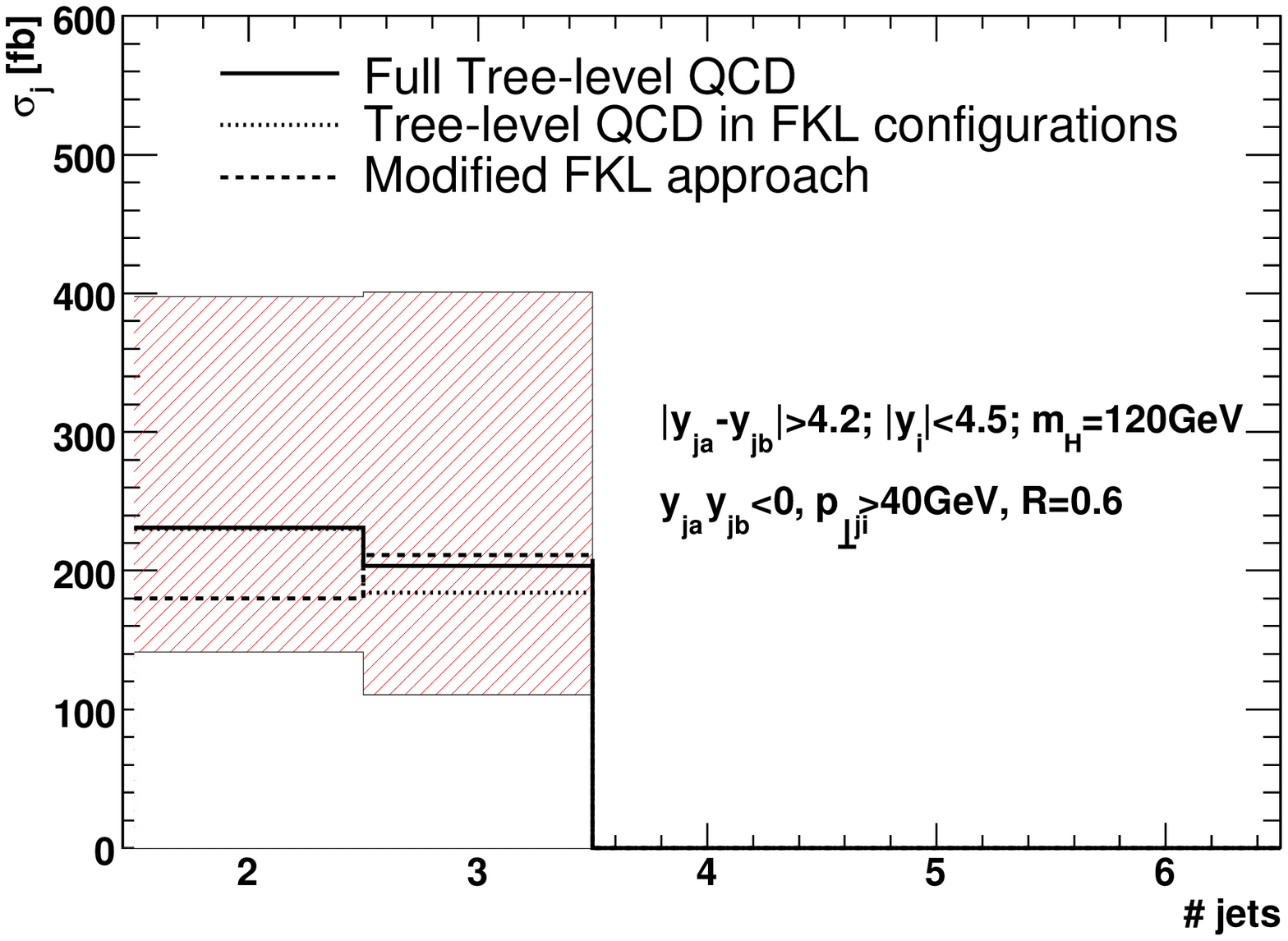}}
\caption{Comparison of the cross-sections for Higgs boson production via GGF in association with two and three partons obtained using the modified FKL approach, and the full tree level results from MadGraph. Uncertainty corresponds to scale variation by a factor of two.}\label{compare}
\end{figure}
Each of the above requirements goes beyond any logarithmic order in the high energy expansion, and the resulting matrix elements are in a sense closer to fixed order perturbation theory than to the original FKL description, as we will see.

\begin{figure}
\scalebox{0.4}{\includegraphics{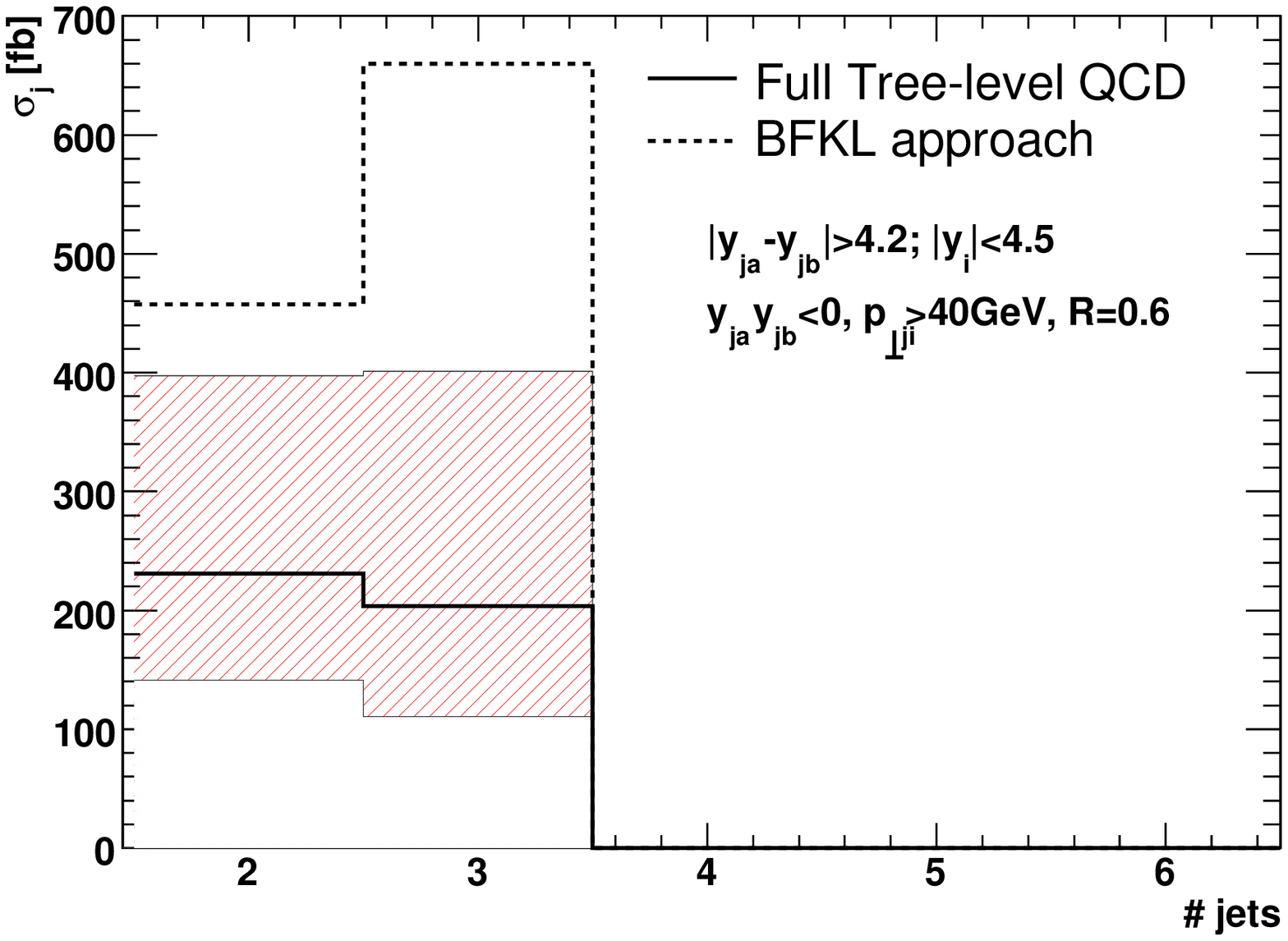}}
\caption{Comparison of the cross-sections for Higgs boson production via GGF in association with two and three partons obtained using a traditional BFKL approach (with 4-momentum conservation implemented), and the full tree level results from MadGraph. Uncertainty corresponds to scale variation by a factor of two.}\label{BFKL}
\end{figure}

In order to judge whether the approximate matrix elements work well, one may compare them with the known tree level results for Higgs boson production in association with two and three partons, which we have obtained from MadGraph~\cite{Alwall:2007st}. This comparison is made in figure~\ref{compare} for WBF-like signal cuts (see~\cite{Andersen:2008gc}). These have been applied inclusively i.e. at least one pair of jets is required to pass the cuts. One sees that the modified FKL results are well within the scale variation of the tree level results. Alternatively, the results obtained using a traditional BFKL implementation (but also including energy and momentum conservation) are shown in figure~\ref{BFKL}. One sees that this lies outside the scale-variation associated with the tree level results. Without 4-momentum conservation, the situation is far worse \cite{Andersen:2008gc}, and indeed worsens at higher orders due to the ability to create multiple gluons at no cost in energy.

\begin{figure}
\scalebox{0.4}{\includegraphics{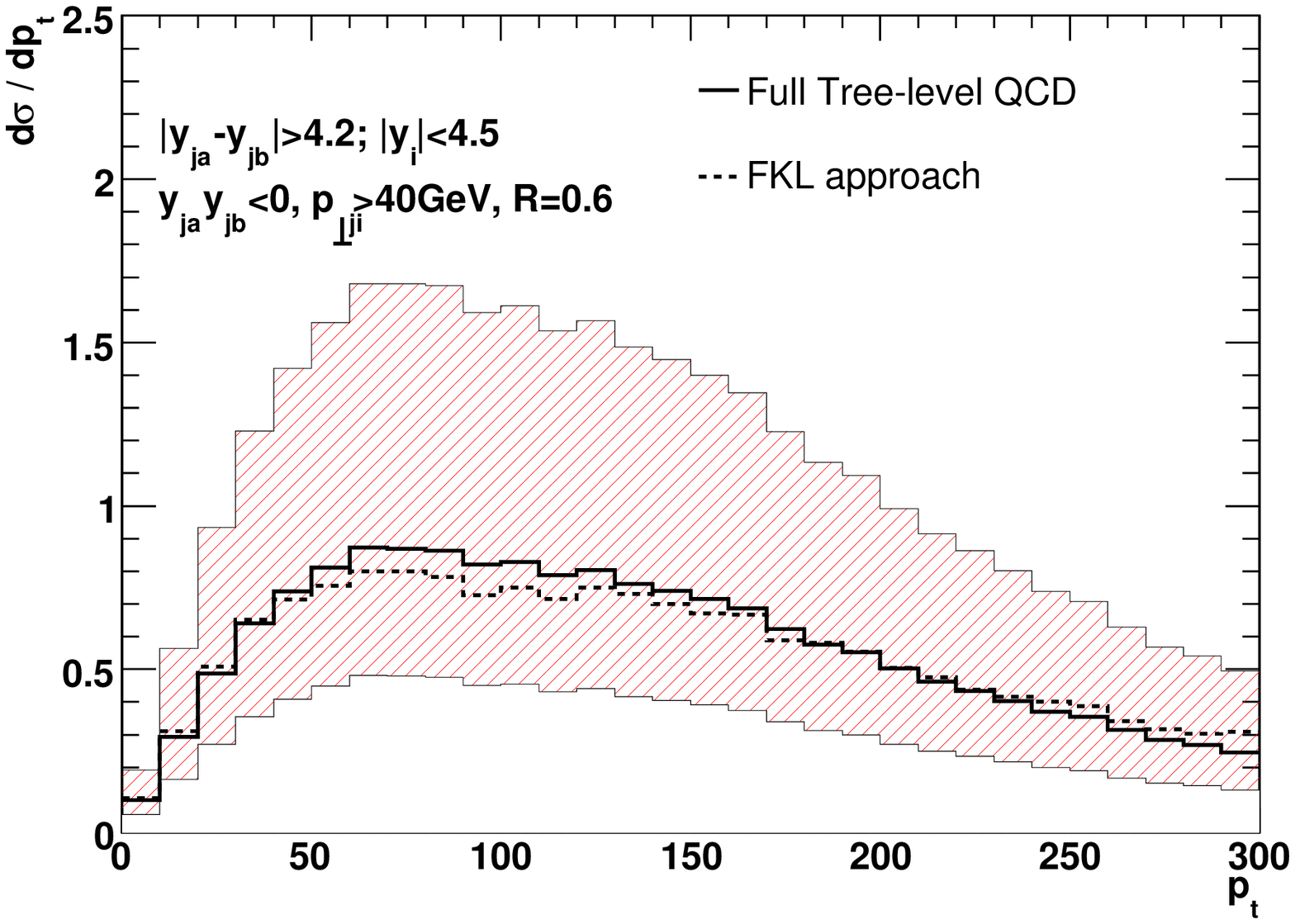}}
\caption{Distribution of the Higgs boson transverse momentum for the $hjjj$ final state.}\label{ptHfig}
\end{figure}

\begin{figure}
\scalebox{0.4}{\includegraphics{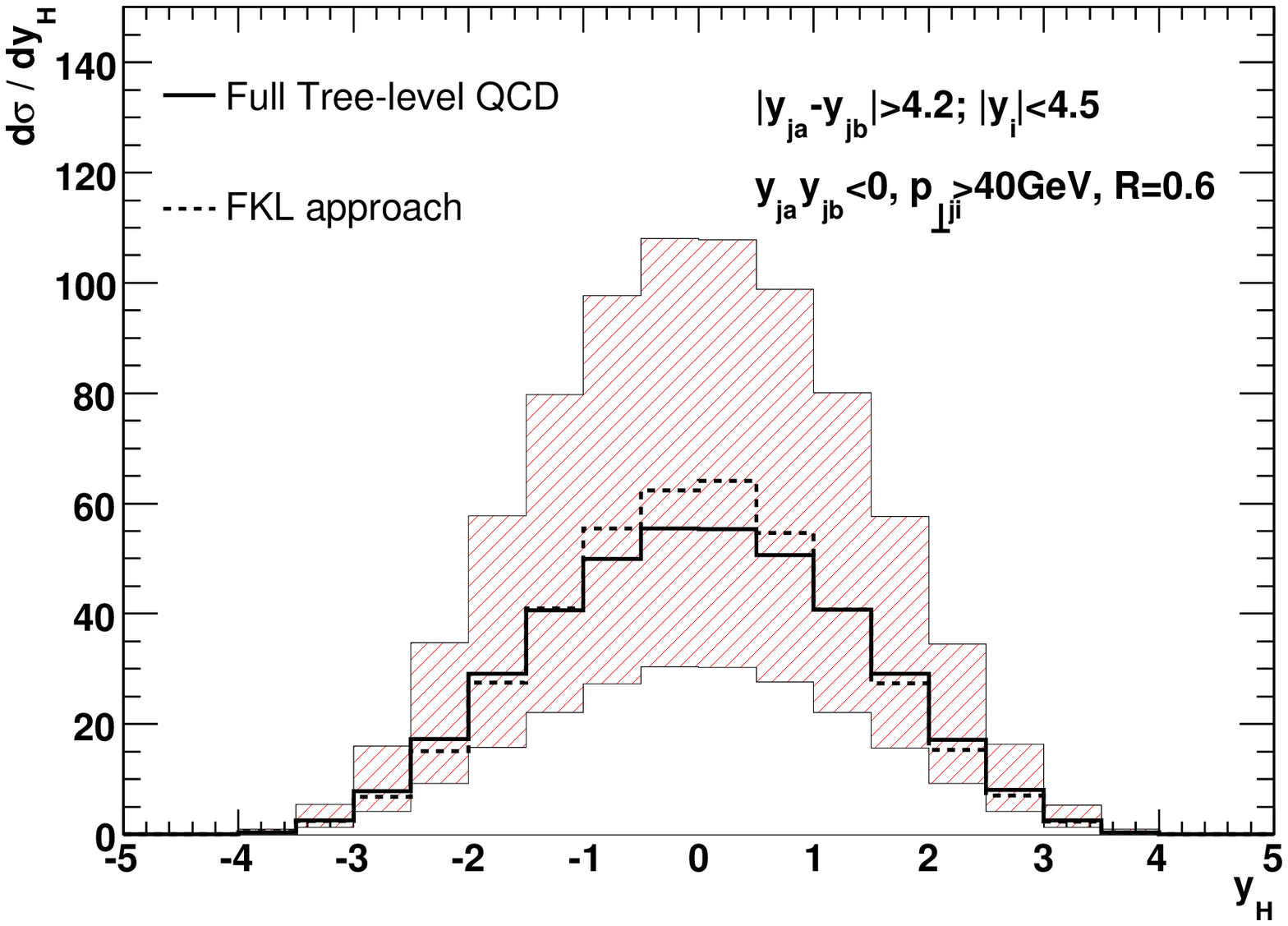}}
\caption{Distribution of the Higgs boson rapidity, for the $hjjj$ final state.}\label{yHfig}
\end{figure}

\begin{figure}
\scalebox{0.4}{\includegraphics{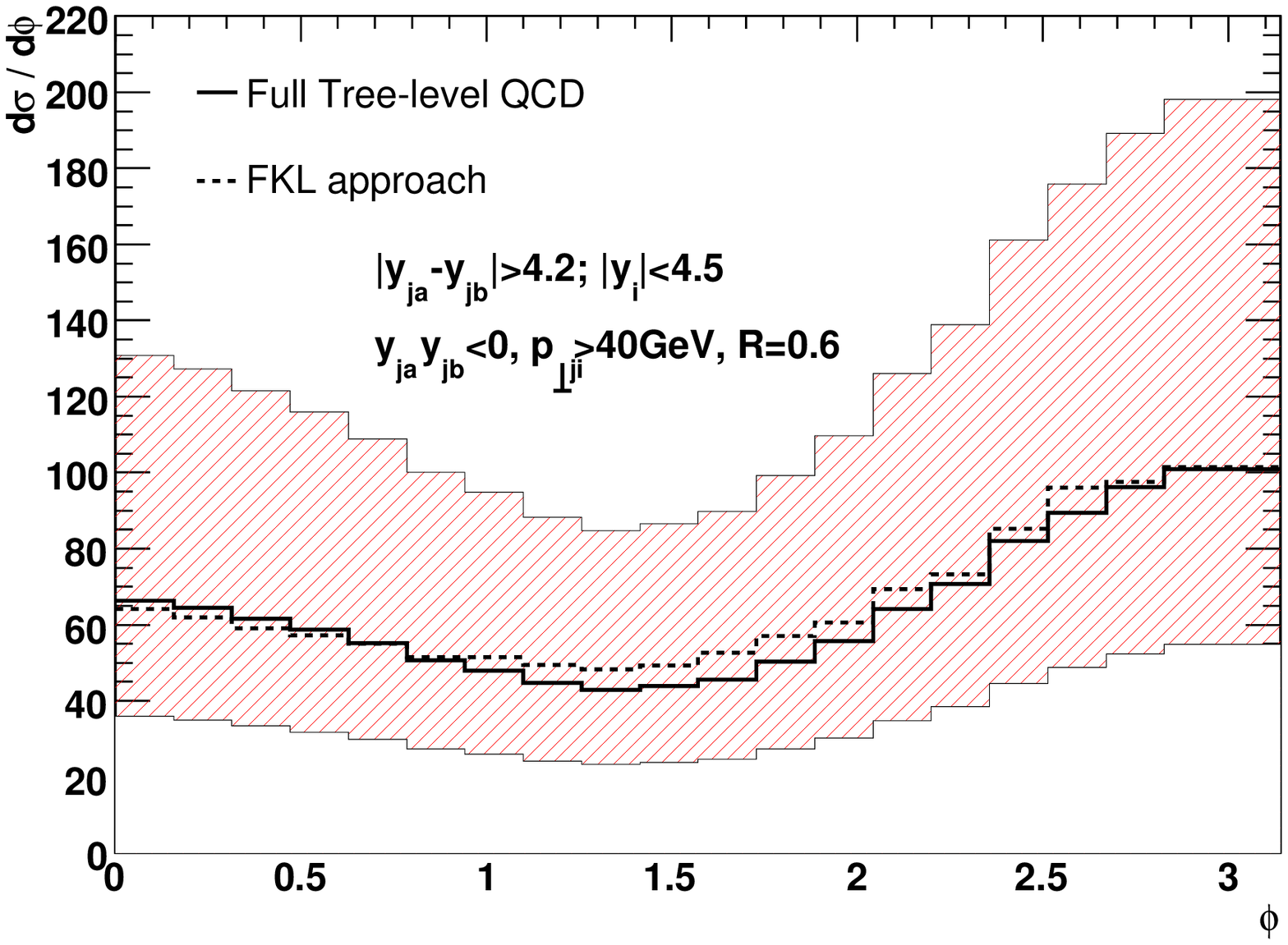}}
\caption{Distribution of the azimuthal angle between the tagging jets, for the $hjjj$ final state.}\label{phi3}
\end{figure}

Kinematic distributions as well as total cross-sections are well approximated using the modified FKL approach. As examples, we show the transverse momentum and rapidity distributions of the Higgs boson in figures~\ref{ptHfig} and~\ref{yHfig}, for the case of Higgs boson production in association with three partons. One sees that the results obtained from the modified FKL approach agree closely with the full tree level results, and are certainly well within the scale variation uncertainty.
Another example is given in figure~\ref{phi3}, which shows the distribution of azimuthal angle between the tagging jets. The shape of this distribution is intimately related to the CP nature of the Higgs boson~\cite{Hankele:2006ja}. Again, very good agreement is observed between the modified FKL description and the full tree level result.

\begin{figure*}
\begin{tabular}{cc}
\scalebox{0.4}{\includegraphics{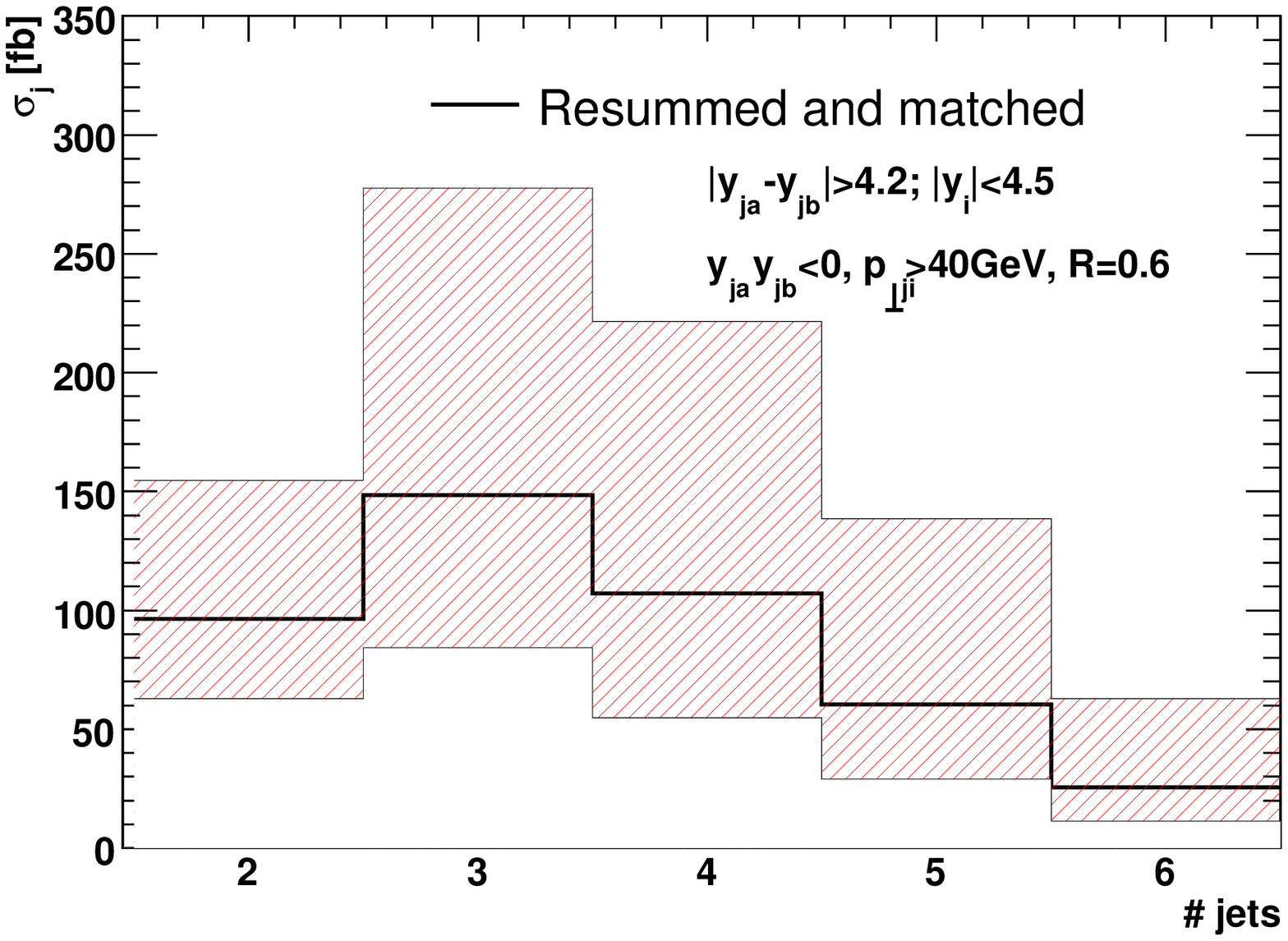}}&\scalebox{0.4}{\includegraphics{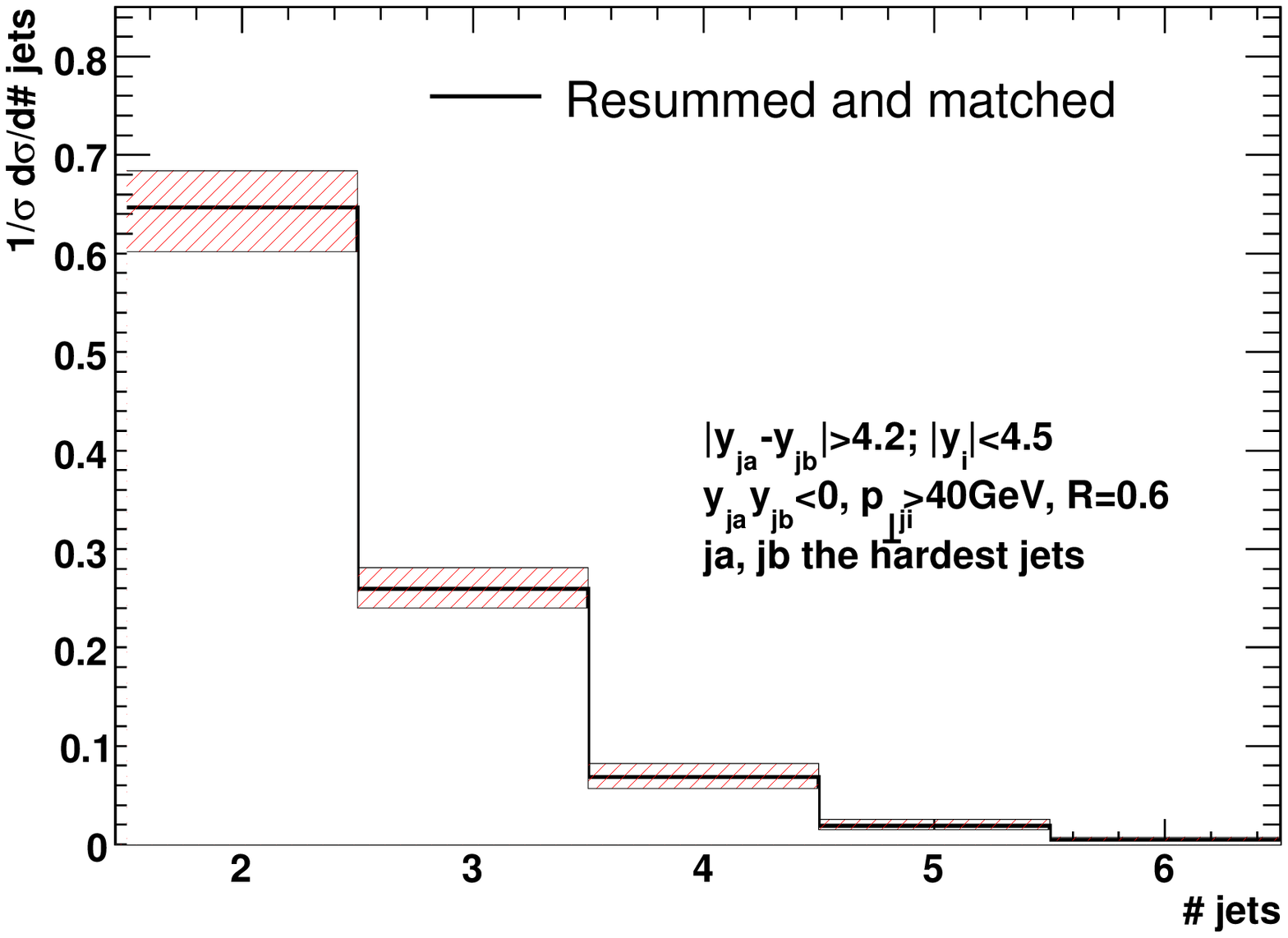}}\\
(a)&(b)
\end{tabular}
\caption{Resummed distribution in the number of hard jets i.e. those jets satisfying $p_t>40$GeV, in the cases where (a) the WBF-like cuts are applied inclusively; (b) the hardest jets are required to pass the cuts. Jets have been clustered with a $k_t$ algorithm, and uncertainty corresponds to scale variation by a factor of two.}\label{njets}
\end{figure*}

Having validated the approximation by comparing to fixed order results, one may then use the modified FKL formalism to estimate multiparton matrix elements with any number of final state partons. We have implemented the framework in a Monte Carlo event generator for Higgs boson plus multijet production. The tree level matrix elements for $hjj$ and $hjjj$ production are also included, with a matching procedure to avoid any double counting~\cite{Andersen:2008gc}, and code is available at 
\begin{verbatim}
http://andersen.web.cern.ch/andersen/MJEV/
index.html
\end{verbatim}
The known tree level results (i.e. for two and three final state partons) have also been included using a suitable matching procedure to avoid double counting~\cite{Andersen:2008ue,Andersen:2008gc}, and here we show example results. In figure~\ref{njets}(a) we show the resummed distribution in the number of hard jets, obtained using inclusively applied WBF-like signal cuts, as above. In figure~\ref{njets}(b) we show the same distribution, but where the hardest jets are explicitly required to pass the cuts. One sees in both cases that there are a significant number of events with more than three hard jets. Furthermore, when the hardest jets are used as the tagging jets, the impact of higher order jet radiation is reduced.

\begin{figure*}
\begin{tabular}{cc}
\scalebox{0.4}{\includegraphics{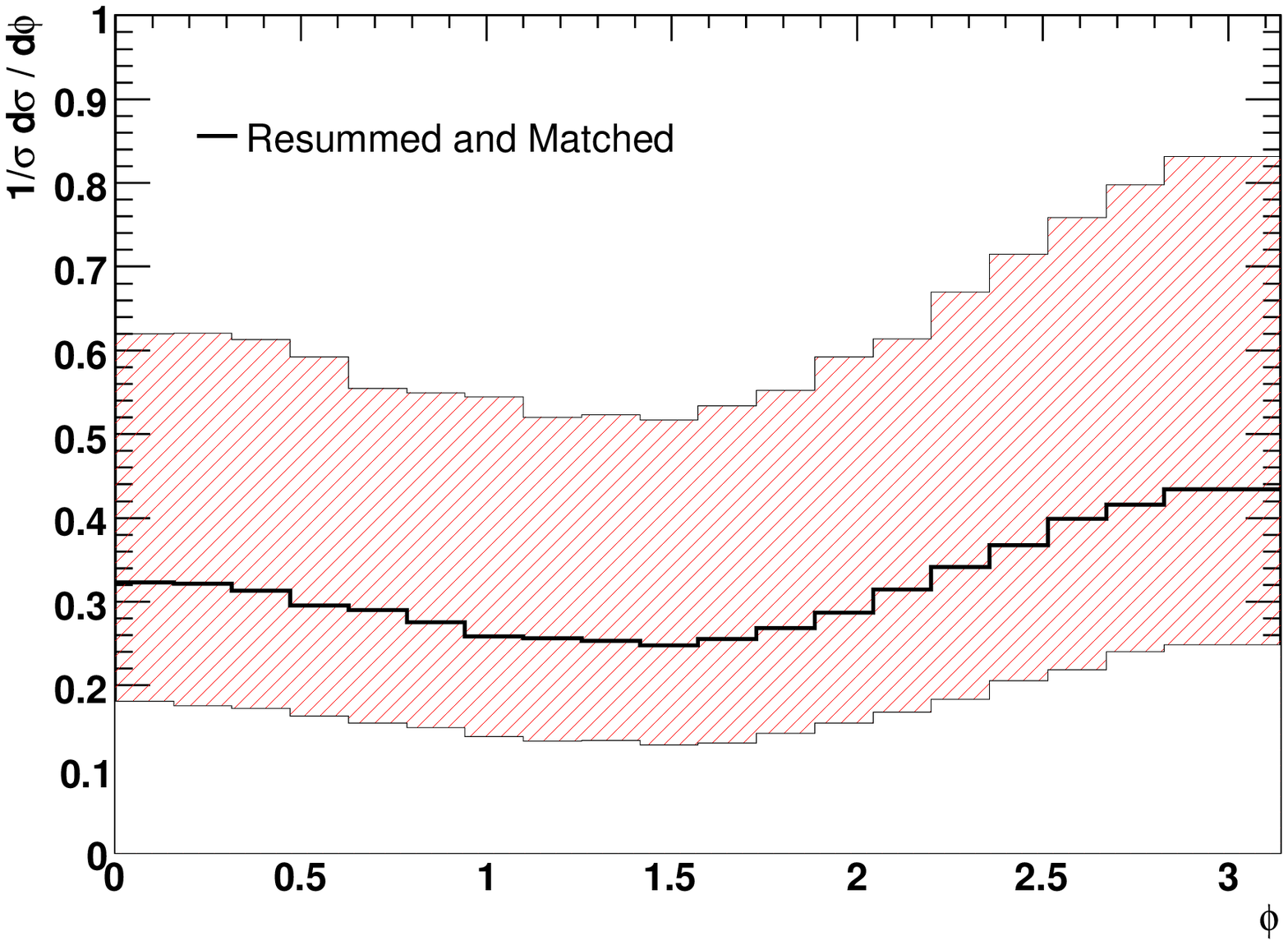}}&\scalebox{0.4}{\includegraphics{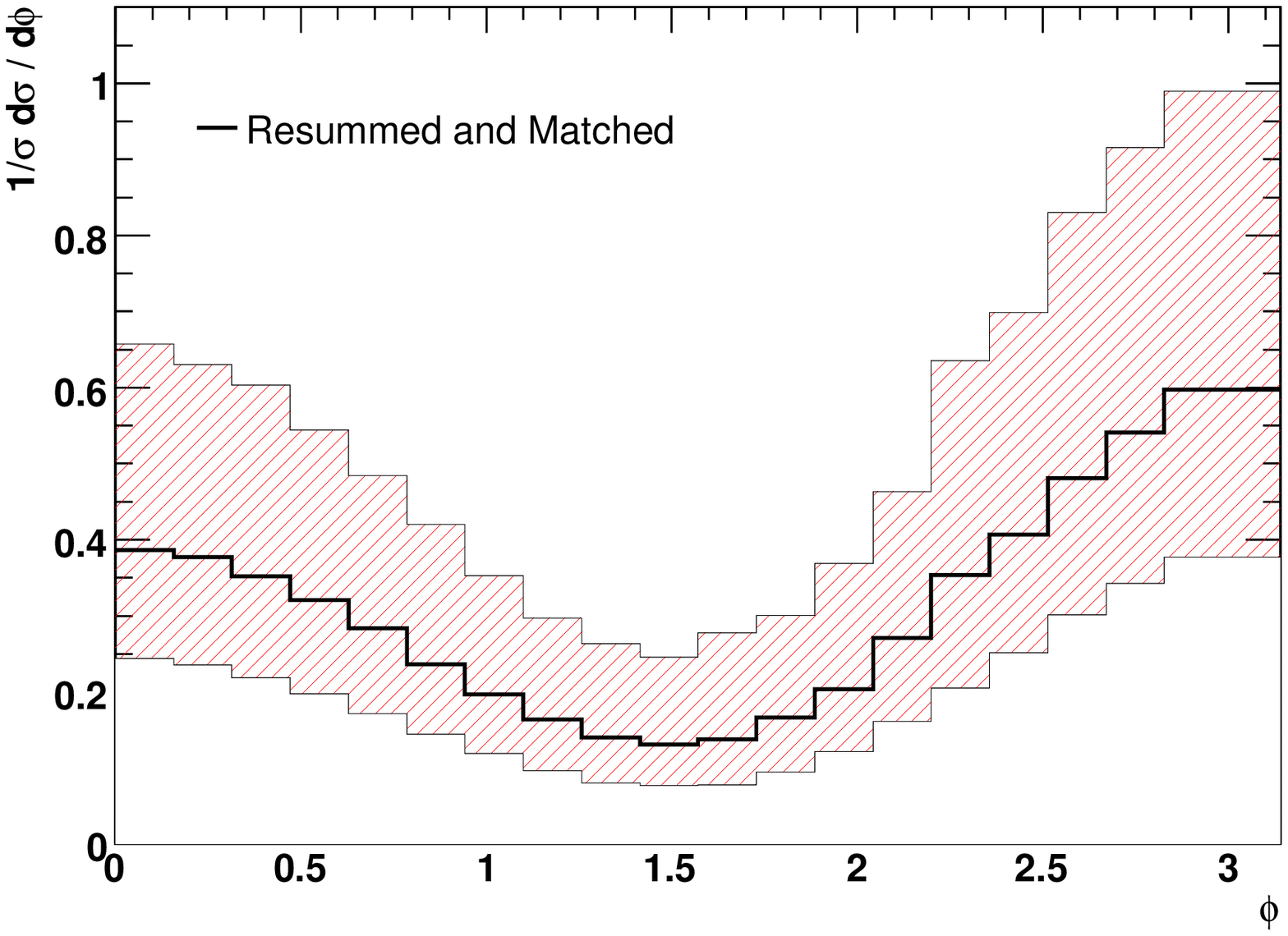}}\\
(a)&(b)
\end{tabular}
\caption{Resummed distribution of the azimuthal angle between the tagging jets, in the case where (a) the WBF-like signal cuts are applied inclusively; (b) the hardest jets are required to pass the cuts.}\label{phiresum}
\end{figure*}

The azimuthal angle $\phi_{j_aj_b}$ between the tagging jets is depicted in figure~\ref{phiresum} (i.e. the fully resummed equivalent of figure~\ref{phi3}). The choice of cuts has a significant impact on how much decorrelation is observed, and the hardest jet cuts lead to a stronger observed correlation than the inclusive cuts (consistent with figure~\ref{njets}). Further phenomenological investigation is necessary in order to determine whether the CP nature of the Higgs boson can be fully determined i.e. whether a strong enough correlation can be preserved without cutting into the cross-section too much. A useful quantity to consider is the parameter~\cite{Hankele:2006ja}
\begin{equation}
 A_\phi=\frac{\sigma(\phi_{j_aj_b}<\pi/4)-\sigma(\pi/4<\phi_{j_aj_b}<3\pi/4)+\sigma(\phi_{j_aj_b}>3\pi/4)}{\sigma(\phi_{j_aj_b}<\pi/4)+\sigma(\pi/4<\phi_{j_aj_b}<3\pi/4)+\sigma(\phi_{j_aj_b}>3\pi/4)},
\label{Aphi}
\end{equation}
where $\sigma$ is the fully integrated cross-section. This parameter is positive in the case of a scalar Higgs boson, and negative for a pseudo-scalar. Thus, a measurement of the correlation parameter $A_{\phi}$ allows one to infer whether or not a discovered Higgs boson is of Standard Model character or not. Values of $A_\phi$ obtained in different calculations are shown in table~\ref{Aphivals}. The results labelled LO correspond to the tree level calculations for $hjj$ and $hjjj$ production, obtained using MadGraph. Results are shown for WBF-like signal cuts (see~\cite{Andersen:2008gc}), which are applied both inclusively (left-hand columns) and in the case where the hardest jets are used as the tagging jets (right-hand columns). One sees that the degree of correlation observed is highly dependent on the cuts used, and that much more decorrelation is observed for the inclusive cuts. Furthermore, the emission of one extra hard parton already leads to significant decorrelation, which is then exacerbated by the resummation (albeit less severely when the hardest jets are used as the tagging jets). Finally, one recovers a high degree of correlation by requiring that only two hard jets are present, even when the full resummation is included.

  \begin{table}
      \begin{tabular}{c|c}
	Inclusive cuts&$A_\phi$\\
	\hline
	LO 2-jet&$0.456$\\
	Resummed, $=2$-jet & $0.444$\\
	LO 3-jet&$0.203$\\
	Resummed&$0.123$
      \end{tabular}
      \begin{tabular}{c|c}
	Hardest cuts&$A_\phi$\\
	\hline
	LO 2-jet&$0.456$\\
	Resummed, $=2$-jet & $0.436$\\
	LO 3-jet&$0.374$\\
	Resummed&$0.372$
      \end{tabular}
      \label{Aphivals}
      \caption{Values of the correlation parameter of eq.~(\ref{Aphi}) using tree-level matrix elements (first and third rows), and the resummed approach described here. Results are shown for the two choices of cuts discussed in the text. }
  \end{table}
\section{Conclusions}
We have described a framework for estimating scattering amplitudes with many hard final state partons, and applied this to Higgs production via gluon-gluon fusion at the LHC. The approach is based on the FKL factorisation formalism, which is then modified to incorporate known features of perturbation theory. The results have been validated by comparing with known tree level results for the two and three parton final states, and a Monte Carlo generator for GGF is available. 

The approach described here is not limited to Higgs boson production, but is also applicable to other processes such as $W$ production in association with jets, and pure QCD jet production. Furthermore, the underlying FKL factorisation formula can be improved, although the leading order description is already working rather well. 

It would also be interesting to interface the modified FKL approach with a parton shower algorithm. The former technique is complementary to the latter in a well-defined sense, and should estimate the final state jet topology well rather than the jet substructure (which is approximated well by the parton shower).
\section*{Acknowledgments}
This work was done in collaboration with Jeppe Andersen and Vittorio Del Duca, and was supported by the Dutch Foundation for Fundamental Matter Research (FOM). 


\input{white_chris.bbl}
%

%% file: sm_ucl.tex
\begin{center}
{\Large \bf Studying the Underlying Event at the Tevatron}

\vspace*{1cm}
Deepak Kar, On behalf of the CDF Collaboration

\vspace*{0.5cm}
\end{center}

\begin{abstract}

CDF Run II data for the underlying event associated with Drell-Yan lepton pair production are studied as a function of the
lepton-pair transverse momentum. The data are compared with a previous analysis on the behavior of the underlying event in high transverse momentum jet production and also with several other QCD Monte-Carlo models. The goal is to provide data that can be used to tune and improve the QCD Monte-Carlo models of the underlying event, which is especially important now for the startup of the Large Hadron Collider.

\end{abstract}



\section{Introduction: the Underlying Event}

In order to find `new' physics at a hadron-hadron collider it is essential to have Monte-Carlo models that simulate accurately the `ordinary' QCD hard-scattering events. To do this one must not only have a good model of the hard scattering part of the process, but also of the
underlying event.

A typical $2$-to-$2$ hard scattering event is a proton-antiproton collision at the hadron colliders as shown in the Fig. 1, all happening inside the radius of a proton. In addition to the two hard scattered outgoing partons, which fragment into jets - there is initial and final state radiation (caused by bremsstrahlung and gluon emission), multiple parton interaction (additional $2$-to-$2$ scattering within the same event), `beam beam remnants' (particles that come from the breakup of the proton and antiproton, from the partons not participating in the primary hard scatter). We define the `underlying event' \cite{UE} as everything except the hard scattered components, which includes the `beam-beam remnants' (or the BBR) plus the multiple parton interaction (or the MPI). However, it is not possible on an event-by-event basis to be certain which particles came from the underlying event and, which particles originated from the hard scattering. The `underlying event' ({\it i.e.} BBR plus MPI) is an unavoidable background to most collider observables.  For example, at the Tevatron both the inclusive jet cross section and the b-jet cross section, as well as isolation cuts and the measurement of missing energy depend sensitively on the underlying event. A good understanding of it will lead to more precise measurements at the Tevatron and the LHC.

\begin{figure}[htbp]
  \centering
    \includegraphics[width=0.5\textwidth]{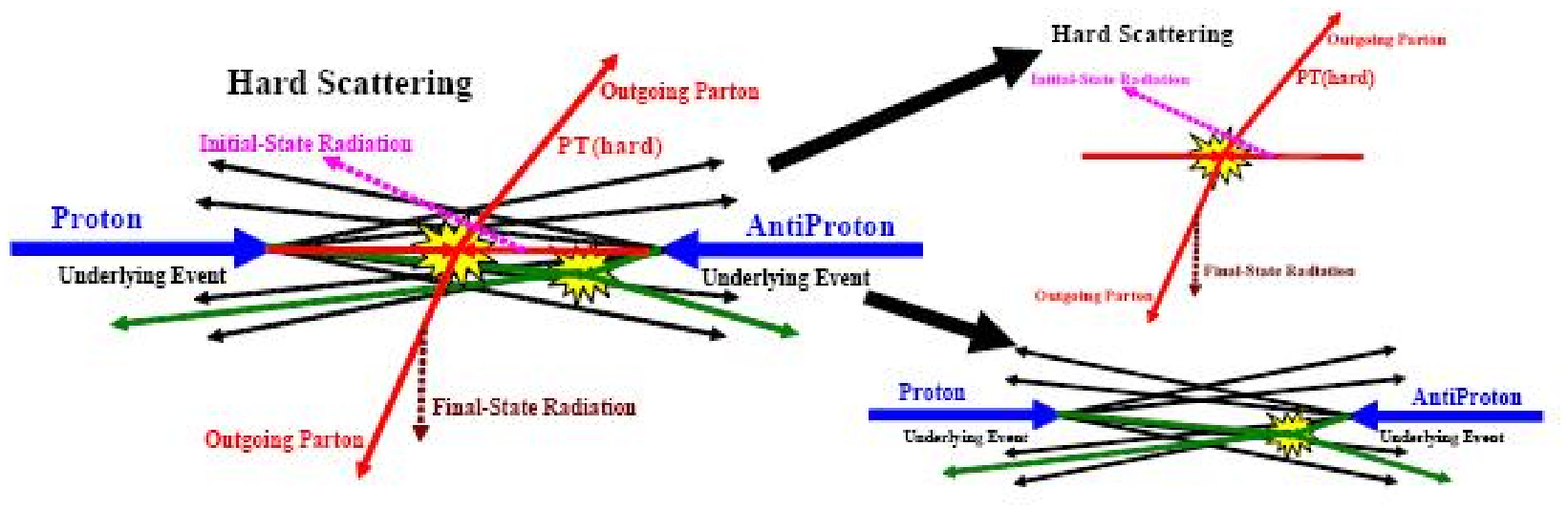}
    \caption[]{A typical 2-2 hard scattering process} 
 \end{figure}

Experimentally it is possible to take advantage of the topological structure of hadron-hadron collisions to study the underlying event. 
The direction of the Z boson is used to isolate regions of $\eta-\phi$ space that are sensitive to the underlying event.
The angle $\Delta\phi = \phi - \phi_{Z}$ is the relative azimuthal angle between charged particles coming from the underlying event and the direction of Z boson, as in Fig. 2 (left). We split the central region defined between $|\eta|<1$ as follows,

\begin{itemize}

	\item $\vert\Delta\phi\vert < 60^{\circ}$ as the toward region.
	\item $60^{\circ}<|\Delta\phi|<120^{\circ}$ as the transverse region. And,
	\item $\vert\Delta\phi\vert > 120^{\circ}$ as the away region.

\end{itemize}

As illustrated in Fig. 2 (right), we define MAX and MIN transverse regions which help to separate the hard component (initial and final-state radiation) from the beam-beam remnant component. MAX (MIN) refer to the transverse region containing largest (smallest) number of charged
particles or to the region containing the largest (smallest) scalar $p_T$sum of charged particles, on an event by event basis. One expects that the transMAX region will pick up the hardest initial or final-state radiation while both the transMAX and transMIN regions should receive beam-beam remnant contributions. Hence one expects the transMIN region to be more sensitive to the beam-beam remnant component of the underlying event, while the transMAX minus the transMIN ({\it i.e.}, transDIF) is very sensitive to hard initial and final-state radiation. This idea, was first suggested by Bryan Webber and Pino Marchesini \cite{MaxMin1}, and implemented in a paper by Jon Pumplin \cite{MaxMin2}.

\begin{figure}
     \centering
     \subfigure{
             \includegraphics[width=.135\textwidth]{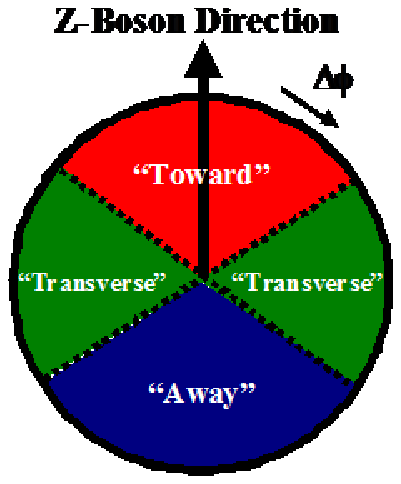}}
     \subfigure{
             \includegraphics[width=0.15\textwidth]{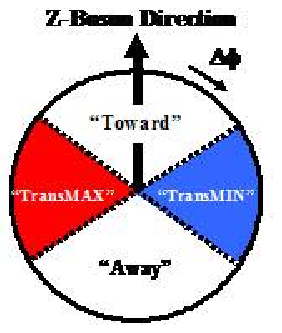}}\\
       \caption{On the left, dividing the central region, with relative to the Z boson direction and on the right, TransMAX and transMIN regions.}
\end{figure}

For hard scattered jets the transverse regions are most sensitive to underlying events, since they are perpendicular to the plane of
$2$-to-$2$ hard scattering. For Drell-Yan its easy to identify and remove leptons (since they are the colorless components) from the transverse and toward (which can not be done for dijet events, as the leading jet is itself in toward region) regions and use them to study the underlying event. So we can see that Drell-Yan events are a clean probe of the underlying events.


\section{Comparing data with QCD Monte Carlo Models}

\subsection{The underlying event as a function of lepton pair $p_T$}

We looked at the charged particles in the range $p_T > 0.5~GeV/c$ and $|\eta| <1$, at the region of Z-boson, defined as $70~GeV/c^2 < M_{ll}<110~GeV/c^2$, in the `toward', `away' and `transverse' regions, as defined in Fig. 2. The electron and muon selections are based on the standard CDF high $p_{T}$ electron and muon selection criteria. The lepton pairs are formed by oppositely charged leptons, with the requirement that both the leptons came from the same primary collision. We use a time of flight \cite{TOF} cosmic filter to eliminate cosmic muons, as the time difference between the muons recorded in the upper and lower half of the detector is expected to be very small for muons not coming from cosmic rays. The $Z\rightarrow e^{+}e^{-}$ and $Z\rightarrow \mu^{+}\mu^{-}$ data sample contains backgrounds mainly from QCD jets and W+jets. Studies \cite{BG} have shown that these backgrounds are negligible at the region of Z boson. We use the ratio of the generator level Monte Carlo result \cite{PL}  and the detector level Monte Carlo result as our correction factor for correcting the data back to the particle level. We analyzed data corresponding to the luminosity of approximately 2.7 $fb^{-1}$. 

The underlying event observables are found to be reasonably flat with the increasing lepton pair transverse momentum in the transverse and toward regions, but goes up in the away region to balance the lepton pairs.

In Fig. 3, we looked at the two observables corresponding to the underlying event, the number of charged particle density and the charged transverse momentum sum density in the transverse region, compared with {\sc pythia} \cite{PYTHIA} tunes A and AW \cite{Tunes1}, {\sc herwig} \cite{HERWIG} without MPI and a previous CDF analysis on leading jet underlying event results. We overlay the results for all the regions in Fig. 4. In Fig. 5 and 6, we look at the same observables in transMAX, transMIN, and transDIF regions. Overall, we can see that {\sc pythia} tune AW does a good job of reproducing the data. {\sc herwig} (without MPI) does not produce enough activity in the transverse region for either process. There is no final-state radiation in Z-boson production so that the lack of MPI becomes more evident. {\sc herwig} (with {\sc jimmy} \cite{Jim} MPI) agrees with tune AW for the scalar pT sum density in the toward and transMIN regions. However, it produces too much charged particle density in these regions. {\sc herwig} (with {\sc jimmy} MPI) fits the $p_T$ sum density, but it does so by producing too many charged particles (\textit{i.e.} it has too soft of a $p_T$ spectrum in these regions). This can be seen in Fig. 7 which shows the data for Z-boson events on the average charged particle $p_T$ and the average maximum charged particle $p_T$, for the transverse region compared with the QCD Monte Carlo models. So the $p_T$ distributions in the transverse region are too soft, resulting in an average $p_T$ and average $p_T$ maximum that are too small. Comparing {\sc herwig} (without MPI) with {\sc herwig} (with {\sc jimmy} MPI) clearly shows the importance of MPI in these regions.

We also compared them with leading jet underlying event results \cite{Jet UE} and observed reasonably close agreement  - which may indicate the universality of underlying event modeling. However, at low $p_T$, we see a difference. If the leading jet has no transverse momentum then there
are no charged particles, we just get min-bias events. There are a lot of low transverse momentum jets and for $p_T$ (jet\#1) $< 30~GeV/c$ the leading jet is not always the jet resulting from the hard 2-to-2 scattering. This produces is a 'bump' in the transverse density at low $p_T$ .

\begin{figure}
     \centering
     \subfigure{
             \includegraphics[width=.3\textwidth, angle =270]{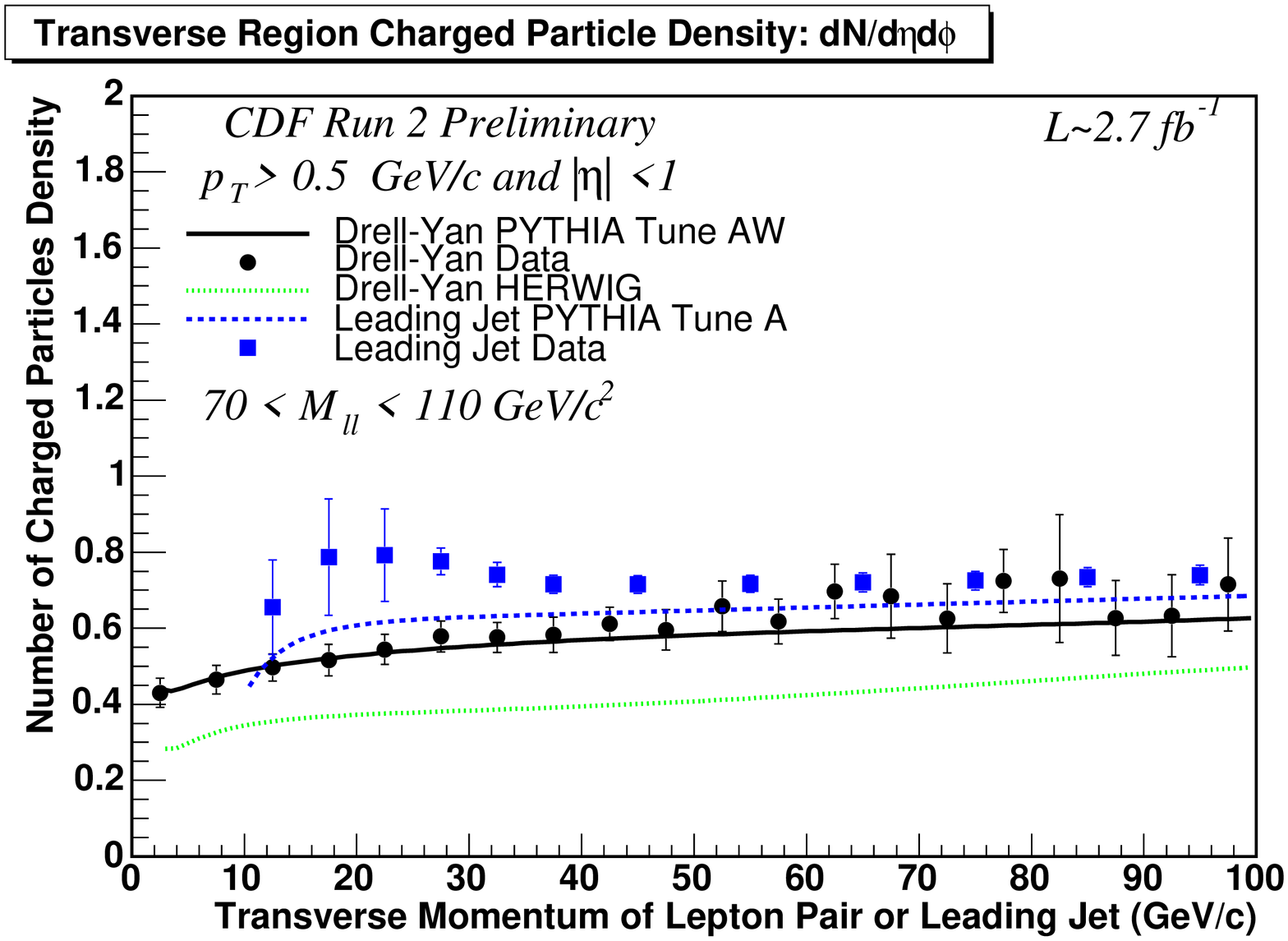}}
     \hspace{.3in}
     \subfigure{
              \includegraphics[width=.3\textwidth, angle=270]{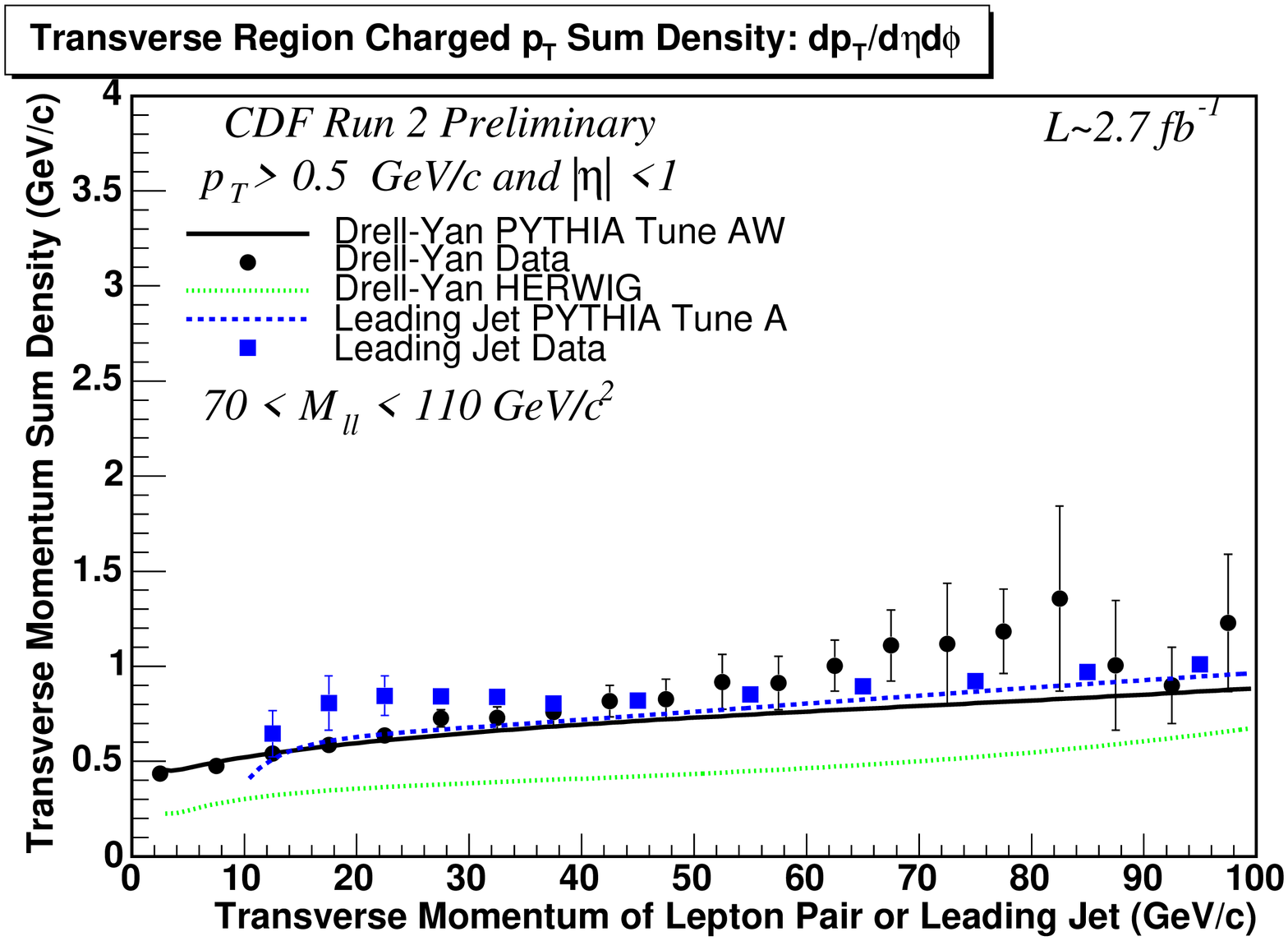}}\\
       \caption{Drell-Yan underlying event plots, charged particle multiplicity at the top and the charged $p_T$ sum at the bottom.}
\end{figure}

\begin{figure}
     \centering
     \subfigure{
             \includegraphics[width=.4\textwidth, angle =0]{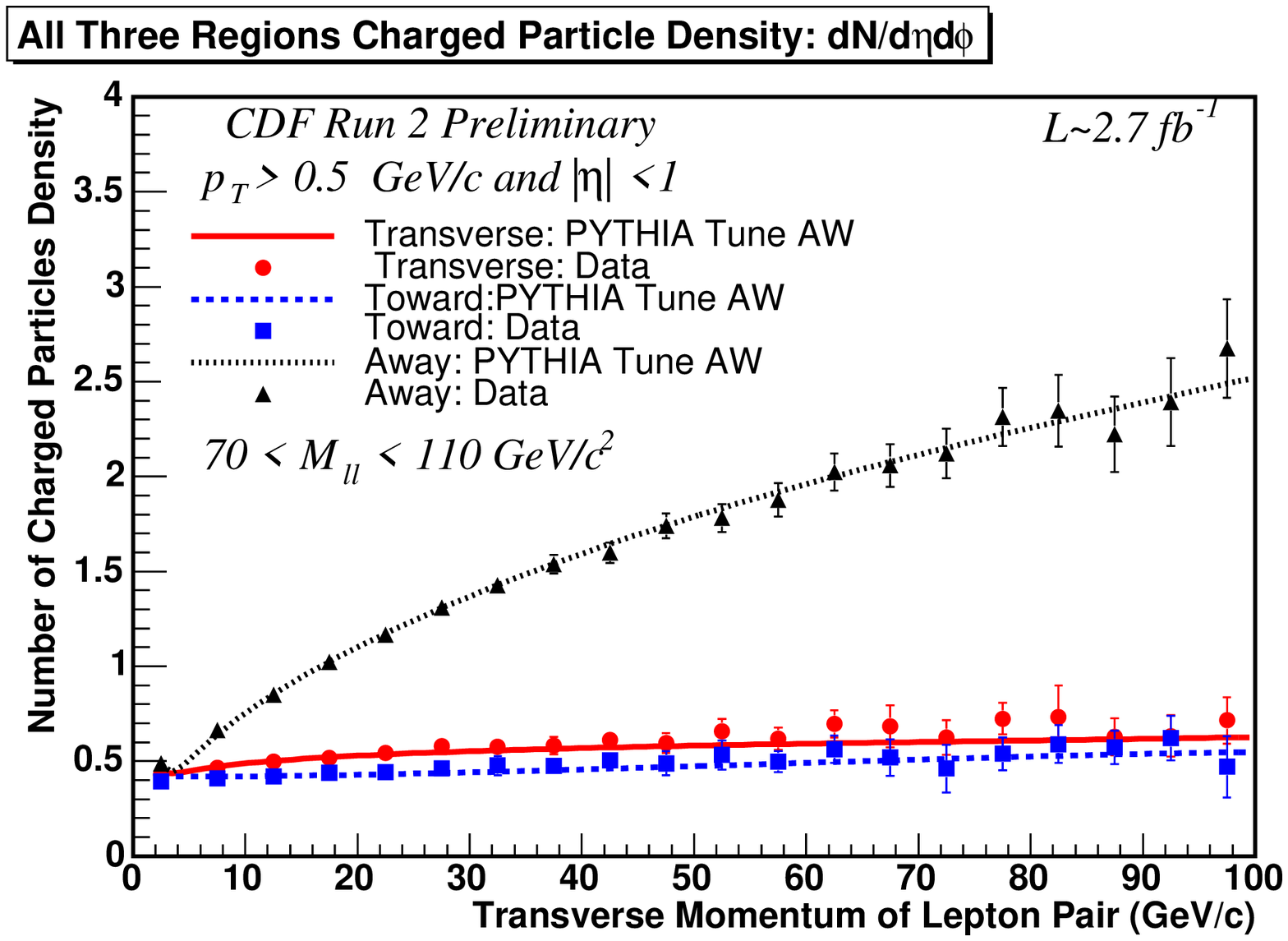}}
     \hspace{.3in}
     \subfigure{
              \includegraphics[width=.4\textwidth, angle=0]{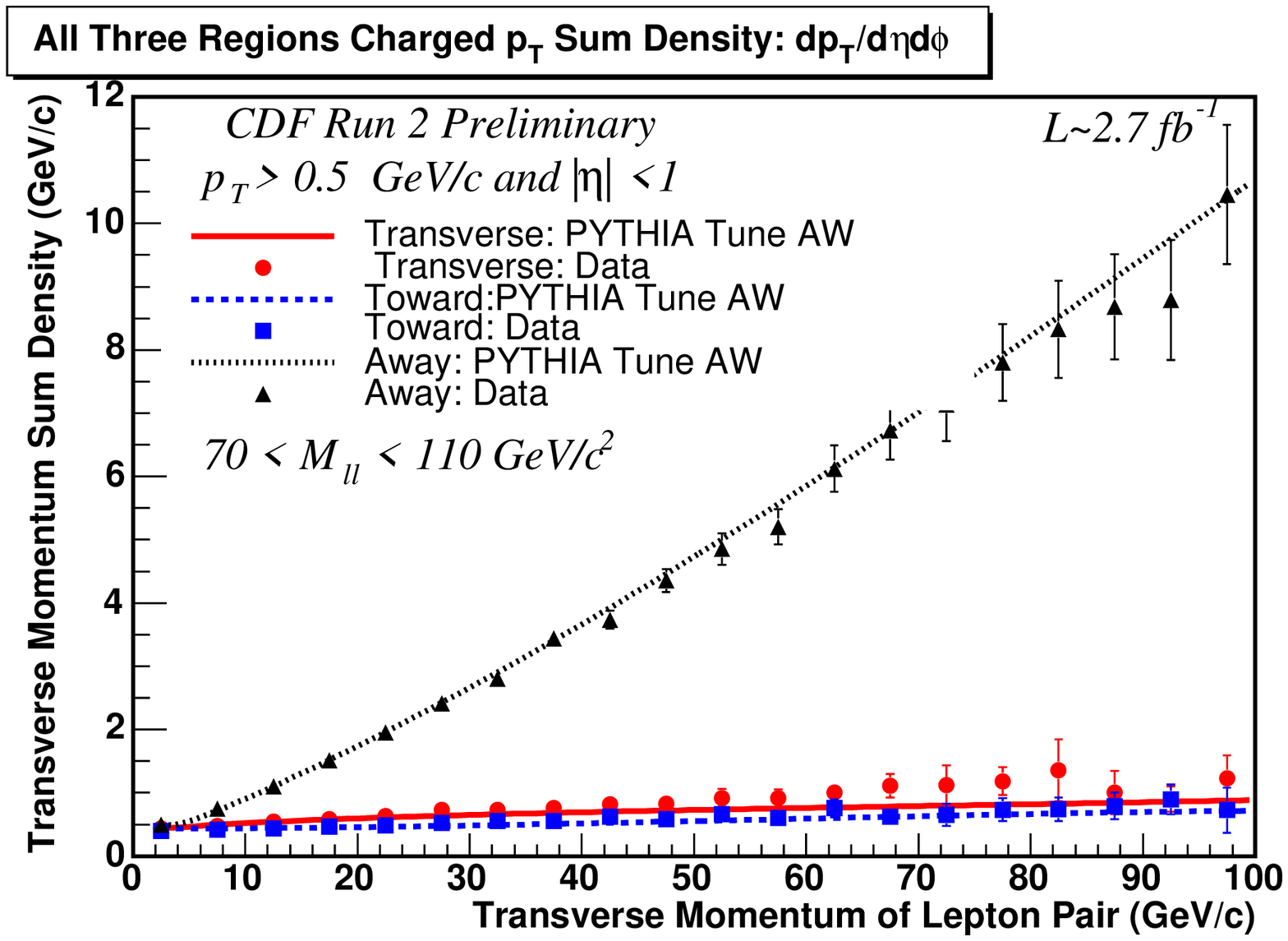}}\\
       \caption{Overlaying all three regions, charged particle multiplicity at the top and the charged $p_T$ sum at the bottom.}
\end{figure}

\begin{figure}
     \centering
     \subfigure{
             \includegraphics[width=.4\textwidth, angle =0]{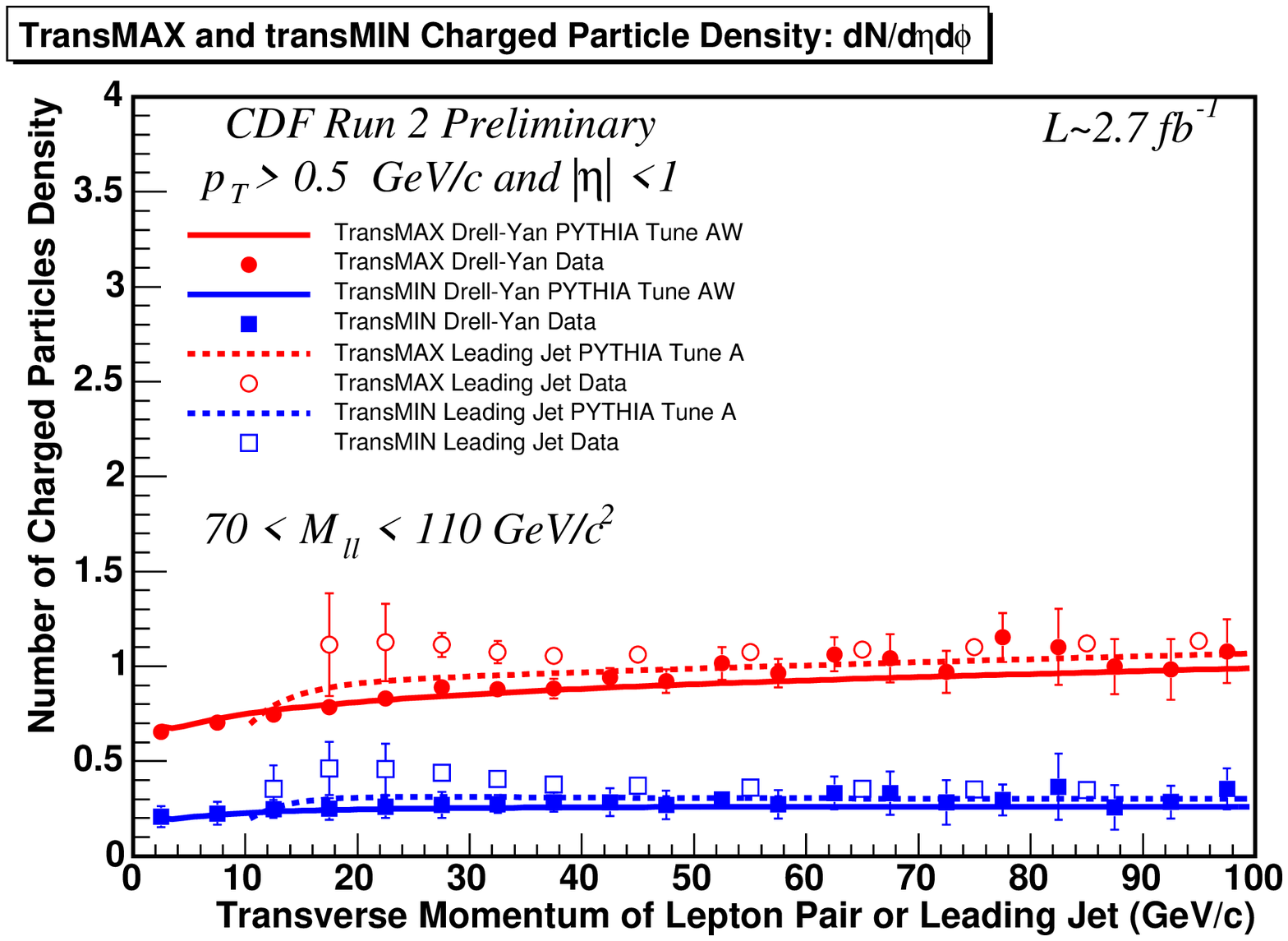}}
     \hspace{.3in}
     \subfigure{
              \includegraphics[width=.4\textwidth, angle=0]{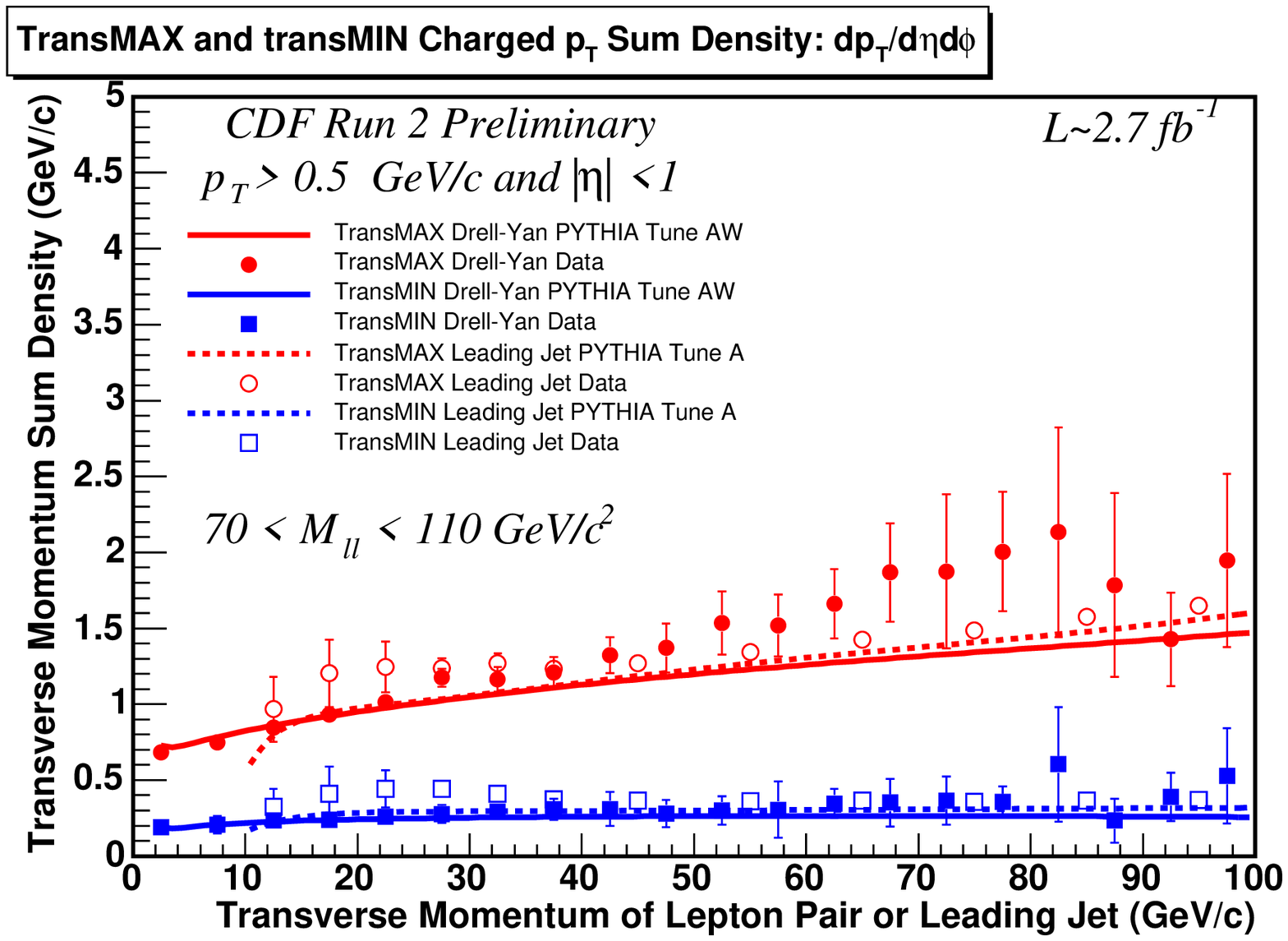}}\\
       \caption{TransMAX and transMIN regions, charged particle multiplicity at the top and the charged $p_T$ sum at the bottom.}
\end{figure}

\begin{figure}
     \centering
     \subfigure{
             \includegraphics[width=.4\textwidth, angle =0]{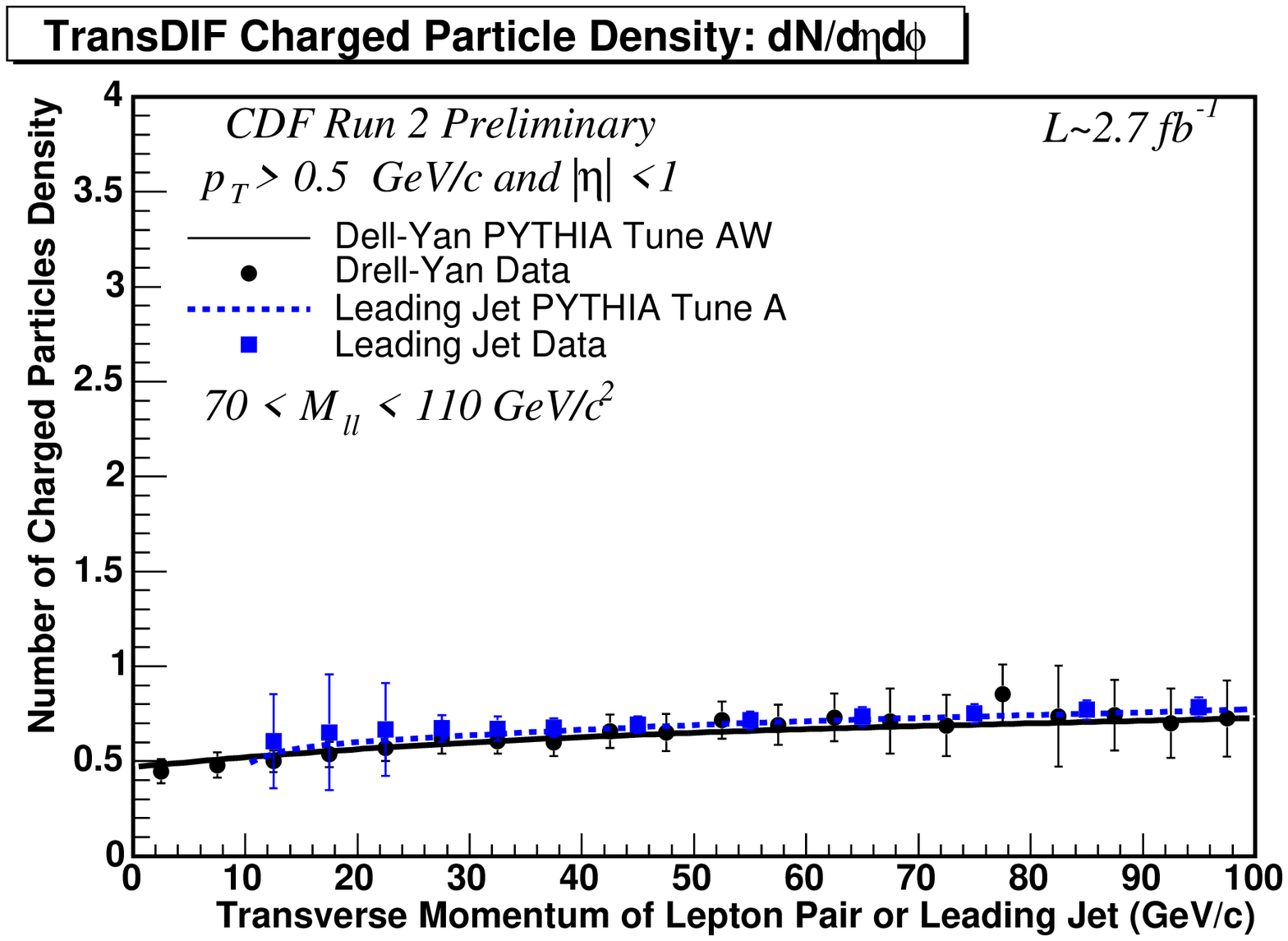}}
     \hspace{.3in}
     \subfigure{
              \includegraphics[width=.4\textwidth, angle=0]{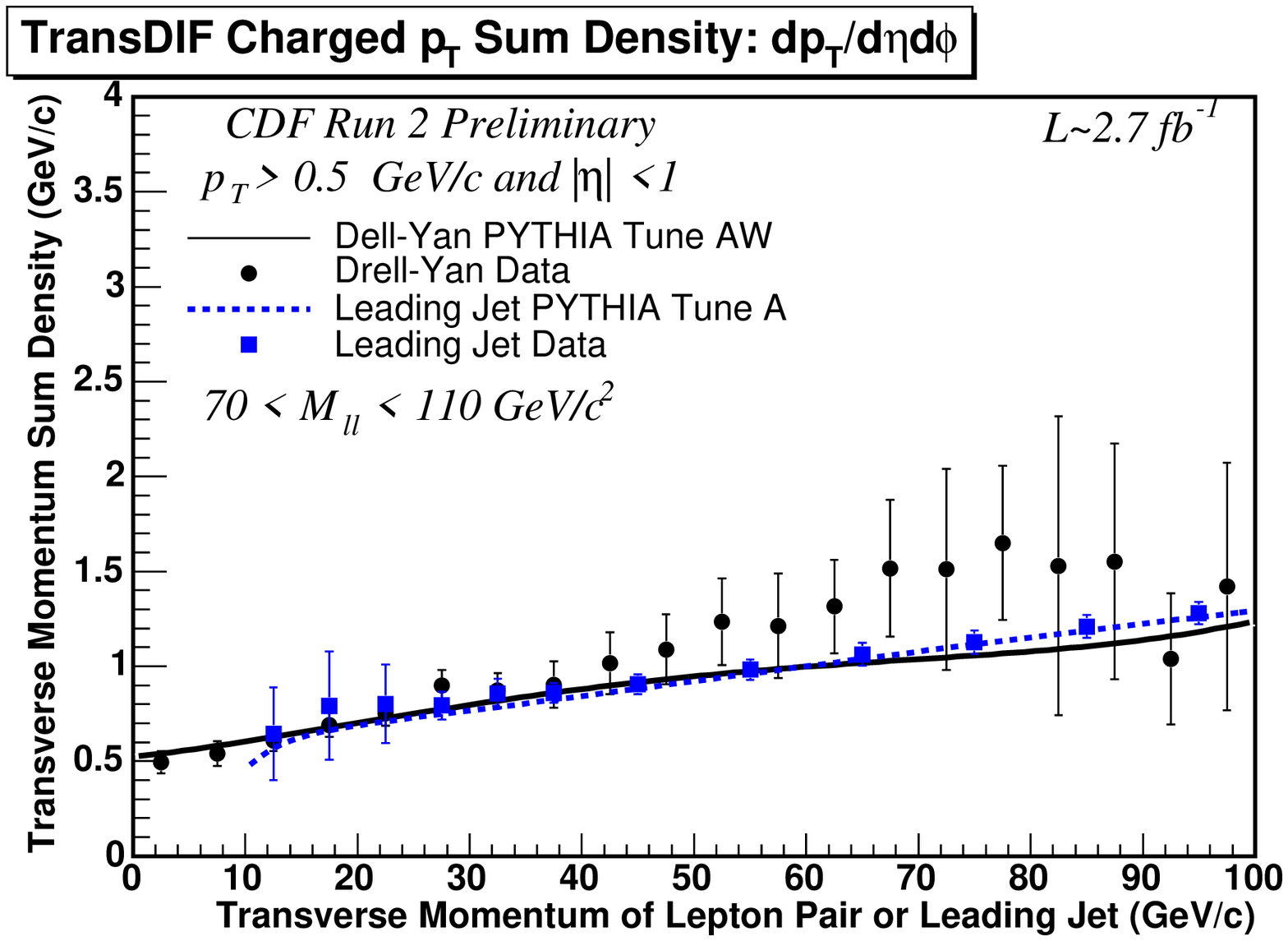}}\\
       \caption{TransDIF regions, charged particle multiplicity at the top and the charged $p_T$ sum at the bottom.}
\end{figure}

\begin{figure}
     \centering
     \subfigure{
             \includegraphics[width=.4\textwidth, angle =0]{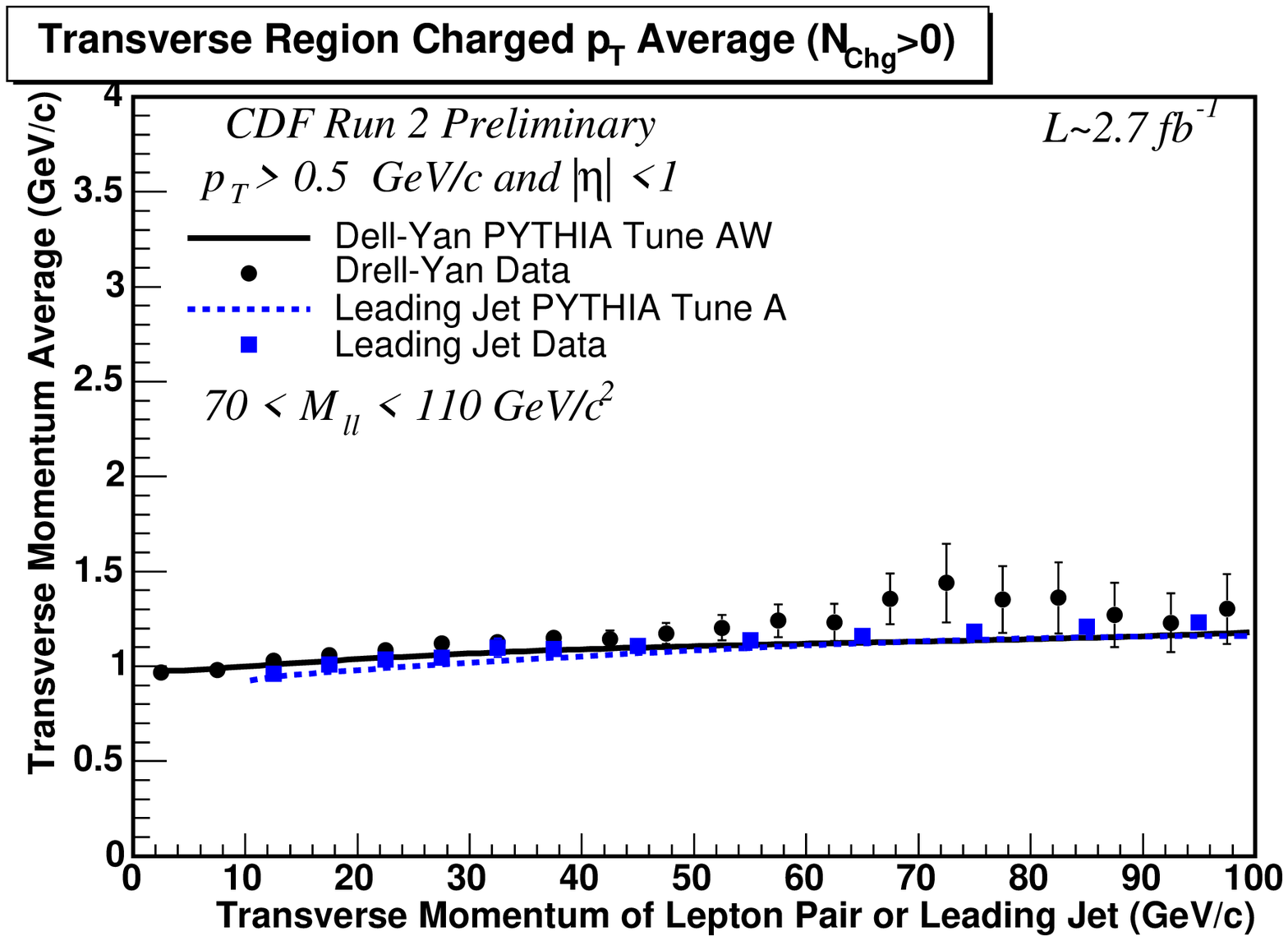}}
     \hspace{.3in}
     \subfigure{
              \includegraphics[width=.4\textwidth, angle=0]{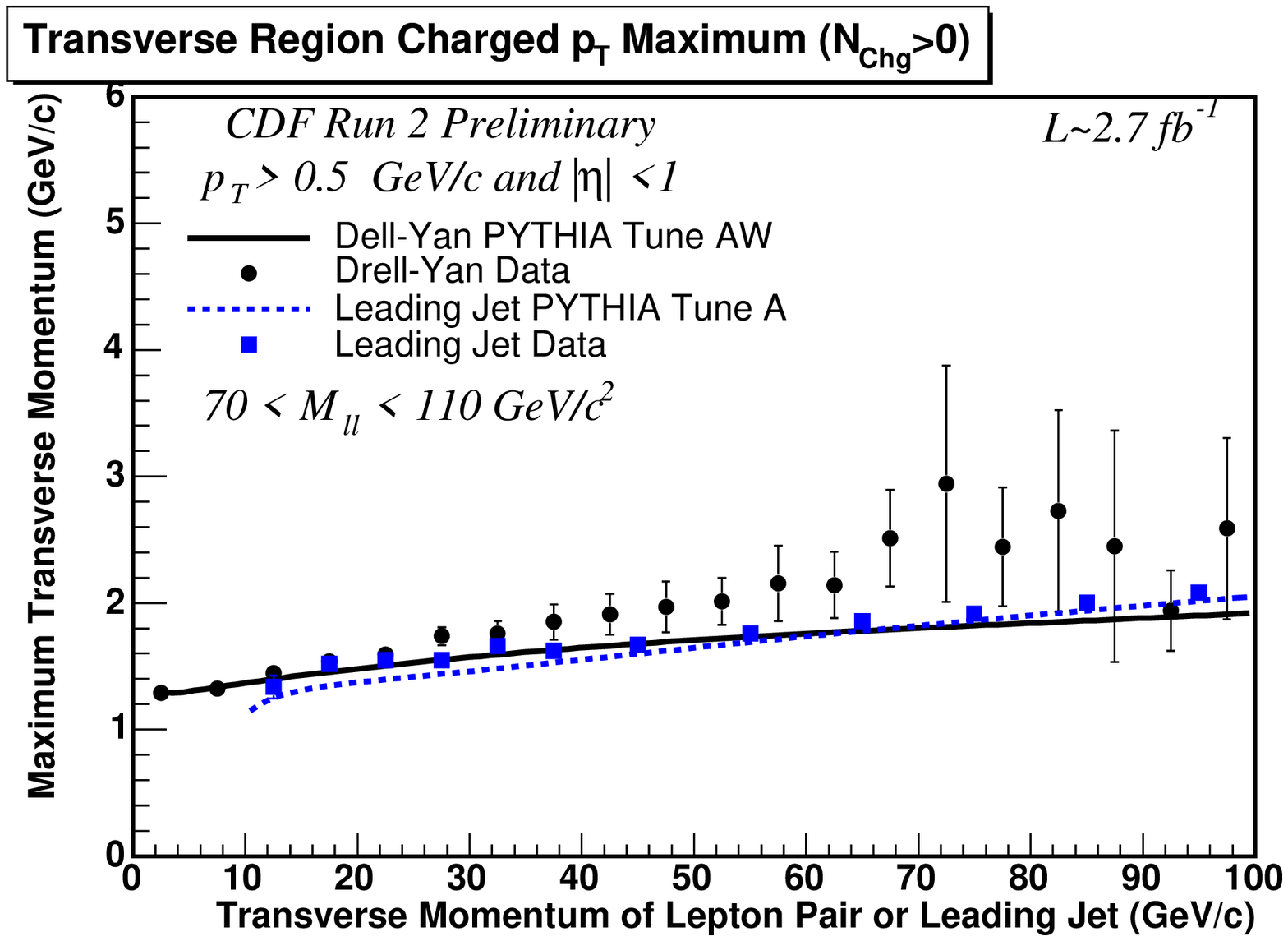}}\\
       \caption{Drell-Yan underlying event plots, charged particle $<p_T>$ at the top and the charged $p_T$ maximum at the bottom.}
\end{figure}


\subsection{Correlation studies}

The rate of change of $<p_T>$ versus charged multiplicity is a measure of the amount of hard versus soft
processes contributing to collisions and it is sensitive to the modeling of the multiple parton
interactions \cite{Corr}. This variable is the most poorly reproduced variable by the available
Monte-Carlo generators. If only the soft beam-beam remnants contributed to
min-bias collisions then $<p_T>$ would not depend on charged multiplicity. If one has two
processes contributing, one soft (beam-beam remnants) and one hard (hard 2-to-2 parton-parton
scattering), then demanding large multiplicity would preferentially select the hard process
and lead to a high $<p_T>$. However, we see that with only these two processes $<p_T>$ increases
much too rapidly as a function of multiplicity. Multiple-parton interactions
provides another mechanism for producing large multiplicities that are harder than the beam-beam
remnants, but not as hard as the primary 2-to-2 hard scattering. 

\begin{figure}
     \centering
     \subfigure{
             \includegraphics[width=.4\textwidth]{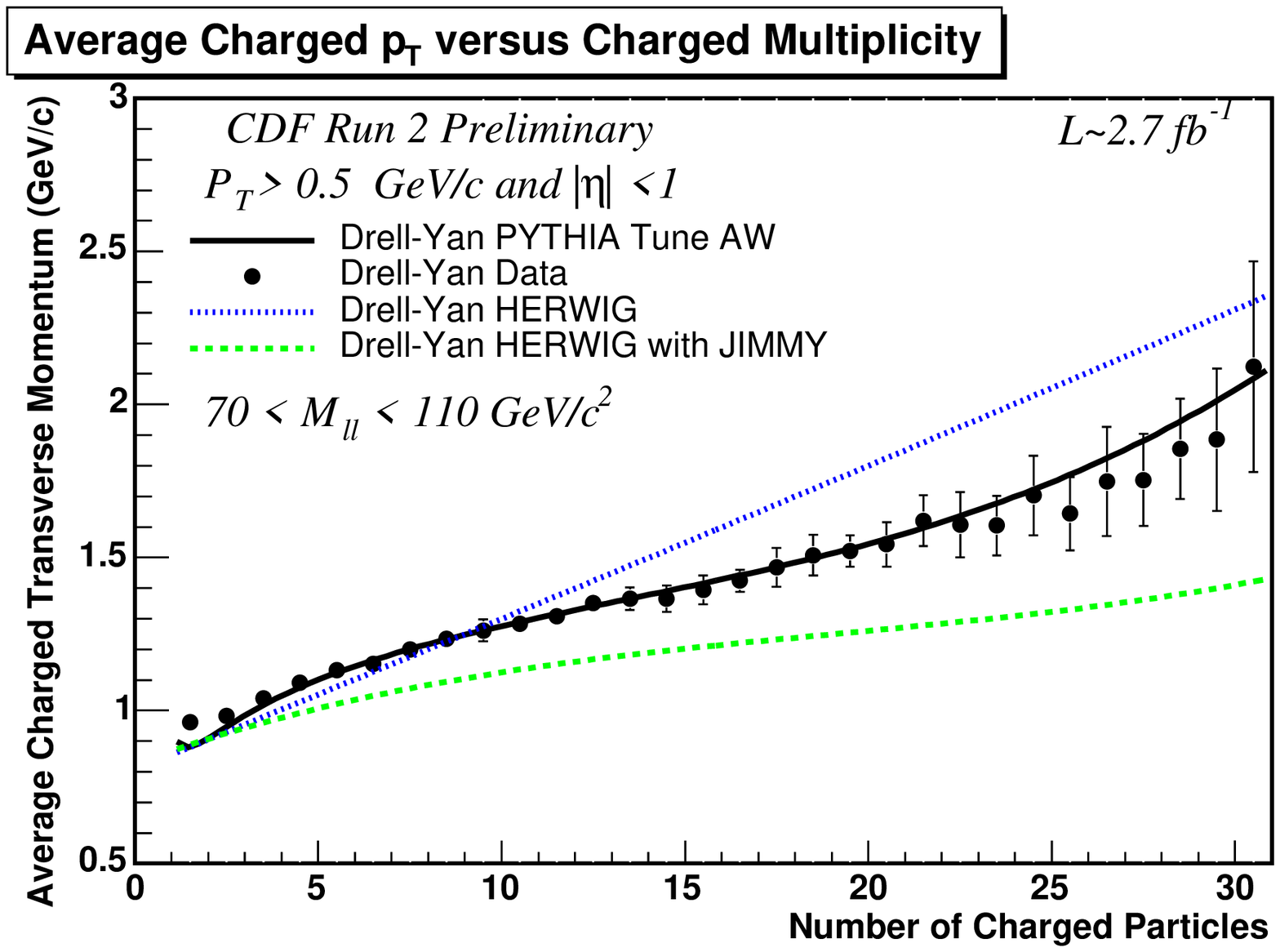}}
     \subfigure{
              \includegraphics[width=.4\textwidth]{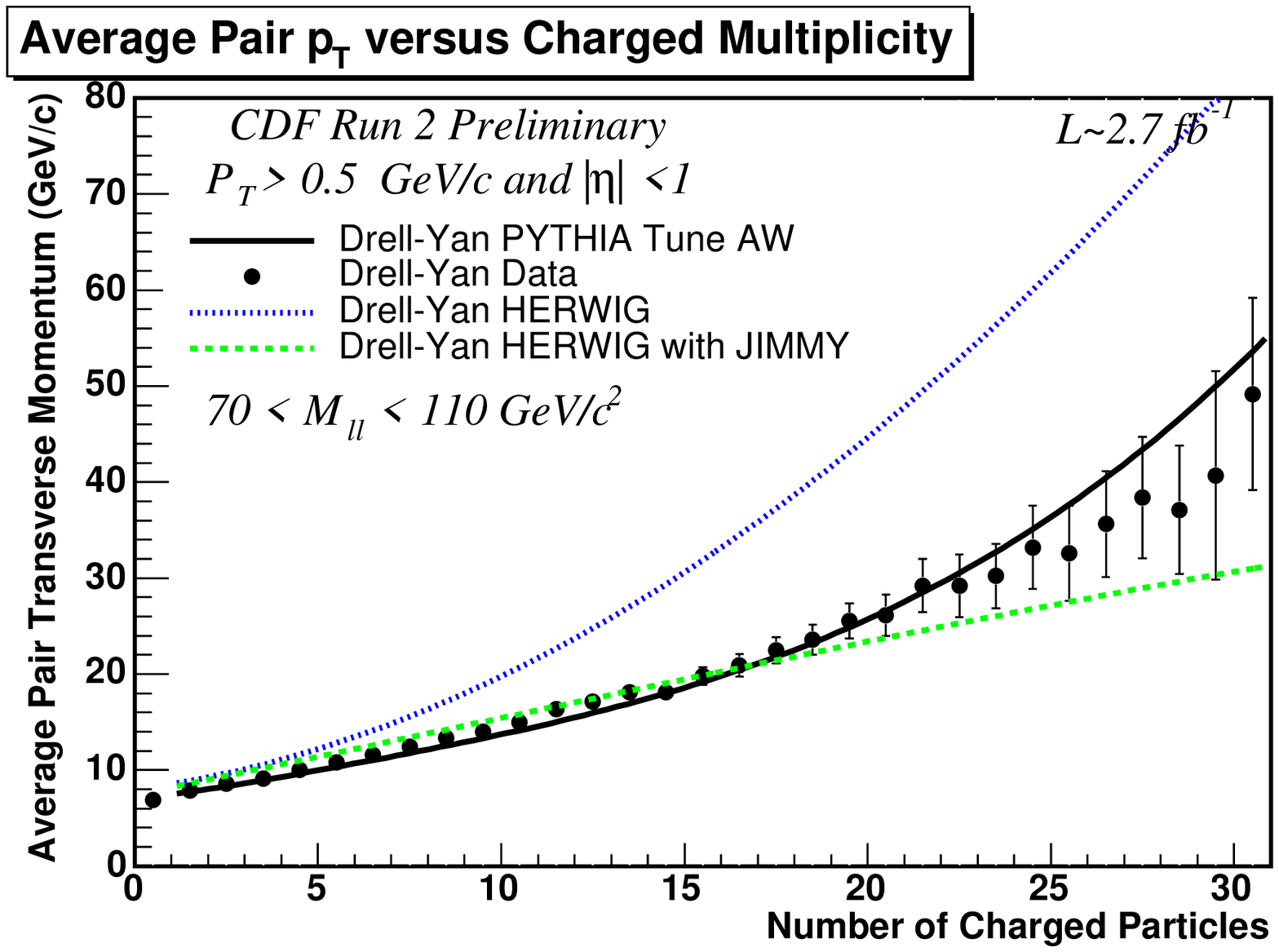}}\\
     \subfigure{
                \includegraphics[width=.4\textwidth]{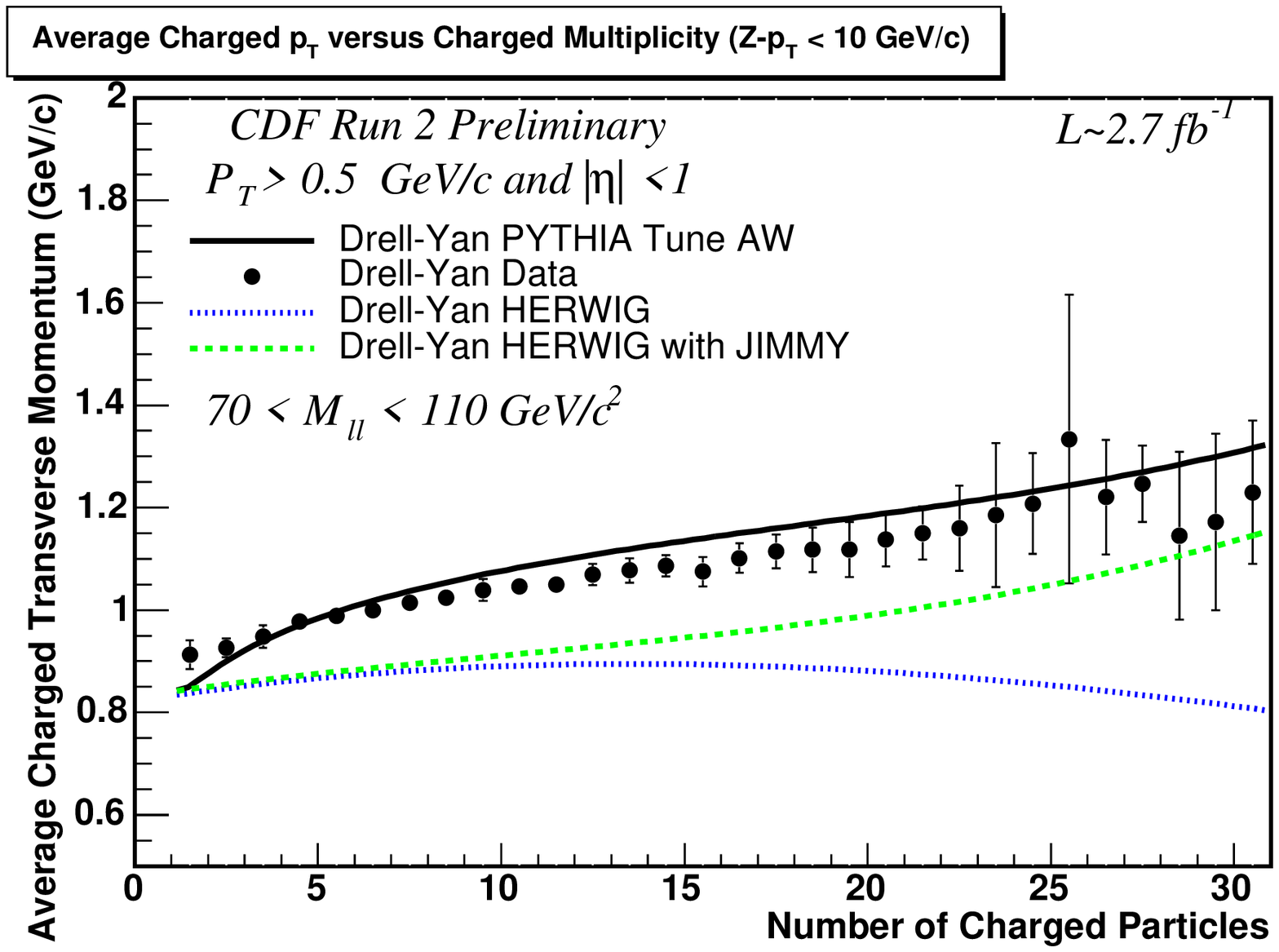}}
     \caption{Charged multiplicity against charged transverse momentum average correlation plots.}
\end{figure}

Fig. 8(top) shows the data corrected to the particle level on the average $p_T$ of
charged particles versus the multiplicity for charged particles with $p_T > 0.5~GeV/c$ and $|\eta| < 1$ for
Z-boson events from this analysis. {\sc herwig} (without MPI) predicts the $<p_T>$ to rise too
rapidly as the multiplicity increases. For {\sc herwig} (without MPI) large multiplicities come from events with a
high $p_T$ Z-boson and hence a large $p_T$ `away-side' jet. This can be seen clearly in Fig. 8(middle)
which shows the average $p_T$ of the Z-boson versus the charged multiplicity. Without MPI
the only way of getting large multiplicity is with high $p_T$(Z) events. For the models with MPI
one can get large multiplicity either from high $p_T$(Z) events or from MPI and hence $<p_T(Z)>$
does not rise as sharply with multiplicity in accord with the data. {\sc pythia} tune AW describes
the Z-boson data fairly well.
Fig. 8(bottom) shows the data corrected to the particle level on the average $p_T$ of
charged particles versus the multiplicity for charged particles with $p_T > 0.5~GeV/c$ and $|\eta| < 1$ for
Z-boson events in which $p_T(Z) < 10 ~GeV/c$. Regardless of all the improvements in the comprehension of low-$p_T$ production, the models are still unable to reproduce second order quantities such as final state particle correlations. We see that $<p_T>$ still increases as the multiplicity increases although not as fast. If we require $p_T(Z) < 10~GeV/c$, then {\sc herwig} (without MPI) predicts that the $<p_T>$ decreases slightly as the multiplicity increases. This is because without MPI and without the high $p_T$ `away-side' jet which is suppressed by requiring low $p_T(Z)$, large
multiplicities come from events with a lot of initial-state radiation and the particles coming from
initial-state radiation are `soft'. {\sc pythia} tune AW describes the behavior of $<p_T>$ versus the multiplicity fairly well even when we select $p_T(Z) < 10~GeV/c$. This strongly suggests that MPI are playing an important role in both these processes.

\section{Summary and Conclusions}

We are making good progress in understanding and modeling the softer physics. CDF tunes A and AW describe the data very well, although we still do not yet have a perfect fit to all the features of the CDF underlying event data. Future studies should focus on tuning the energy dependence for the event activity in the underlying event, which at the moment seems to be one of the least understood aspects of all the models. The underlying event is expected to be much more active in LHC and it is critical to have sensible underlying event models containing our best physical knowledge and intuition, tuned to all relevant available data.


%% file: dasgupta.tex
\begin{center}
{\Large \bf Hadronisation corrections to jets in the $k_t$ algorithm}
\vspace*{1cm}

Mrinal Dasgupta
\vspace*{0.5cm}

School of Physics and Astronomy, University of Manchester \\
Oxford Road, Manchester M13 9PL, U.K.\\

\vspace*{1cm}

Yazid Delenda

\vspace*{0.5cm}
D\'epartement de Physique, Facult\'e des Sciences\\
        Universit\'e de Batna, Algeria.\\

\vspace*{1cm}
\end{center}

\begin{abstract}
We improve  previously derived analytical estimates of
hadronisation corrections to QCD jets at hadron colliders, firmly
establishing at the two-loop level the link to the well-known power
corrections to LEP event-shape variables. The results of this paper
apply to jets defined in the $k_t$ and anti-$k_t$ algorithms but the
general framework presented here holds also for other algorithms for
which calculations are in progress.
\end{abstract}

\setcounter{section}{0}
\setcounter{figure}{0}
\setcounter{table}{0}
\setcounter{equation}{0}

An understanding of QCD jets and their properties will be integral
to the success of the LHC physics program. In particular one of the
most important issues in current jet studies is the question of jet
energy scale. The shift induced in jet energy by effects such as
perturbative radiation and non-perturbative effects like
hadronisation and the underlying event would contribute to a
smearing of, for instance, mass peaks that may be a signal for new
physics. Thus in order to choose optimal jet definitions that
minimise such smearing one would need to know the dependence of the
above effects on the experimental parameters like jet radius.
Moreover, even in pure QCD studies such as the extraction of parton
distribution functions (pdfs) and the strong coupling $\alpha_s$
from jet observables like the inclusive jet cross-sections, a
knowledge of the non-perturbative contribution is important to
supplement perturbative calculations. A relatively small shift in
the transverse momentum $p_t$ of a jet, induced by hadronisation,
can result in a significant change in the inclusive jet spectrum
since we are dealing with a quantity that has a steeply falling
$p_t$ distribution.

While it is traditional to study the hadronisation contribution via
Monte Carlo models such as those in {\tt HERWIG} and {\tt PYTHIA},
it turns out that in cases like the jet energy there is additionally
valuable analytical insight available \cite{Dasgupta:2007wa}.
Analytical models based on renormalons \cite{Beneke:1998ui} have in
the past been met with great success in the description of LEP and
HERA event-shape variables \cite{Dasgupta:2003iq}, but have not
really been utilised outside that context. In
Ref.~\cite{Dasgupta:2007wa} one such model (due to Dokshitzer and
Webber \cite{Dokshitzer:1995zt}) was used to estimate hadronisation
corrections to jet transverse momentum $p_t$. The result found there
was striking: hadronisation effects have a singular $1/R$ dependence
on the jet radius $R$, at small $R$. This is in complete contrast to
the contribution from the underlying event which varies as $R^2$.
The knowledge of the $R$ dependence of non-perturbative effects in
conjunction with the $\ln R$ behaviour involved in perturbative
estimates can then be used to arrive at conclusions about the
optimal values of $R$ to be used in diverse studies involving jets,
as exemplified in Ref.~\cite{Dasgupta:2007wa}.

While the computations of Ref.~\cite{Dasgupta:2007wa} indicate the
dependence of hadronisation corrections and the underlying event on
$R$, there remains the question of the overall magnitude of these
effects. While for the underlying event one is reliant solely on
Monte Carlo event generators to obtain the overall magnitude, for
the hadronisation correction a tentative link was made in
Ref.~\cite{Dasgupta:2007wa} between the magnitude of the $1/R$
correction and that of $1/Q$ corrections to LEP and HERA event
shapes such as the thrust distribution (see \cite{Dasgupta:2003iq}
for a review). In order to definitely link the magnitude of jet
hadronisation to that of event-shape power corrections one needs to
carry out a calculation at the two-loop level rather than the simple
one-loop estimate reported in \cite{Dasgupta:2007wa}. The
calculation for jets defined in the $k_t$ algorithm \cite{ESkt,inclukt} 
is reported here while the corresponding result for jets in the anti-$k_t$ algorithm
 is already known \cite{Cacciari:2008gp}. Work on the other jet
algorithms is currently in progress.

\section{The single-gluon result}

In the Dokshitzer-Webber model non-perturbative hadronisation
corrections are associated to the emission of a soft gluon with
transverse momentum $k_t \sim \Lambda_{\mathrm{QCD}}$. For the jet
$p_t$ case we work out the change in transverse momentum $\delta
p_t$ induced by the emission of such a gluon and combined with the
gluon emission probability (as given by \emph{perturbative} QCD)
this yields the average shift in $p_t$ induced by hadronisation:
\begin{equation}
\label{eq:oneor_delenda} \langle \delta p_t \rangle^h \sim
\frac{C_j}{ 2 \pi} \int \frac{dk_t}{k_t} d\eta \frac{d\phi}{2\pi}\,
\delta p_t(k) \, \alpha_s(k_t),
\end{equation}
where $k_t$, $\eta$ and $\phi$ respectively denote the transverse
momentum, rapidity and azimuth of the emitted gluon with respect to
the emitting hard parton (jet) and $C_j$ is the colour factor
associated to emission from the given hard parton. The reason we are
able to single out a given hard parton initiating a high-$p_t$ jet
in any hard process is essentially since the leading $1/R$ result
stems from emission collinear to the triggered jet. It is thus
possible to talk in terms of the single-jet limit ignoring the rest
of the details of the hard process.

The only non-perturbative ingredient that is involved above is the
value of $\alpha_s(k_t)$ at scales around or below $\Lambda_{
\mathrm{QCD}}$. If one makes the assumption of a universal
infrared-finite coupling, which replaces the perturbative coupling
that has an unphysical divergence at $\Lambda_{\mathrm{QCD}}$, then
one arrives at a prediction for the hadronisation correction
$\langle \delta p_t \rangle^h$. Using the fact
\cite{Dasgupta:2007wa} that the $\delta p_t(k)$ is essentially the
energy of the gluon emitted outside the jet we perform the integral
over rapidity in Eq.~\eqref{eq:oneor_delenda} to obtain the leading
$1/R$ behaviour. The result is of the form $c \, {\mathcal{A}}/R $,
where $c$ is a number obtained from the rapidity integral and
${\mathcal{A}}$ is the moment of the coupling $\alpha_s(k_t)$ over
the infrared region (we refer the reader to
Refs.~\cite{Dokshitzer:1997iz,Dokshitzer:1998pt} for the precise
details). Since the same coupling moment enters the predictions for
event-shape variables we can take its value from data on event
shapes and hence obtain a numerical prediction for the leading $1/R$
hadronisation correction to jet $p_t$. This was the method adopted
in Ref.~\cite{Dasgupta:2007wa}.

Here we point out a limitation of the above approach
\cite{Nason:1995hd} which is that while we have written down and
used a running coupling $\alpha_s (k_t)$, this quantity only emerges
when one considers not just the emission of a single gluon but in
fact gluon decay as well. To be precise an inclusive integration
over gluon decay products is responsible for building up the
quantity $\alpha_s(k_t)$. Unfortunately, as is known for event-shape
variables, our observable is sensitive to the precise details of
gluon branching and hence one is not free to carry out such an
inclusive integration. One must therefore return to the details of
the gluon branching and identify the correction to the above
inclusive approximation. The analysis at this level has already been
carried out for event-shape variables
\cite{Dokshitzer:1997iz,Dokshitzer:1998pt,Dasgupta:1999mb,Dasgupta:1998xt}
and below we report on it for the jet $p_t$ case.

\section{Non-perturbative effects and gluon decay}

\begin{figure}
\includegraphics{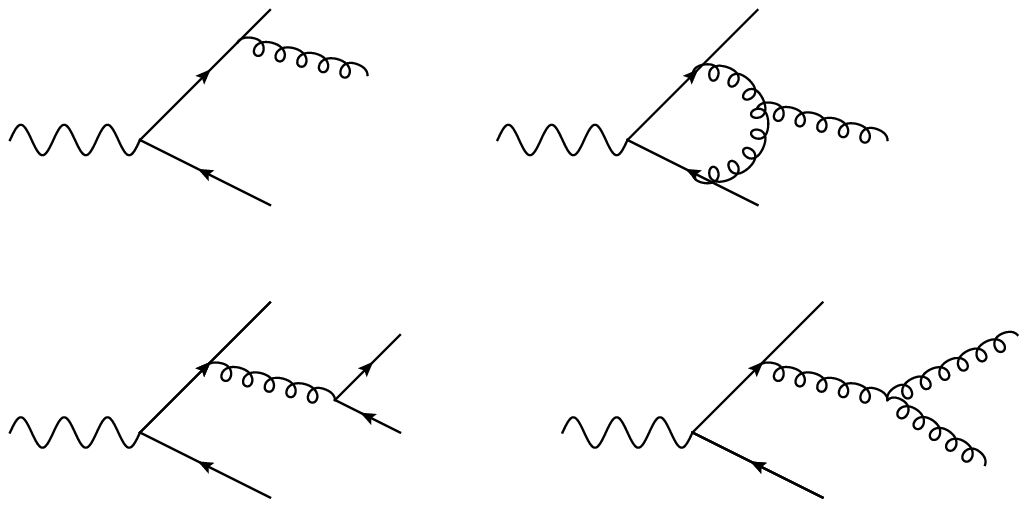}
\caption{Gluon decay and one-loop corrections to single-gluon
emission.}\label{fig:nlo_delenda}
\end{figure}

Now we consider the situation where the emitted gluon with $k_t \sim
\Lambda_{\mathrm{QCD}}$ is allowed to decay and at the same accuracy
account for virtual corrections to single gluon emission, as
depicted in Fig.~\ref{fig:nlo_delenda}.

At this two-loop level the change in $p_t$ (for a quark jet) can be
expressed as:
\begin{multline}
\langle \delta p_t \rangle^h = \frac{C_F}{\pi} \int \frac{d^2
k_t}{\pi k_t^2} \frac{d\alpha}{\alpha} \left \{\alpha_s(0)+4 \pi
\chi (k_t^2) \right \} \delta p_t (k) \\ +4 C_F \int \left (
\frac{\alpha_s}{4 \pi} \right)^2 d \Gamma_2 \frac{M^2}{2!} \delta
p_t (k_1,k_2),\nonumber
\end{multline}
where $\alpha$ is a Sudakov variable, $\alpha_s(0)$ is an
ill-defined quantity which will cancel away subsequently, $\chi$
represents the virtual correction to gluon emission, $d\Gamma_2$ is
the gluon decay phase-space and $M^2$ is the decay matrix element
\cite{Dokshitzer:1997iz,Dokshitzer:1998pt}. We also denote by
$\delta p_t(k_1,k_2)$ the change in $p_t$ due to correlated
two-parton emission while $\delta p_t(k)$ is the corresponding
single-gluon quantity. To correctly account for gluon branching one
thus has to perform the above calculation, the details of which are
reported in Ref.~\cite{Dasgupta:2009tm}.

The analogous two-loop analysis for event-shape variables
\cite{Dokshitzer:1997iz, Dokshitzer:1998pt, Dasgupta:1999mb,
Dasgupta:1998xt} revealed an initially surprising result -- the
two-loop correction simply provided a \emph{universal} rescaling
factor to the one-gluon result, which became known as the
\emph{Milan factor}. Its value for $n_f=3$ (which is the number of
flavours excited in the relevant soft region) was found to be
$\mathcal{M} = 1.49$. Thus the ratio of corrections to two event
shapes $v_1$ and $v_2$ was merely the ratio of the one-loop
coefficients computed previously \cite{Dokshitzer:1995zt}:
\begin{equation*}
\frac{\delta v_2^{\mathrm{NP}}}{\delta v_1^{\mathrm{NP}}}=
\frac{\delta v_2^{\mathrm{NP},1}}{ \delta v_1^{\mathrm{NP},1}},
\end{equation*}
where $\delta v_1^{\mathrm{NP},1}$ denotes the non-perturbative
single-gluon correction for $v_1$ computed as discussed in the
preceding section and likewise for $v_2$. This remarkable result was
understood to arise as a consequence of the fact that all the
variables considered could be expressed as linear sums over the
transverse momenta $k_{ti}$ of emissions, $v =\sum_i k_{ti} c_i$,
where the $c_i$ are rapidity-dependent coefficients
\cite{Dokshitzer:1998pt}.

\begin{figure}
\includegraphics{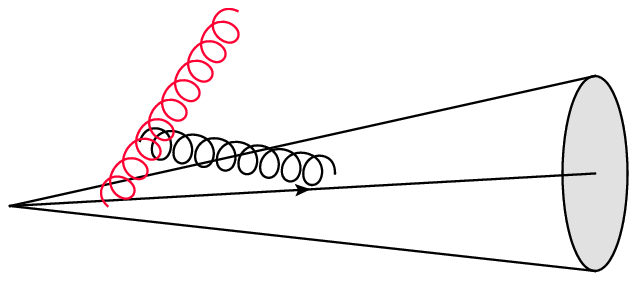}
\caption{The role of clustering in determining whether a soft gluon
ends up within or outside a hard jet.}\label{fig:cluster_delenda}
\end{figure}

In the case of jet $p_t$ this linear dependence is ruined by the
non-trivial action of the jet algorithm in all cases except the case
of jets defined in the anti-$k_t$ algorithm \cite{Cacciari:2008gp}.
The contribution to the jet $\delta p_t$ of a given emission is
found to be of the form $k_t e^\eta\, \Xi_{\mathrm{out}}$, where
$\eta$ denotes the rapidity with respect to the emitting hard jet
and $\Xi_{\mathrm{out}}$ denotes the condition that the emission
ends up outside the jet \emph{after the application of the jet
algorithm}. It should be immediately clear from this that in most
current sensible jet algorithms (both of sequential recombination
and cone type) the condition $\Xi_{\mathrm{out}}$ is non-trivial and
introduces non-linearity in $k_t$. For instance in the $k_t$
algorithm we can consider the situation in
Fig.~\ref{fig:cluster_delenda}, where although one may have a soft
parton separated by more than a certain distance $R$ in rapidity and
azimuth (denoted by the red gluon line) from a given hard parton, it
may be clustered to another soft parton (denoted by the black gluon
line) and hence swept into the final jet. This clustering depends on
the $k_t$ of a soft parton relative to the other partons and hence
the condition $\Xi_{\mathrm{out}}$ derived in \cite{Dasgupta:2009tm}
contains dependence on the $k_t$ of the soft partons, spoiling the
simple linear dependence needed for universality.

An exception to the above situation is to be found in the anti-$k_t$
algorithm for which the condition $\Xi_{\mathrm{out}} = \Theta
(\eta^2+\phi^2-R^2)$ ensures that a given parton is outside the jet
if its angular ($\eta,\phi$) separation from the hard jet is more
than $R$. The linear dependence on $k_t$ is maintained and the Milan
factor $\mathcal{M}=1.49$ is computed as for event shapes.

For the $k_t$ algorithm we have carried out an equivalent
calculation for the leading $1/R$ hadronisation correction (at small
$R$) with the more complicated $\Xi_{\mathrm{out}}$ function
involved there and we found the result $\mathcal{M}_{k_t}=1.01 \,
(n_f=3)$. Thus while at the level of the one-gluon studies of
Ref.~\cite{Dasgupta:2007wa} the $k_t$ and anti-$k_t$ algorithms
received identical hadronisation corrections, a detailed analysis at
the two-loop level breaks this equality. One finds that the ratio of
hadronisation corrections is then:
\begin{equation}
\frac{{\langle \delta p_t \rangle}^h_{k_t}}{{\langle\delta p_t
\rangle}^h_{\mathrm{anti-}k_t}} = \frac{1.01}{1.49} \sim
0.7\,.\nonumber
\end{equation}

Thus one expects somewhat smaller hadronisation corrections for the
$k_t$ algorithm as compared to those for the anti-$k_t$ algorithm
which is also borne out by the Monte Carlo studies with {\tt HERWIG}
and {\tt PYTHIA} reported in Ref.~\cite{Dasgupta:2007wa}. We remind
the reader that these conclusions apply only to the $1/R$
hadronisation corrections that would be dominant at small $R$ and we
neglect finite $R$ corrections which need to be considered alongside
the underlying event contribution which also has a regular $R$
dependence $\sim R^2$.

\section{Conclusions}

We have reported on a study of hadronisation corrections to jet
$p_t$ or energy scale based on two-loop extensions of the one-gluon
estimates reported in Ref.~\cite{Dasgupta:2007wa},  with jets
defined in the $k_t$ algorithm. Studies for other jet algorithms
(SISCone \cite{Salam:2007xv} and Cambridge/Aachen
\cite{Dokshitzer:1997}) are in progress.


\section{Acknowledgments}
We wish to thank the organizers and convenors of the UCL workshop on ``Standard model discoveries with early LHC data'' for a very pleasant and stimulating 
meeting.

\bibliographystyle{plain}
\bibliography{dasgupta}

%